%% file: main.tex
\documentclass[12pt,a4paper,twoside,openright]{book}
\usepackage[utf8]{inputenc}
\usepackage[english]{babel}
\usepackage{amsmath}
\usepackage{amsfonts}
\usepackage{amssymb}
\usepackage{amsthm}
\usepackage[pagebackref=true]{hyperref}
\usepackage{graphicx}
\usepackage{epstopdf}
\usepackage{latexsym}
\usepackage{braket}
\usepackage{feynmp-auto}
\usepackage[utf8]{inputenc}
\usepackage{amsmath}
\usepackage{amsfonts}
\usepackage{amssymb}
\usepackage{slashed}
\usepackage{lscape}
\usepackage[gen]{eurosym}
\usepackage{geometry}
\usepackage{caption}
\usepackage{subfig}
\usepackage{physics}
\usepackage{cite}
\usepackage{indentfirst}
\usepackage{bm}
\usepackage{nomencl}
\usepackage{hyperref}
\newcommand{\ee}{\mathrm{e}}
\newcommand{\ii}{\mathrm{i}}

\newcommand{\on}{\operatorname}
\newcommand{\arccosh}{\on{arccosh}}

\newcommand{\Ai}{\on{Ai}}
\newcommand{\Bi}{\on{Bi}}

\numberwithin{equation}{section}
\graphicspath{{/}}
\usepackage{tikz}
\usepackage{hyperref}
  \DeclareUnicodeCharacter{200B}{{\hskip 0pt}}
\allowdisplaybreaks
\usepackage{xcolor}
\usepackage{pgf}
\usetikzlibrary{intersections,calc,backgrounds,through,decorations.markings,3d,decorations.pathmorphing,trees}

\tikzset{
    photon/.style={decorate, decoration={snake}, draw=red},
    particle/.style={draw=blue, postaction={decorate},
        decoration={markings,mark=at position .5 with {\arrow[draw=blue]{>}}}},
    antiparticle/.style={draw=blue, postaction={decorate},
        decoration={markings,mark=at position .5 with {\arrow[draw=blue]{<}}}},
    gluon/.style={decorate, draw=black,
        decoration={snake,amplitude=4pt, segment length=5pt}}
     }
\usepackage{tabularx}
\usepackage{afterpage}

\newcommand\blankpage{%
    \null
    \thispagestyle{empty}%
    \addtocounter{page}{-1}%
    \newpage}

\begin{document}

\frontmatter
\include{cover.tex}
\afterpage{\blankpage}
\include{cover2.tex}
\pagenumbering{Roman}

\include{abstracteng.tex}
\include{abstractpor.tex}

\include{0ack.tex}
\tableofcontents
\mainmatter
\include{1intro.tex}
\include{2findiffschrod}
\include{3connformulae}
\include{4connformulaecomp}
\include{5applications}
\include{conclusion}

\appendix
\numberwithin{equation}{section}
\include{appnum}
\include{appbox}

\backmatter

\bibliographystyle{plain}

\end{document}

%% file: cover.tex
\setcounter{page}{1} \pagenumbering{Alph}

% Add PDF bookmark 
%%% LOGO
\thispagestyle{empty}
 
 %%% Instituição
\begin{center}
\LARGE \textbf{UNIVERSIDADE DE LISBOA \\ INSTITUTO SUPERIOR TÉCNICO}
%%% espaço sem gráficos
\vspace{10mm}

%%% Optional Image
%\vspace{10mm}
%~\\ \vspace{50mm} % gráficos
%\\ \begin{center} \includegraphics[height=50mm]{Cover/coverimage}  \end{center} % gráficos
 
 %%% Tituloì

\LARGE \textbf{WKB Methods for Finite Difference Schrödinger Equations}
\\ \vspace{10mm}  % NO SUBTITLE
\Large \textbf{Salvatore Baldino} \\
\vspace{2cm}
\begin{center}
\small{\textbf{Supervisor: Doctor Ricardo Pina Schiappa de Carvalho}}
\end{center}
\vspace{10mm}
%\vspace{12mm}
\centering
\large \textbf{Thesis approved in public session to obtain the PhD degree in Mathematics}\\
%\\ \vspace{2mm}
\vspace{10mm}
\Large \textbf{Jury final classification: approved with distinction}
 
\vspace{5mm}

%\large \textbf{\todaythesis\today} \\

\large \textbf{2023} \\
\end{center}
\let\thepage\relax
\pagebreak

%% file: cover2.tex
\setcounter{page}{1} \pagenumbering{Alph}
\vspace{2cm}
\newgeometry{bottom=1cm,top=0.1cm}
% Add PDF bookmark 
%%% LOGO
\thispagestyle{empty}
 %%% Instituição
\begin{center}
\LARGE \textbf{UNIVERSIDADE DE LISBOA \\ INSTITUTO SUPERIOR TÉCNICO}
%%% espaço sem gráficos
\vspace{0mm}

%%% Optional Image
%\vspace{10mm}
%~\\ \vspace{50mm} % gráficos
%\\ \begin{center} \includegraphics[height=50mm]{Cover/coverimage}  \end{center} % gráficos
 
 %%% Tituloì

\LARGE \textbf{WKB Methods for Finite Difference Schrödinger Equations}
\\ \vspace{5mm}  % NO SUBTITLE
\Large \textbf{Salvatore Baldino} \\
\vspace{0.5cm}
\small{\textbf{Supervisor: Doctor Ricardo Pina Schiappa de Carvalho}}
\vspace{3mm}\\
\vspace{0mm}
{\centering
\large \textbf{Thesis approved in public session to obtain the PhD degree in Mathematics}\\
%\\ \vspace{2mm}
\vspace{3mm}
\Large \textbf{Jury final classification: approved with distinction}}
\end{center}
\vspace{-10mm}
{\raggedright
\small
\textbf{Jury}
\textbf{Chairperson: Doctor Gabriel Lopes Cardoso, Instituto Superior Técnico, the University of Lisbon}\\\vspace{0.1cm}
\textbf{Members of the Committee:}\\\vspace{0.1cm}
\textbf{Doctor André Voros, Institut de Physique Théorique, Commissariat a l'énergie atomique et aux énergies alternatives (CEA) Saclay, France}\\\vspace{0.1cm}
\textbf{Doctor Marcos Mari\~no Beiras, Faculté des Sciences, Université de Genève, Switzerland}\\\vspace{0.1cm}
\textbf{Doctor Joao Pimentel Nunes, Instituto Superior Técnico, the University of Lisbon}\\\vspace{0.1cm}
\textbf{Doctor Ricardo Pina Schiappa de Carvalho, Instituto Superior Técnico, the University of Lisbon}\\\vspace{0.1cm}
\textbf{Doctor Davide Masoero, Grupo de Física Matemática, the University of Lisbon}\\\vspace{0.1cm}}
\begin{center}
\vspace{-10mm}\Large \textbf{Funding Institutions - Fundação para a Ciência e a Tecnologia}\\
%\large \textbf{\todaythesis\today} \\
\vspace{3mm}
\large \textbf{2023} \\
\end{center}
\let\thepage\relax
\pagebreak

%% file: abstracteng.tex
\restoregeometry

\section*{Abstract}

The object of study of this thesis is the finite difference one dimensional Schr\"odinger equation. This equation appears in different fields of theoretical physics, from integrable models in low energy physics (Toda lattice) to string theories, as the equation describing the quantization of the Seiberg-Witten curve.

In this thesis, we develop WKB techniques for the finite difference Schr\"odinger equation, following the construction of the WKB approach for the standard differential Schr\"odinger equation. In particular, we will develop an all-order WKB algorithm to get arbitrary $\hbar$-corrections and construct a general quantum momentum, underlining the various properties of its coefficients and the quantities that will be used when constructing the quantization condition. In doing so, we discover the existence of additional periodic factors that need to be considered in order to obtain the most general solution to the problem at hand. We will then proceed to study the simplest non trivial example, the linear potential case and the Bessel functions, that provide a solution to the linear problem. 

After studying the resurgence properties of the Bessel functions from an analytical and numerical point of view, we will then proceed to use those results in order to build general connection formulae, allowing us to connect the local solutions defined on two sides of a turning point into a smooth solution on the whole real line. With those connection formulae, we will analyse a selection of problems, constructing the discrete spectrum of various finite difference Schr\"odinger problems and comparing our results with existing literature.

Our work is a starting point for a more general analysis of the finite difference Schr\"odinger equation, and sets the stage for further applications of techniques developed in the ordinary Schr\"odinger equation to generalize the results that have been obtained in standard quantum mechanics to this new problem.

\textbf{Keywords:} WKB method - Resurgence - Quantization conditions - Finite difference equations - Quantum Mechanics

%% file: abstractpor.tex
\section*{Resumo}

O objeto de estudo desta tese é a equação de Schr\"odinger unidimensional de diferenças finitas. Esta equação aparece em diferentes campos da f\'isica te\'orica, desde modelos integr\'aveis ​​em física de baixas energias (Toda lattice) até teorias de cordas, como a equação que descreve a quantização da curva de Seiberg-Witten.

Nesta tese, desenvolvemos técnicas WKB para a equação de Schr\"odinger de diferenças finitas, seguindo a construção do método WKB para a equação de Schr\"odinger diferencial. Em particular, desenvolveremos um algoritmo WKB de todas as ordens para obter correções $\hbar$ arbitrárias e construir um momento quântico geral, estudando as várias propriedades de seus coeficientes e as quantidades que serão usadas ao construir as condições de quantização. Ao fazer isso, descobrimos a existência de fatores periódicos adicionais que precisam ser considerados para obter a solução mais geral para o problema em questão. Passaremos então a estudar o exemplo não trivial mais simples, o caso do potencial linear e as funções de Bessel, que fornecem uma solução para o problema linear.

Depois de estudar as propriedades de ressurgência das funções de Bessel do ponto de vista analítico e numérico, passaremos a usar esses resultados para construir fórmulas gerais de conexão, permitindo-nos conectar as soluções locais definidas nos dois lados de um turning point em uma solução suave em toda a linha real. Com essas fórmulas de conexão, analisaremos uma seleção de problemas, construindo o espectro discreto de vários problemas de Schr\"odinger de diferenças finitas e comparando nossos resultados com a literatura existente.

Nosso trabalho é um ponto de partida para uma análise mais geral da equação de Schr\"odinger de diferenças finitas e prepara o terreno para futuras aplicações de técnicas desenvolvidas na equação de Schr\"odinger ordinária para generalizar os resultados que foram obtidos na mecânica quântica ordinária para este novo problema.

\textbf{Palavras-chave:} Método WKB - Ressurgência - Condições de quantização - Equações de diferenças finitas - Mecânica quântica

%% file: 0ack.tex
\chapter*{Acknowledgments}

The PhD work of Salvatore Baldino has been supported by Fundação para a Ciência e a Tecnologia (FCT) through the PhD scholarships SFRH/BD/130088/2017 and COVID/BD/151897/2021. The author acknowledges the support of Professor Jie Gu, for providing the code to compute high order quantum corrections to the cycles in Chapter \ref{chapter5}. In addition, the author wishes to thank the members of the jury of his PhD discussion for insightful comments, corrections and suggestions about the thesis.

%% file: 1intro.tex
\chapter*{Introduction}

\addcontentsline{toc}{chapter}{Introduction}
\markboth{INTRODUCTION}{INTRODUCTION}
\renewcommand{\theequation}{\Roman{equation}}
The finite difference Schr\"odinger equation has appeared in various incarnations in recent years, in different fields of mathematical physics. Historically, it appeared in Toda lattices \cite{toda2012theory,flaschka1976canonically,kac1975explicitly}, a model consisting of $N$ particles moving in one dimension with canonical coordinates $(p_i,q_i)$ and Hamiltonian
\begin{align}
H=\frac12\sum_{i=1}^{N}p_i^2+\sum_{i=1}^{N}e^{q_i-q_{i+1}}.
\end{align}
The classical model is integrable, in the sense that one can construct $N$ integrals of motion, thus reducing the determination of the problem to the determination of a set of boundary conditions. In \cite{pasquier1992periodic}, it was shown that in the quantum theory the spectrum is obtained by solving the finite difference equation
\begin{align}
\Lambda(u)Q(u)=\ii^N Q(u+\ii\hbar)+\ii^{-N}Q(u-\ii\hbar).
\end{align}
Here $\Lambda(u)$ and $Q(u)$ are matrices, and in particular $Q(u)$, $Q(v)$ and $\Lambda(u)$ commute for all values of $u$ and $v$. Furthermore, $\Lambda(u)$ is a polynomial of order $N$ in $u$ whose coefficients are the conserved quantities in involution: those quantities are the spectral parameters of the problem. This equation is a finite difference equation, and energies at which the asymptotic behaviour of $Q$ is of a certain form will be part of the spectrum of the quantum Toda lattice.

While in the Toda lattice example the finite difference Schr\"odinger equation is a tool that can be used to obtain the solution of a problem that is in principle ruled by differential equations, there has been work on studying finite difference equations in their own merit. As examples, \cite{dingle1968wkb,dingle1968wkb2} studied the finite difference equation using a WKB approach, limited to leading order in $\hbar$. In \cite{fedotov2021wkb} a similar approach was taken, but focusing more on the study of the finite difference equation as a deformation of the correspondent ordinary Schr\"odinger equation.

Finite difference linear equations arise in the context of matrix models for $N\times N$ matrices. As shown in \cite{marino2004houches}, the partition function of matrix models can be solved in terms of $N$ orthogonal polynomials, defined as polinomials $p_n(\lambda)$ of rank $n$ and normalized as $p_n(\lambda)=\lambda^n+...$, with scalar product
\begin{align}
\frac{1}{2\pi}\int p_n(\lambda)p_m(\lambda)\ee^{-\frac{1}{g_s}W(\lambda)}\dd\lambda=h_n\delta_{nm}.
\end{align}
Here $W(\lambda)$ is the potential of the matrix model (expressed in terms of its eigenvalues) and $g_s$ is the string coupling. It is easy to prove that those polynomials satisfy recursion relations of the form
\begin{align}
(\lambda+s_n)p_n(\lambda)=p_{n+1}(\lambda)+\frac{h_n}{h_{n-1}}p_{n-1}(\lambda),
\label{eq:matrixmodels}
\end{align}
where $s_n$ depends on the potential $W$ (as an example, it is zero if $W$ is even). Once the orthogonal polynomials are obtained, the partition function $Z$ of the matrix model can be expressed as
\begin{align}
Z=\prod_{i=0}^{N-1}h_i.
\end{align}
The typical procedure to obtain the orthogonal polynomials for \eqref{eq:matrixmodels} is to take a continuum limit, sending $N\to\infty$. Employing finite difference techniques, one could aim to solve \eqref{eq:matrixmodels} at finite $N$. In this thesis, we will focus on a different type of equation, but in future the method can be generalized in order to solve equations as \eqref{eq:matrixmodels}.

Natural applications of finite difference equations can be found in lattice quantum systems. An important example in quantum computation is the problem of \textit{quantum walk} \cite{childs2009universal,farhi1998quantum,watrous2001quantum}, defined by the finite difference equation
\begin{align}
\psi(x-\Delta x,\hbar)-2\psi(x,\hbar)+\psi(x+\Delta x,\hbar)+V(x,\hbar)\psi(x)=E\psi(x,\hbar)
\end{align}
This is not the same example as our finite difference equation, as the step size $\Delta x$ is unrelated to $\hbar$, and is real. Our algorithms can be adapted to such a case with some work. Furthermore, in the context of systems of electrons we can consider the \textit{almost Mathieu operator} (reviewed as example in \cite{simon2000schrodinger,cedzich2021almost}), where we can study the spectrum of the operator
\begin{align}
H^\alpha_\omega u(n)=u(n+1)+u(n-1)+2\lambda\cos(2\pi(\omega+n\alpha))u(n),
\end{align}
with $\alpha,\omega$ parameters. This operator is important in the study of the quantum Hall effect. As we can see, quantum mechanics can benefit greatly from the study of finite difference operators.

The most recent application of the finite difference Schr\"odinger equation comes from string theory. In particular, the Hamiltonian
\begin{align}
H=\cosh p+V_N(x)
\end{align}
where $V_N(x)$ is a polynomial of order $N$ naturally arises in the context of Seiberg-Witten theories \cite{seiberg1994electric}, where it is used for the quantization of the spectral curve describing $\mathcal N=2$ Yang-Mills theory with gauge group SU($N$) \cite{klemm1996nonperturbative,klemm1995simple}. In particular, in \cite{gorsky1995integrability,martinec1996integrable} it was shown that this curve agrees with the spectral curve of the periodic Toda lattice. More recently, in \cite{grassi2019solvable} the spectral problem was studied using tools from the topological strings/spectral theory (TS/ST) correspondence, showing that while the Toda lattice spectrum is indeed part of the spectrum of the finite difference Hamiltonian, the spectrum is actually richer.

The Wentzel–Kramers–Brillouin (WKB) method \cite{wentzel1926verallgemeinerung,kramers1926wellenmechanik,brillouin1926mecanique} originated as an approximation method to obtain solutions to the classical spectral problem of quantum mechanics, with the Hamiltonian
\begin{align}
H=p^2+V(x).
\end{align}
The method consists in approximating the solution to the spectral problem with the ansatz
\begin{align}
\psi(x,\hbar)=\left(\pm\frac{1}{\ii\hbar}\int\sqrt{E-V(x)}\dd x\right),
\end{align}
and provides both the spectrum and the wavefunction. This is an approximation in $\hbar$, and it can be easily proven that remainder terms in the exponential are of order $o(\hbar^0)$. In following work, the approximation was made more precise, arriving to the all-orders WKB method \cite{dunham1932wentzel}, in which the classical quantum momentum $\sqrt{E-V(x)}$ was substituted by a more general quantum momentum $P(x,\hbar)$, including corrections in $\hbar$ in a power series. The theory of resurgence \cite{ecalle1981fonctions,costin2008asymptotics,sauzin2007resurgent,sauzin2014introduction} was then applied to this power series to show that particular integrals between points in which the approximation is singular (the turning points) have important resurgence relations between them, summarized in the Dillinger-Delabaere-Pham formula \cite{delabaere1997exact} that gives a geometrical prescription to compute resurgence relations between the periods. Various tools have been developed, like the theory of connection formulae \cite{silverstone1985jwkb} to write an exact wavefunction on the whole $x$ space, and works like \cite{balian2005quartic} in which the contribution of non real turning points was examined, obtaining non perturbative corrections to already existing connection formulae. The WKB method has become a very powerful method to exactly solve one dimensional QM problems.

This thesis is a first step in applying the methods of standard WKB to a finite difference equation, inspired by the recent work \cite{grassi2019solvable} and following the method outlined in \cite{silverstone1985jwkb} and \cite{marino2021advanced}. In particular, this thesis will focus on the computation of connection formulae, to bridge the description of the wavefunction between two sides of real turning points. The thesis is structured as follows. Chapter \ref{chap:1} gives a general analysis of the finite difference Schr\"odinger equation, starting from the analysis of its classical mechanics counterpart and then presenting the important features of the finite difference equation, and defining the WKB approach together with the algorithm to compute quantum corrections to all orders, also introducing ways to resum the power series that is obtained from the WKB analysis. Chapter \ref{chapter3} deals with the linear problem, solved by the Bessel functions, and presents a complete study of resurgence properties of the Bessel functions and their asymptotic expansion. Chapter \ref{chapter4} presents a method to compute the connection formulae, making parallels to the method used in standard quantum mechanics. Finally, chapter \ref{chapter5} presents applications of those ideas to simple problems.

%% file: 2findiffschrod.tex
\chapter{Finite difference Schr\"odinger equation}

\label{chap:1}
\renewcommand{\theequation}{\arabic{chapter}.\arabic{section}.\arabic{equation}}
\section{The equation}

\subsection{The Hamiltonian system}

Our goal in this thesis is to study the spectral problem associated to the Hamiltonian studied in \cite{grassi2019solvable}, given by
\begin{align}
H=\Lambda\left(\cosh\left(\frac{p}{\sqrt{m\Lambda}}\right)-1\right)+V(x),
\label{eq:fin_dif_ham}
\end{align}
where $m$ and $\Lambda$ are constants and $V(x)$ is a polynomial in $x$. In physical terms, $m$ has the units of mass and $\Lambda$ has the units of energy. While $m$ is a parameter present in ordinary quantum mechanics, $\Lambda$ has no analogue and is useful to define a limit to standard quantum mechanics: in fact, for $\Lambda\to\infty$ the Hamiltonian becomes
\begin{align}
H=\frac{p^2}{2m}+V(x),
\end{align}
that is the Hamiltonian for the motion of one particle in ordinary quantum mechanics. We will set $m=1$ and $\Lambda=1$, restoring $\Lambda$ whenever we make a limit to ordinary quantum mechanics. We quantize Hamiltonian \eqref{eq:fin_dif_ham} as in standard quantum mechanics in position representation,
\begin{align}
\hat x=x,\quad \hat p=-\ii\hbar\frac{\dd}{\dd x}.
\end{align}
The eigenvalues $E$ and eigenstates $\psi(x,\hbar)$ of the quantized Hamiltonian $\hat H$ are then given by solving the eigenvalue equation
\begin{align}
\hat H\psi(x,\hbar)=E\psi(x,\hbar).
\label{eq:spec_eq}
\end{align}
In order to understand the action of the Hamiltonian on $\psi(x,\hbar)$, we compute
\begin{align}
\ee^{\pm\hat p}\psi(x,\hbar)=\sum_{n=0}^\infty\frac{(\mp\ii\hbar)^n}{n!}\frac{\dd^n}{\dd x^n}\psi(x,\hbar)=\psi(x\mp\ii\hbar,\hbar).
\end{align}
\eqref{eq:spec_eq} then becomes
\begin{align}
\psi(x+\ii\hbar,\hbar)+\psi(x-\ii\hbar,\hbar)=2(E-V(x)+1)\psi(x,\hbar).
\label{eq:fin_dif_main}
\end{align}
This is the first striking difference between ordinary quantum mechanics and our deformed quantum mechanics. In ordinary quantum mechanics, the ordinary quantization procedure gives rise in position representation to the spectral problem
\begin{align}
-\frac{\dd^2}{\dd x^2}\psi(x,\hbar)=2(E-V(x))\psi(x,\hbar)
\label{eq:differ}
\end{align}
that is a second order homogeneous differential equation, we now have a \textit{second order finite difference} homogeneous equation\footnote{The \textit{order} of the finite difference equation can be defined in the following way: suppose to have a finite difference equation of the form
\begin{align}
\sum_{j=-m^n}c_j(x,\hbar)\psi_j(x+j\ii\hbar,\hbar)=0,
\end{align}
where the $c_j$ are arbitrary functions of $x$ and $\hbar$: such an equation is a finite difference equation of order $n+m$. In our case, $n=m=1$, so our difference equation is of second order.}. There are some striking differences between a differential equation and a finite difference equation of the same order, that we will analyse in the following subsections.

Before starting the analysis, we want to remark an important point on the ordinary quantum mechanics limit. While it is tempting to interpret the limit $\hbar\to0$ as the limit to standard quantum mechanics, this is actually incorrect due to the fact that removing $\hbar$ from the equation provides a \textit{classical limit}, not a limit to ordinary quantum mechanics. The correct way to take the limit is to restore the $\Lambda$ dependence, obtaining
\begin{align}
\psi\left(x+\frac{\ii\hbar}{\sqrt{m\Lambda}},\hbar\right)+\psi\left(x-\frac{\ii\hbar}{\sqrt{m\Lambda}},\hbar\right)=2\left(\frac{E-V(x)}\Lambda+1\right)\psi(x,\hbar).
\end{align}
The limit $\Lambda\to\infty$ can then be used to correctly recover \eqref{eq:differ}.

\subsection{Classical analysis}
\label{subsec:clas_an}
Before analysing equation \eqref{eq:fin_dif_main}, the classical analysis of the Hamiltonian system defined by \eqref{eq:fin_dif_ham} can give some insight on the main characteristics that we can expect upon quantization. The Hamilton equations derived from \eqref{eq:fin_dif_ham} read
\begin{align}
\dot p=-\frac{\partial}{\partial x}V(x),\quad \dot x=\sinh p,
\label{eq:ham_jac}
\end{align}
where the dot indicates derivative with respect to time. While the first equation of \eqref{eq:ham_jac} is the familiar momentum derivative of quantum mechanics under a conservative potential, the second equation is novel.

The relation between $p$ and $x$ at initial energy $E$ is given by
\begin{align}
p(x)=\pm\arccosh(E-V(x)+1).
\label{eq:impulse_def}
\end{align}
The $\pm$ sign is determined by initial conditions (the direction in which the particle moves at the beginning of motion). In this thesis, by $\arccosh$ we will mean the main determination of the inverse of the $\cosh$ function, determined by the condition of $\arccosh(y)$ being real and positive whenever $y>1$. With this convention, we see that $p(x)$ is real if and only if $V(x)<E$, and this condition defines the classically allowed region. Points $x^*$ such as $E=V(x^*)$ are then denoted as turning points in the classical analysis, as those are the points for which the particle ``bounces off" the potential, and the motion is confined in classically allowed regions delimited by two such turning points.

\begin{figure}
\centering
\includegraphics[width=0.8\textwidth]{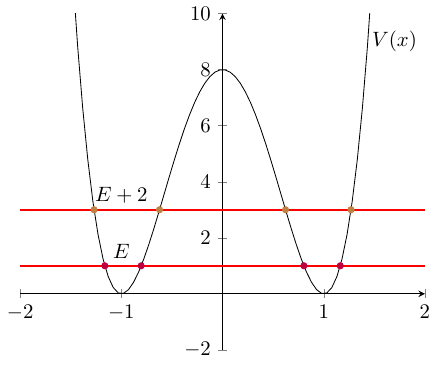}
\includegraphics[width=0.8\textwidth]{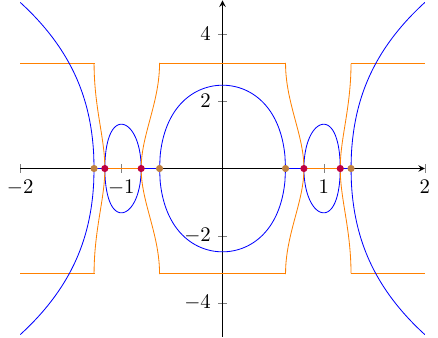}
\caption{Turning points for the example potential $V(x)=8(x^2-1)^2$. On the top, we have the a plot of the potential, shaped as a double well. The points marked in purple are turning points for the condition $E=V(x^*)$. The points marked in brown are turning points for the condition $E=V(x^*)+2$. On the bottom, we have real (blue) and imaginary (orange) parts of $p(x)$ in the same potential. Between two points of the same colour, only the real part of $p(x)$ varies, while between points of different colour only the imaginary part changes. We have included both $\pm$ determinations of \eqref{eq:impulse_def}.}
\label{fig:turning_classical}
\end{figure}

It is interesting to consider other regions. If $E-V(x)<-2$ we can observe another phenomenon, due to the fact that $\arccosh(y)$ has a constant imaginary part for $y<-1$, and zero real part for $-1<y<1$. We can interpret regions with $y<-1$ as similar to classically allowed regions, with $p(x)$ changing in its real part and keeping a fixed, non zero imaginary part. We denote those regions as ``imaginary allowed regions". Regions in which $-1<E-V(x)+1<1$ present a $p(x)$ that is purely imaginary, where its real part goes to the value that is then fixed in the imaginary allowed regions. We will denote those regions as classically forbidden regions. We see that points $x^*$ such as $E=V(x^*)-2$ act as delimitations between the two new regions, and can be considered turning points. This interpretation will be confirmed by the WKB analysis, where the meaning of turning point can be made more precise. In figure \ref{fig:turning_classical}, we picture an example of those phenomena.

As always, we conclude this part by restoring $\Lambda$ and taking the $\Lambda\to\infty$ limit, that in this case is a limit to standard classical mechanics. While the points delimiting the classically allowed regions remain the same, the other points change, as the new equation becomes $V(x^*)=E+2\Lambda$. The limit $\Lambda\to\infty$ can be understood by referring to figure \ref{fig:turning_classical}: the upper horizontal line moves to $+\infty$ in this limit, so those turning points disappear. In this case, we return to the standard situation of classical mechanics, with only one kind of turning point, a classically allowed and a classically forbidden zone (the imaginary allowed zone disappears). This behaviour is also evident by restoring units in $p(x)$ and taking the $\Lambda\to\infty$ limit:
\begin{align}
p(x)=\pm\sqrt{m\Lambda}\arccosh\left(\frac{E-V(x)}\Lambda+1\right)=\pm\sqrt{2m(E-V(x))}.
\end{align}
The leading term of the expansion in $\Lambda^{-1}$ is precisely the standard definition of the momentum.

\subsection{Linear finite difference equations: general analysis}

\label{subsec:basis}

We now turn our attention to the finite difference Schr\"odinger equation \eqref{eq:fin_dif_main}. It will be useful to state some properties of finite difference equations in preparation of the analysis of the specific equation. In this part, we will follow \cite{levy1992finite}.

As stated, the equation we are interested in is a linear homogeneous finite difference equation of second order. Solving such a difference equation can be done explicitly once a boundary condition is known. While for differential equations we typically need the value of the function and its derivatives at a point, in order to define the solution to a difference equation uniquely we need to know the initial condition on a strip of width $2\ii\hbar$. Once such a condition is known, the function can be reconstructed anywhere by algebraic means.

To see how the procedure works, let us consider a concrete example for $\hbar$ real and positive. Let $\psi_0(x,\hbar)$ be the boundary condition, fully known on the strip $-\hbar\leq\Im x<\hbar$,\footnote{The choice of such a strip has been done for definiteness, but one can start on a completely different strip. The only important factor is the width of $2\hbar$.}, and consider the evaluation at a point $y$ with $\Im y>\hbar$. Let $n$ be such as
\begin{align}
2n-1\leq \Im\hbar<2n+1.
\end{align}
In order to know the solution on the strip $\hbar\leq\Im x<3\hbar$, one can rearrange the finite difference equation and translate the argument to obtain
\begin{align}
\psi(x,\hbar)=(E-V(x-\ii\hbar)+1)\psi(x-\ii\hbar,\hbar)-\psi(x-2\ii\hbar,\hbar).
\label{eq:progression}
\end{align}
When $x$ is in the strip $\hbar\leq\Im x<3\hbar$, the RHS of \eqref{eq:progression} is entirely in the first strip, where the solution $\psi$ is known and coincides with $\psi_0$. To reach the strip of $y$, one has to iterate this procedure $n$ times. The evaluation for $\Im y<-\hbar$ is analogous.

While this procedure allows for the construction of a generic solution involving only a finite number of algebraic steps, we will take an alternative point of view. A general solution to the finite difference equation can be built by considering a pair of functions $\psi_+(x,\hbar)$ and $\psi_-(x,\hbar)$ satisfying the following condition: there exist no function $P(x,\hbar)$ periodic in $x$ of period $\ii\hbar$ such as $\psi_+(x,\hbar)=P(x,\hbar)\psi_-(x,\hbar)$. Then any solution $\psi(x,\hbar)$ of the finite difference equation can be written as
\begin{align}
\psi(x,\hbar)=A(x,\hbar)\psi_+(x,\hbar)+B(x,\hbar)\psi_-(x,\hbar),
\end{align}
where $A(x,\hbar)$ and $B(x,\hbar)$ are periodic functions in $x$ of period $\ii\hbar$. The reason for this form of the solutions is that periodic functions obviously factorize in the finite difference equation. The situation is analogous to linear differential equations, with constants substituted by periodic functions. In the rest of the thesis, unless otherwise specified, by periodic function we will always intend a function periodic in the $x$ argument of period $\ii\hbar$.

Periodic functions can be expanded in Fourier series by defining the periodic term
\begin{align}
q=\exp\left(-\frac{2\pi}{\hbar}x\right).
\end{align}
The general solution can then be written as
\begin{align}
\psi(x,\hbar)=\sum_{n=-\infty}^\infty a_n(\hbar)q^{-n}\psi_+(x,\hbar)+\sum_{n=-\infty}^\infty b_n(\hbar)q^{-n}\psi_-(x,\hbar),
\label{eq:gen_sol}
\end{align}
where $a_n$ and $b_n$ are Fourier coefficients for the expansion of $A$ and $B$ respectively. In this point of view, one can see the space of solutions to the finite difference equation to be spanned by $\cup_{n=-\infty}^\infty\{q^n\psi_+(x,\hbar),q^n\psi_-(x,\hbar)\}$, so even a second order linear finite difference equation has an infinite dimensional space of solutions. This is the main point of departure between linear differential and linear finite difference equations: even if the solution to finite difference equations can be constructed algebraically from an initial condition, this initial condition is more complicated and needs the specification of an infinite number of constants.

In this thesis, we will impose an additional restriction on the types of solutions that we study. As we aim to work with the same Hilbert space of standard quantum mechanics (in order to keep the interpretation of $\abs{\psi(x,\hbar)}^2$ as a probability density in $x$), we will mainly be interested in normalizable spectra: we will restrict $V(x)$ to even potentials, going to $+\infty$ at $x\to\pm\infty$, and look for solutions $\psi(x,\hbar)$ that belong to $\mathbb L^2(\mathbb R)$ in the $x$ variable. We will see in the next chapters how this additional condition greatly restricts the space of available solutions.

\section{The WKB approach}

\subsection{Computing the quantum momentum}

We now introduce the main method with which we build solutions to the difference equation \eqref{eq:fin_dif_main}.

In the WKB approach, the starting point is the ansatz\footnote{An alternative definition is to define $\psi(x,\hbar)=\exp\left(-\frac{1}{\ii\hbar}S_{\text T}(x,\hbar)\right)$. Due to the fact that the finite difference equation is even in $\hbar$, those two definitions are equivalent.}
\begin{align}
\psi(x,\hbar)=\exp\left(\frac{1}{\ii\hbar}S_{\text T}(x,\hbar)\right).
\end{align}
Inserting this ansatz in \eqref{eq:fin_dif_main}, we obtain
\begin{align}
\begin{aligned}
\exp\left(\frac{1}{\ii\hbar}(S_{\text T}(x+\ii\hbar,\hbar)-S_{\text T}(x,\hbar))\right)+&\exp\left(\frac{1}{\ii\hbar}(S_{\text T}(x-\ii\hbar,\hbar)-S_{\text T}(x,\hbar))\right)=\\&=2Q(x),\label{eq:fin_dif_start}
\end{aligned}
\end{align}
where we have defined $Q(x)=E-V(x)+1$. The function $S_{\text T}(x,\hbar)$ is usually denoted as the \textit{quantum action}, and can be expanded in series of $\hbar$ as\footnote{The usual series expansion is
\begin{align*}
S_{\text T}(x,\hbar)=\sum_{n=0}^\infty S_n(x)\hbar^n.
\end{align*}
In this thesis, we have decided to go with a different expansion, for reasons that will be evident in the WKB algorithm. Obviously, the two ansatzes can be easily related.}
\begin{align}
S_{\text T}(x,\hbar)=\sum_{n=0}^\infty\frac{S_n(x)}{n!}(\ii\hbar)^n.
\end{align}
The term $S_0(x)$ is denoted as \textit{classical action}. It is useful to introduce the \textit{quantum momentum} $P_{\text T}(x,\hbar)$ as (primes indicate derivatives with respect to $x$)
\begin{align}
S'_{\text T}(x,\hbar)=P_{\text T}(x,\hbar),
\end{align}
that can also be expanded in a similar $\hbar$ expansion as
\begin{align}
P_{\text T}(x,\hbar)=\sum_{n=0}^\infty\frac{P_n(x)}{n!}(\ii\hbar)^n,
\label{eq:mom_exp}
\end{align}
with $P_n(x)=S_n'(x)$. The quantum momentum obeys the integral equation
\begin{align}
\exp\left(\frac{1}{\ii\hbar}\int_x^{x+\ii\hbar}P_{\text T}(t,\hbar)\dd t\right)+\exp\left(\frac{1}{\ii\hbar}\int_x^{x-\ii\hbar}P_{\text T}(t,\hbar)\dd t\right)=2Q(x).
\label{eq:quan_mom}
\end{align}
The WKB algorithm consists in expanding the quantum momentum in an $\hbar$ series, plugging the expansion in \eqref{eq:quan_mom} and continuing the expansion in $\hbar$ until we get algebraic equations for determining $P_n(x)$ in terms of $P_m(x)$ with $m<n$ and their derivatives. The only exception to this rule is the classical action $P_0(x)$, that is determined in terms of $Q(x)$ by expanding the integral to first order in $\hbar$:
\begin{align}
\exp\left(P_0(x)\right)+\exp\left(-P_0(x)\right)=2Q(x)\implies P_0(x)=\arccosh Q(x).
\label{eq:clas_mom_def}
\end{align}
As before, we are always using the main determination of $\arccosh$: $\arccosh x$ is real and positive for $x>1$.

In order to obtain the rest of the coefficients, we have to perform some preliminary computations. First, we have to expand the integral as
\begin{align}
\frac{1}{\ii\hbar}\int_x^{x\pm\ii\hbar}P_{\text T}(t,\hbar)\dd t=\sum_{m=0}^\infty(\pm1)^{m+1} \frac{P_{\text T}^{(m)}(x,\hbar)}{(m+1)!}(\ii\hbar)^n.
\end{align}
Here by $P_{\text T}^{(m)}(x,\hbar)$ we mean the $m-$th derivative of $P_{\text T}$ with respect to $x$. Then we plug expansion \eqref{eq:mom_exp} in the integral expansion to obtain
\begin{align}
\frac{1}{\ii\hbar}\int_x^{x\pm\ii\hbar}P_{\text T}(t,\hbar)\dd t=\sum_{n=0}^\infty\sum_{m=0}^\infty\frac{(\pm1)^{m+1}}{(m+1)!n!}P^{(m)}_n(x)(\ii\hbar)^{n+m}.
\end{align}
We rearrange terms to obtain
\begin{align}
\frac{1}{\ii\hbar}\int_x^{x\pm\ii\hbar}P_{\text T}(t,\hbar)\dd t=\sum_{n=0}^\infty\frac{(\ii\hbar)^n}{n!}\frac{1}{n+1}\sum_{m=0}^n(\pm1)^{m+1}{n+1\choose m+1}P_{n-m}^{(m)}(x).
\end{align}
By defining
\begin{align}
\mathcal I_n^{(\pm)}(x)=\frac{1}{n+1}\sum_{m=0}^n(\pm1)^{m+1}{n+1\choose m+1}P_{n-m}^{(m)}(x),
\label{eq:d_def}
\end{align}
we have reduced the expansion of the integral to
\begin{align}
\frac{1}{\ii\hbar}\int_x^{x\pm\ii\hbar}P_{\text T}(t,\hbar)\dd t=\sum_{n=0}^\infty\frac{(\ii\hbar)^n}{n!}\mathcal I_n^{(\pm)}(x),
\end{align}
in a form that is ready to be exponentiated. The exponentiation can be performed through the use of Bell polynomials\footnote{Bell polynomials are widely used in combinatorics, obeying the relation
\begin{align}
\exp\left(\sum_{j=1}^\infty x_j\frac{t^j}{j!}\right)=\sum_{n=0}^\infty B_n(x_1,...,x_n)\frac{t^n}{n!},
\end{align}
where $x_1,...,x_n$ is a succession. The reader can find properties and alternative definitions of the Bell polynomials in \cite{comtet2012advanced}.}, obtaining
\begin{align}
\exp\left(\frac1{\ii\hbar}\int_x^{x\pm\ii\hbar}P_{\text T}(t,\hbar)\dd t\right)=\exp\left(\pm P_0(x)\right)\sum_{n=0}^\infty\frac{(\ii\hbar)^n}{n!}B_n(\mathcal I_1^{(\pm)}(x),...,\mathcal I_n^{(\pm)}(x)).
\end{align}
Equation \eqref{eq:quan_mom} then becomes
\begin{align}
\begin{aligned}
&\ee^{P_0(x)}\sum_{n=0}^\infty\frac{(\ii\hbar)^n}{n!}B_n(\mathcal I_1^{(+)}(x),...,\mathcal I_n^{(+)}(x))+\\+&\ee^{-P_0(x)}\sum_{n=0}^\infty\frac{(\ii\hbar)^n}{n!}B_n(\mathcal I_1^{(-)}(x),...,\mathcal I_n^{(-)}(x))=2Q(x).
\end{aligned}
\end{align}
Order by order, this equation becomes
\begin{align}
\begin{cases}
&\ee^{P_0(x)}+\ee^{-P_0(x)}=2Q(x),\\
&\ee^{P_0(x)}B_n(\mathcal I_1^{(+)}(x),...,\mathcal I_n^{(+)}(x))+\ee^{-P_0(x)}B_n(\mathcal I_1^{(-)}(x),...,\mathcal I_n^{(-)}(x))=0.
\end{cases}
\label{eq:rec_1}
\end{align}
As stated before, the first equation is solved by $P_0(x)=\arccosh Q(x)$. The other equations can be solved inductively, by identifying $P_n(x)$ and writing a formula for it. The first step in doing so is to identify $P_n(x)$ in $I_n^{(\pm)}(x)$: as evident from \eqref{eq:d_def}, $P_n(x)$ only appears with no derivatives. If we define $\hat {\mathcal I}_n^{(\pm)}=\mathcal I_n^{(\pm)}\mp P_n(x)$, then $\hat {\mathcal I}_n^{(\pm)}$ does not contain $P_n$ or its derivatives. Analogously, using the properties of the Bell polynomial $B_n(x_1,...,x_n)$ we can see that $B_n(x_1,...,x_n)-x_n$ does not depend on $x_n$. We can use this property to ``extract" the $P_n$ dependency from the Bell polynomial. We can take the second equation of $\eqref{eq:rec_1}$ and extract the $P_n$ dependence to obtain
\begin{align}
\ee^{P_0}(\hat B_n({\mathcal I}_1^{(+)},...,{\mathcal I}_{n-1}^{(+)},\hat {\mathcal I}_n^{(+)})+P_n)+\ee^{-P_0}(\hat B_n({\mathcal I}_1^{(-)},...,{\mathcal I}_{n-1}^{(-)},\hat {\mathcal I}_n^{(-)})-P_n)=0.
\end{align}
We can now rearrange and solve for $P_n$, as hatted quantities do not depend on $P_n$. We obtain
\begin{align}
\begin{aligned}
P_n=-\frac{1}{2\sinh P_0}(&\ee^{P_0}B_n({\mathcal I}_1^{(+)},...,{\mathcal I}_{n-1}^{(+)},\hat {\mathcal I}_n^{(+)})+\ee^{-P_0} B_n({\mathcal I}_1^{(-)},...,{\mathcal I}_{n-1}^{(-)},\hat {\mathcal I}_n^{(-)})).
\label{eq:rec_sol}
\end{aligned}
\end{align}
As in the standard WKB approach, $P_n$ is obtained by solving an algebraic equation that involves $P_m$ with $m<n$ and their derivatives. The first difference is in the fact that the equation for the finite difference case involves derivatives beyond the first one, and in general more combinatorics has to be performed. As a consequence, computations involving high orders are prohibitively costly. In our implementation in Mathematica 12.1, we have managed to compute up to $P_7$.

Now that we have computed the quantum momentum, we can obtain the quantum action by simple integration. This integration will depend on the choice of a base point $x_0$. We will observe later how the choice of this base point is very important in creating connection formulae. For now, let us leave $x_0$ unspecified and write the relation
\begin{align}
S_{\text T}(x,\hbar)=\int_{x_0}^xP_{\text T}(t,\hbar)\dd t.
\end{align} 

We conclude this introduction to the method with some examples. The first terms of the quantum momentum obtained through the algorithm are given by
\begin{align}
&P_0(x)=\arccosh Q(x),\label{eq:clas_mom}\\
&P_1(x)=-\frac14\frac{\dd}{\dd x}\log(1-Q(x)^2),\\
&P_2(x)=\frac{2(Q(x)^4+Q(x)^2-2)Q''(x)-Q(x)(13+2Q(x))Q'(x)^2}{12(Q(x)+1)^{\frac52}(Q(x)-1)^{\frac52}}\\
&P_3(x)=\frac14\frac{\dd}{\dd x}\left(\frac{3(1+4Q(x)^2)Q'(x)^2}{2(Q(x)^2-1)}-\frac{3Q(x)Q''(x)}{(Q(x)^2-1)^2}\right).\label{eq:third_mom}
\end{align}
Those terms coincide with the terms computed in \cite{dingle1968wkb}, with the addition that our method provides an algorithm to compute every correction to the quantum momentum.

\subsection{Basis of solutions}

In \eqref{eq:clas_mom_def}, we have made a choice in writing $P_0(x)$ in terms of $Q(x)$. As it is evident from the properties of the $\cosh$ function, this is not the only choice. Here we analyse the consequences of making a different choice, and see how this ties with the discussion in subsection \ref{subsec:basis}.

If $P_0(x)$ solves $\cosh P_0(x)=Q(x)$, then obviously also $\pm P_0(x)+2\pi\ii n$ (with $n\in\mathbb Z$) solves the same equation. We now compute what happens to the rest of the tower of solutions when $P_0$ is substituted by $\pm P_0+2\pi\ii n$. We will do that by first analysing the substitution $P_0\to P_0+2\pi \ii n$.

With this translation, it is evident that the ${\mathcal I}_n^{(\pm)}$ factors never change. From \eqref{eq:d_def}, we see that ${\mathcal I}_n^{(\pm)}$ with $n>0$ never contains $P_0(x)$ but only derivatives of the classical momentum. As the recursive solution \eqref{eq:rec_sol} only involves ${\mathcal I}_n^{(\pm)}$ with $n>0$, the ${\mathcal I}_n^{(\pm)}$ are unchanged. The only dependence on $P_0$ without derivatives comes from $\sinh P_0$, $\ee^{P_0}$ and $\ee^{-P_0}$, but the functions of $P_0$ are periodic with period $2\pi\ii$. Equation $\eqref{eq:d_def}$ is then completely unchanged if we translate $P_0$ to $P_0+2\pi\ii n$, so the other coefficients of the quantum momentum do not change.

We now analyse $P_0\to-P_0$. This transformation is a little trickier, and we will have to use induction. Our inductive hypothesis is that the transformation $P_0\to-P_0$ induces a transformation $P_n\to(-1)^{n+1}P_n$ in the rest of the tower: the odd terms of the series are unchanged, while the even terms change sign. As base case we use $P_0$ and $P_1$, for which the result can be obtained through a straightforward computation. We now assume the inductive hypothesis true up to $P_n$, and prove the result for $P_{n+1}$. First, for $m\leq n$ we have
\begin{align}
{\mathcal I}_m^{(\pm)}(x)\to\frac{1}{m+1}\sum_{k=0}^m(\pm1)^{k+1}(-1)^{m-k+1}{m+1\choose k+1}P_{m-k}^{(k)}(x),
\end{align}
where the factor $(-1)^{m-k+1}$ comes from the inductive hypothesis, as $m-k\leq n$. We then have
\begin{align}
{\mathcal I}_m^{(\pm)}(x)\to(-1)^m{\mathcal I}_m^{(\mp)}(x).
\label{eq:inv_d}
\end{align}
This result also applies to $\hat {\mathcal I}_{n+1}^{(\pm)}(x)$, as it does not contain $P_{n+1}(x)$ by definition:
\begin{align}
\hat {\mathcal I}_{n+1}^{(\pm)}(x)\to(-1)^{n+1}\hat {\mathcal I}_{n+1}^{(\mp)}(x).
\label{eq:inv_d_hat}
\end{align}
We now have to use properties of the Bell polynomials. The property that we will use is
\begin{align}
B_n(-x_1,x_2,-x_3,...,(-1)^{n-1}x_{n-1},(-1)^nx_n)=(-1)^nB_n(x_1,x_2,x_3,...,x_{n+1},x_n).
\end{align}
With this property, \eqref{eq:inv_d} and \eqref{eq:inv_d_hat}, we obtain
\begin{align}
B_n({\mathcal I}^{(\pm)}_1,...,{\mathcal I}^{(\pm)}_{n},\hat {\mathcal I}^{(\pm)}_{n+1})\to(-1)^nB_n({\mathcal I}^{(\mp)}_1,...,{\mathcal I}^{(\mp)}_{n},\hat {\mathcal I}^{(\mp)}_{n+1}).
\end{align} 
Applying this transformation to \eqref{eq:rec_sol} (with $P_{n+1}$ instead of $P_n$), we obtain that the signs of the exponentials change in such a way that the inner bracket gets transformed into itself times $(-1)^{n+1}$, and $\sinh P_0$ provides the additional minus sign that is needed to prove the inductive hypothesis. We can then conclude that $P_0\to-P_0$ induces the transformation $P_n\to(-1)^{n+1}P_n$. Lastly, transforming $P_0$ into $-P_0+2\pi\ii n$ can be done trivially by composing the two described transformations.

What we have proven here can be expressed in short through the quantum momentum $P_{\text T}(x,\hbar)$. We have proven that if $P_{\text T}(x,\hbar)$ solves \eqref{eq:quan_mom}, then $\pm P_{\text T}(x,\hbar)+2\pi\ii n$ (with $n\in\mathbb Z$) also solves \eqref{eq:quan_mom}. We can now connect our WKB ansatz to the analysis of subsection \ref{subsec:basis}. By defining
\begin{align}
\psi_{+,x_0}(x,\hbar)=\exp\left(-\frac{1}{\ii\hbar}\int_{x_0}^xP_{\text T}(t,-\hbar)\dd t\right),\quad \psi_{-,x_0}(x,\hbar)=\exp\left(\frac{1}{\ii\hbar}\int_{x_0}^xP_{\text T}(t,\hbar)\dd t\right),
\label{eq:basis}
\end{align}
we have found the functions $\psi_+$ and $\psi_-$ that we can use to build the general solution, in which we have also specified a base point $x_0$. The translation $P_{\text T}(x,\pm\hbar)\to P_{\text T}(x,\pm\hbar)+2\pi\ii n$ has the effect
\begin{align}
\psi_{\pm,x_0}(x,\hbar)\to\exp\left(\mp\frac{2\pi n}{\hbar}(x-x_0)\right)\psi_{\pm,x_0}(x,\hbar)=\exp\left(\pm\frac{2\pi n}{\hbar}x_0\right)q^{\pm n}\psi_{\pm,x_0}(x,\hbar),
\end{align}
providing the necessary functions to build solutions as in \eqref{eq:gen_sol}, with an additional $x_0$ dependent term that can be reabsorbed in the constants $a_n(\hbar)$ and $b_n(\hbar)$.

\subsection{The even-odd relation}

The quantum momentum coefficients naturally split in two subsets: the even coefficients $P_{2n}(x)$ and the odd coefficients $P_{2n+1}(x)$, respectively forming the even and odd parts $P(x,\hbar)$ and $P_{\text o}(x,\hbar)$ of the quantum momentum (where even and odd refer to the inversion $\hbar\to-\hbar$). Analogously, we define $S(x,\hbar)$ and $S_{\text o}(x,\hbar)$ as the even and odd parts of the quantum action, integral of the respective quantum momenta. We have already seen from the first coefficients \eqref{eq:clas_mom} to \eqref{eq:third_mom} some hints of different properties between the two sets: namely, the even terms are multivalued functions, while the odd terms are rational functions that can be written as total derivatives in $x$. We will see in this section of those properties generalize to the whole tower, and compute a relation between the two sets. In standard WKB, we have a similar situation: the even and odd parts of the momentum are related by
\begin{align}
P_{\text o}^{\text{ standard}}(x,\hbar)=\frac{\ii\hbar}{2}\frac{\dd}{\dd x}\log P^{\text{ standard}}(x,\hbar).
\label{eq:stand_even_odd}
\end{align}
We call such relation an \textit{even-odd relation}. We now compute the analogous relation for our finite difference equation.

We start from \eqref{eq:quan_mom}, that we rewrite by splitting even and odd part as
\begin{align}
\exp\left(\int_x^{x+\ii\hbar}\frac{P}{\ii\hbar}\right)\exp\left(\int_x^{x+\ii\hbar}\frac{P_{\text o}}{\ii\hbar}\right)+\exp\left(\int_x^{x-\ii\hbar}\frac{P}{\ii\hbar}\right)\exp\left(\int_x^{x-\ii\hbar}\frac{P_{\text o}}{\ii\hbar}\right)=2Q,\label{eq:first_evod}
\end{align}
where we have suppressed arguments and integration variables for brevity. We invert $\hbar\to-\hbar$, using the fact that $Q$ does not depend on $\hbar$, and use the definite parity of $P$ and $P_{\text o}$. We obtain
\begin{align}
\exp\left(-\int_x^{x+\ii\hbar}\frac{P}{\ii\hbar}\right)\exp\left(\int_x^{x+\ii\hbar}\frac{P_{\text o}}{\ii\hbar}\right)+\exp\left(-\int_x^{x-\ii\hbar}\frac{P}{\ii\hbar}\right)\exp\left(\int_x^{x-\ii\hbar}\frac{P_{\text o}}{\ii\hbar}\right)=2Q\label{eq:sec_evod}.
\end{align}
Subtracting \eqref{eq:sec_evod} to \eqref{eq:first_evod}, we obtain
\begin{align}
\exp\left(\frac{1}{\ii\hbar}\int_x^{x+\ii\hbar}P_{\text o}\right)\sinh\left(\frac{1}{\ii\hbar}\int_x^{x+\ii\hbar}P\right)=-\exp\left(\frac{1}{\ii\hbar}\int_x^{x-\ii\hbar}P_{\text o}\right)\sinh\left(\frac{1}{\ii\hbar}\int_x^{x-\ii\hbar}P\right).
\end{align}
We can rewrite this equation to obtain
\begin{align}
\int_{x-\ii\hbar}^{x+\ii\hbar}P_{\text o}(t,\hbar)\dd t=-\ii\hbar\log\frac{\sinh\frac{1}{\ii\hbar}\int_x^{x+\ii\hbar}P(t,\hbar)\dd t}{\sinh\frac{1}{\ii\hbar}\int_{x-\ii\hbar}^{x}P(t,\hbar)\dd t}.
\end{align}
Deriving with respect to $x$ gives
\begin{align}
P_{\text o}(x+\ii\hbar,\hbar)-P_{\text o}(x-\ii\hbar,\hbar)=-\ii\hbar\frac{\dd}{\dd x}\log\frac{\sinh\frac{1}{\ii\hbar}\int_x^{x+\ii\hbar}P(t,\hbar)\dd t}{\sinh\frac{1}{\ii\hbar}\int_{x-\ii\hbar}^{x}P(t,\hbar)\dd t}.
\label{eq:even_odd_first}
\end{align}
There is a further simplification that we can perform by recognizing that we can write the logarithmic term as
\begin{align}
\begin{aligned}
&\log\frac{\sinh\frac{1}{\ii\hbar}\int_x^{x+\ii\hbar}P(t,\hbar)\dd t}{\sinh\frac{1}{\ii\hbar}\int_{x-\ii\hbar}^{x}P(t,\hbar)\dd t}=\\=&\log\sinh\frac{1}{\ii\hbar}\int_x^{x+\ii\hbar}P(t,\hbar)\dd t-\log\sinh\frac{1}{\ii\hbar}\int_{x-\ii\hbar}^{x}P(t,\hbar)\dd t.
\end{aligned}
\end{align}
We can interpret this term as the difference between the values of the same function evaluated at different points, with such function being $\log\sinh\frac{1}{\ii\hbar}\int_{x-\frac{\ii\hbar}2}^{x+\frac{\ii\hbar}2}P(t,\hbar)\dd t$. The evaluations are made at $x+\frac{\ii\hbar}2$ for the first term and $x-\frac{\ii\hbar}2$ for the second term. This difference is implemented by the operator $2\sinh\left(\frac{\ii\hbar}2\frac{\dd}{\dd x}\right)$, so we can rewrite the $\log$ term as
\begin{align}
\log\frac{\sinh\frac{1}{\ii\hbar}\int_x^{x+\ii\hbar}P(t,\hbar)\dd t}{\sinh\frac{1}{\ii\hbar}\int_{x-\ii\hbar}^{x}P(t,\hbar)\dd t}=2\sinh\left(\frac{\ii\hbar}2\frac{\dd}{\dd x}\right)\log\sinh\frac{1}{\ii\hbar}\int_{x-\frac{\ii\hbar}2}^{x+\frac{\ii\hbar}2}P(t,\hbar)\dd t.
\end{align}
Similarly, the LHS of \eqref{eq:even_odd_first} can be written as
\begin{align}
P_{\text o}(x+\ii\hbar,\hbar)-P_{\text o}(x-\ii\hbar,\hbar)=2\sinh\left(\ii\hbar\frac{\dd}{\dd x}\right)P_{\text o}(x,\hbar).
\end{align}
With the operators, the relation becomes
\begin{align}
\sinh\left(\ii\hbar\frac{\dd}{\dd x}\right)P_{\text o}(x,\hbar)=-\ii\hbar\frac{\dd}{\dd x}\sinh\left(\frac{\ii\hbar}2\frac{\dd}{\dd x}\right)\log\sinh\frac{1}{\ii\hbar}\int_{x-\frac{\ii\hbar}2}^{x+\frac{\ii\hbar}2}P(t,\hbar)\dd t.
\end{align}
The last step is to invert the operator $\sinh\left(\ii\hbar\frac{\dd}{\dd x}\right)$. As the operators commute, we can work as if they were ordinary functions like $\sinh x$ and $\sinh \frac x2$\footnote{The operator $\sinh\left(\ii\hbar\frac{\dd}{\dd x}\right)$ is not invertible, as periodic functions of period $2\ii\hbar$ belong to its kernel. Thus, if we perform the inversion, in theory we would obtain a result up to the addition of periodic functions. What we do here is to work at a formal level, performing expansions in $\hbar$: as both the LHS and the RHS contain terms of order $\hbar^{-1}$, the final result is still valid.}. In particular, we can use the identity $\sinh x/\sinh\frac{x}{2}=\frac12\left(\cosh\frac{x}{2}\right)^{-1}$ to obtain
\begin{align}
P_{\text o}(x,\hbar)=-\frac{\ii\hbar}2\frac{\dd}{\dd x}\left(\cosh\left(\frac{\ii\hbar}2\frac{\dd}{\dd x}\right)\right)^{-1}\log\sinh\frac{1}{\ii\hbar}\int_{x-\frac{\ii\hbar}2}^{x+\frac{\ii\hbar}2}P(t,\hbar)\dd t.
\label{eq:even_odd_fin}
\end{align}
This is the final form of the even-odd relation. As said, it has to be taken as a formal expansion: in order to use it, one has to perform an $\hbar$ expansion of the LHS and the RHS, to obtain relations between the even and odd terms of the quantum momentum. This relation is more complicated than the standard even-odd relation \eqref{eq:stand_even_odd}, but shares the similarity of having $P_{\text o}$ as a total derivative of a function of $P$ (as in performing the expansion of the integral the first term is the quantum momentum itself, not its integral). We have tested this relation by computing up to $P_7(x)$ through the standard algorithm and the even-odd relation, obtaining a perfect match.

The first two examples of this relation are
\begin{align}
&P_1(x)=-\frac12\frac{\dd}{\dd x}\log\sinh P_0(x),\\
&P_3(x)=-\frac1{32}\frac{\dd}{\dd x}\left(\frac{3P_0'(x)^2}{(\sinh P_0(x))^2}+\frac{\cosh P_0(x)}{\sinh P_0(x)}(48 P_2(x)+P_0''(x))\right).
\end{align}

\subsection{Properties of the coefficients}

We now devote some time to show some properties of the coefficients that will be useful later. We have not proven all those coefficients, but shown by explicit computation that they are valid up to $\hbar^{10}$, and it is reasonable to assume that they are always valid.

First, we analyse the form of the various coefficients. By looking at equations from \eqref{eq:clas_mom} to \eqref{eq:third_mom}, we can infer the following ansatz: all $P_n(x)$ with the exception of $P_0(x)$ are of the form
\begin{align}
P_n(x)=\frac{\text{Pol}_n(Q(x),...,Q^{(n)}(x))}{(Q(x)-1)^{\frac{n+1}2}(Q(x)+1)^{\frac{n+1}2}},\label{eq:form}
\end{align}
where $\text{Pol}_n$ are polynomial functions in $Q$ and its derivatives, up to $Q^{(n)}$. Furthermore, the polynomials are never zero whenever $Q(x)=1$ or $Q(x)=-1$ if we also assume that at those points $Q'(x)$ never vanishes (in case it does vanish, it is sufficient for polynomial potentials to change the energy slightly, as in that case we would get a non vanishing value of $Q'(x)$. The limited number of values of $E$ for which $Q'(x)$ vanishes at points $Q(x)=\pm1$ can be addressed in principle on a case-by-case basis). This is a very important property, that allows us to see that our expansion is actually \textit{singular} at the points for which $Q(x)=\pm1$, even if the original equation is not. Points in which our expansion is singular but the true solution is not are denoted as \textit{turning points}, as in ordinary WKB (where the turning points appear whenever $Q(x)=0$). By restoring the ordinary notation, we see that
\begin{align}
Q(x)=1\implies E=V(x),\quad Q(x)=-1\implies E=V(x)-2.
\end{align}
The turning points of the WKB ansatz coincide exactly with the turning points of the classical analysis of subsection \ref{subsec:clas_an}. Those points will be very important in the analysis of the asymptotic approximations to the solution, as crossing those points causes a change in the asymptotic behaviour of our solutions.

It is worth noting that \cite{kashani2016quantization} also presents a WKB analysis of the finite difference problem. The author concludes that turning points are only the points for which $Q(x)=1$. This discrepancy with our result can be traced to a matter of terminology, as the author defines as turning points the points at which $P_0(x)=0$. In our thesis, it will be more useful to use our broader definition of turning points.

By combining \eqref{eq:form} and \eqref{eq:even_odd_fin}, we can see that the odd terms of the quantum momentum are rational functions of $Q$ and its derivatives, so whenever $Q$ is a polynomial (as in this thesis) the odd terms will be rational functions of $x$, with poles in $Q(x)=\pm1$. Also, we note that $P_1(x)$ has non vanishing residues of value $-1/4$ at the turning points, while all other terms have vanishing residues. This will allow us to do integrals in the complex plane around contours that surround two consecutive turning points, entirely neglecting the odd part of the momentum. In the next section, we will see how this is relevant in our analysis.

\section{Turning points and quantum periods}

After defining an algorithm to find the quantum momentum, we now turn to the problem of evaluating the wave function. As we have seen the evaluation depends on the choice of a base point $x_0$, and changing this base point will add phases to our wave function. When the base points are the turning points, those phases are denoted as \textit{quantum periods}, that will be the principal ingredients for solving our spectral problem. For this section, we will carefully follow what happens in \cite{marino2021advanced} and \cite{kawai2005algebraic}, adapting the discussion to our case.

\subsection{The wavefunction: choosing a base point}

The wavefunction $\psi(x,\hbar)$ can be split into a contribution from the even and the odd quantum momentum as
\begin{align}
\psi(x,\hbar)=\exp\left(\frac{1}{\ii\hbar}S_{\text o}(x,\hbar)\right)\exp\left(\frac{1}{\ii\hbar}S(x,\hbar)\right).
\label{eq:wkb_wavefun}
\end{align}
The important part of the quantum action is $S$, as the odd part can be recovered through the even part using
\begin{align}
S_{\text o}(x,\hbar)=-\frac{\ii\hbar}2\frac{\dd}{\dd x}\left(\cosh\left(\frac{\ii\hbar}2\frac{\dd}{\dd x}\right)\right)^{-1}\log\sinh\frac{S\left(x+\frac{\ii\hbar}2,\hbar\right)-S\left(x-\frac{\ii\hbar}2,\hbar\right)}{\ii\hbar}.
\label{eq:ev_odd_act}
\end{align}
Once we manage to assign a value to $S(x,\hbar)$, the evaluation of the odd part can then be done using \eqref{eq:ev_odd_act}. Due to this relation, from now on when we refer to quantum action and momentum we will be referring to $S$ and $P$ respectively instead of $S_{\text T}$ and $P_{\text T}$, keeping the odd parts out.

The quantum action components are related to the quantum momenta components as
\begin{align}
S_n(x)=\int_{x_0}^xP_n(t)\dd t.
\label{eq:mom_to_act}
\end{align}
It will be very convenient in discussing the asymptotics of the solution to choose $x_0$ as a turning point of the equation. With this choice, the integral \eqref{eq:mom_to_act} becomes undefined, due to property \eqref{eq:form}: for all factors $P_n$ with $n\neq0$, the integral is divergent.

\begin{figure}
\centering
\includegraphics[width=\textwidth]{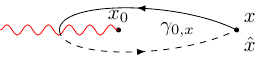}
\caption{Integration path for the integrals of $P_n$, with $x_0$ the turning point and the wavy line being the branch cut. The dashed line is the part of the path that runs on the second sheet after crossing the branch cut. We stress that $\gamma_{0,x}$ is an open path, as its endpoints are different.}
\label{fig:path_integration}
\end{figure}

We can regularize \eqref{eq:mom_to_act} as
\begin{align}
S_n(x)\to\frac12\int_{\gamma_{0,x}}P_n(t)\dd t,
\label{eq:reg_prescr}
\end{align}
where $\gamma_{0,x}$ is a complex contour that is defined by using the square root singularity in each $P_n(x)$ (we recall that the integral of $P_0(x)$ is non singular, so we can assume that each $P_n$ has only square root singularities at $x_0$). We will assume $x>x_0$, and show how to relax the assumption in the next subsection. The contour is defined by first putting the branch cut on the real line, not intersecting $x$. The path starts at $x$, crosses the branch and goes on the second sheet, and then ends at $\hat x$, that is the point on the second sheet that is projected onto $x$. The situation is pictured in figure \ref{fig:path_integration}.
The integrals regularized in this way are convergent, as they encounter no singularity in their integration path, and given the fact that with this definition $S_n'(x)=P_n(x)$, this is still a valid definition for the quantum action.

\subsection{Changing the base point: the quantum periods}

We now come to the problem of changing the base points, that will introduce the fundamental quantities that enter the quantization conditions: the \textit{quantum periods}.

\begin{figure}
\centering
\includegraphics[width=\textwidth]{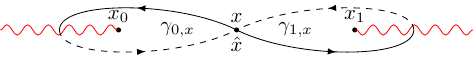}
\caption{Two different choices for the base point, $x_0$ or $x_1$, both turning points.}
\label{fig:twopaths}
\end{figure}

The introduction of the quantum periods is very simple. Suppose to have two turning points, $x_0$ and $x_1$, with the point $x$ in which we evaluate the quantum action between the points. We have two natural choices of paths: $\gamma_{0,x}$ and $\gamma_{1,x}$. The path $\gamma_{1,x}$ is defined in the same way as $\gamma_{0,x}$, replacing $x_0$ with $x_1$ everywhere. The only difference that we make is that, in defining the path for the turning point to the right of $x$, we will first go into the negative imaginary half plane, and then cross the branch. The choice between paths is illustrated in figure \ref{fig:twopaths}.

\begin{figure}
\centering
\includegraphics[width=\textwidth]{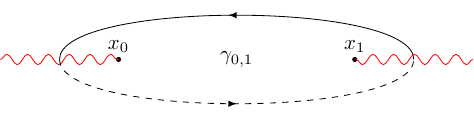}
\caption{The path of integration $\gamma_{0,1}$, running around the two singularities $x_0$ and $x_1$. Contrary to $\gamma_{0,x}$ and $\gamma_{1,x}$, this is a closed path as it crosses the square root branch cuts twice.}
\label{fig:acontour}
\end{figure}

The choice of paths is equivalent, and the difference between choices defines the quantum period. We define $\gamma_{0,1}$ as the complex path enclosing $x_0$ and $x_1$ that is obtained as $\gamma_{0,1}=\gamma_{0,x}-\gamma_{1,x}$. This path is pictured in figure \ref{fig:acontour}.

The base change can be easily understood by first forgetting momentarily about the regularization of the integrals. Elementary integration rules tell us that
\begin{align}
\int_{x_0}^x P_n(t)\dd t=\int_{x_0}^{x_1}P_n(t)\dd t+\int_{x_1}^xP_n(t)\dd t.\label{eq:naivebase}
\end{align}
All the integrals of \eqref{eq:naivebase} are ill defined, as they are divergent. The regularization then instructs us to rewrite the relation as
\begin{align}
\int_{\gamma_{0,x}}P_n(t)\dd t=\oint_{\gamma_{0,1}}P_n(t)\dd t+\int_{\gamma_{1,x}}P_n(t)\dd t.
\end{align}
The integral $\oint_{\gamma_{0,1}}P_n(t)\dd t$ is a component of what is called the \textit{quantum period}, defined as\footnote{Let us stress that the regularization scheme is employed only for $n\neq0$, as for $P_0$ the integral is already convergent and the branch structure is different. This is analogous to ordinary WKB.}
\begin{align}
\Pi^{(i,j)}(\hbar)=\int_{x_i}^{x_j}P_0(t)\dd t+\frac12\sum_{n=1}^\infty\oint_{\gamma_{i,j}}\frac{P_{2n}(t)}{(2n)!}(\ii\hbar)^{2n}\dd t,\label{eq:quant_per_gen}
\end{align}
where we have substituted the turning points $x_0$ and $x_1$ with arbitrary turning points $x_i$ and $x_j$, keeping the order $x_i<x_j$ implicit. In literature, the quantities
\begin{align}
V_{(i,j)}=\exp\left(-\frac{1}{\ii\hbar}\Pi^{(i,j)}\right)
\end{align}
are denoted as \textit{Voros symbols} \cite{voros1983return}, and those functions of the periods appear explicitly in our quantization conditions. Obviously, the properties of the Voros symbols are easily obtainable from the properties of the quantum periods.

\begin{figure}
\centering
\includegraphics[width=\textwidth]{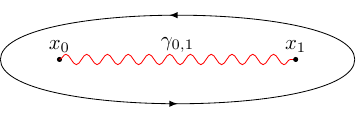}
\caption{The path of integration $\gamma_{0,1}$ where the alternative choice of branch cuts has been made. This contour is also closed, as it never crosses the branch cuts.}
\label{fig:bcontour}
\end{figure}

When setting the branch cuts emanating from the turning points in defining $\gamma_{0,x}$, we have chosen to put them in such a way that $x$ is not on top of them. This restriction was only imposed for ease of plotting, and lifting it comes at no cost. In applications, we will encounter situations in which we make a choice of branch cuts independently of the value of $x$, so $x$ can be on top of branch cuts. The definitions of the various open paths of integration are unchanged, and the final path for the definition of the quantum period in this situation is illustrated in figure \ref{fig:bcontour}.

\section{Resumming the WKB wavefunction}
\label{sec:bor_an}
In the WKB approach for differential equations (that we denote as ordinary WKB), \eqref{eq:wkb_wavefun} generates a formal series in $\hbar$ with zero radius of convergence, even if the integrals are regularized with prescription \eqref{eq:reg_prescr}. We will see in an example in the next chapter that this also holds for a particular example in the WKB approach for finite difference equations (that we denote as deformed WKB). In this thesis, we did not focus on proving the equivalent of those properties, due to the difficulty of computing the coefficients using the algorithm. Nevertheless, we will conjecture that what happens in ordinary WKB also applies in this deformed ansatz.

\eqref{eq:wkb_wavefun} provides a wavefunction of the form
\begin{align}
\psi(x,\hbar)=\exp\left(\frac{1}{\ii\hbar}S_0(x)\right)\exp\left(S_1(x)\right)\sum_{n=0}^\infty\frac{(\ii\hbar)^n}{n!}B_n\left(\frac{S_2(x)}2,...,\frac{S_{n+1}(x)}{n+1}\right).\label{eq:to_resum}
\end{align}
We will conjecture that the coefficients of $(\ii\hbar)^n$ have a factorial growth for each $x$ in the complex plane: more precisely, analogously to \cite{kawai2005algebraic}, we will conjecture that for each open set $U$ on the real line and each compact set in $U^*=\left\{x\in U|Q(x)\neq\pm1\right\}$, there are real positive constants $A_K$ and $S_K$ satisfying
\begin{align}
\sup_{x\in K}\abs{S_n(x)}\leq A_KC^n_Kn!.
\label{eq:fac_growth}
\end{align}
Due to this, the coefficients of $(\ii\hbar)^n$ in \eqref{eq:to_resum} will grow factorially, so for each value of $x$ the convergence radius of the series in \eqref{eq:to_resum} is zero. Thus, as it is, \eqref{eq:to_resum} does not describe a function at any $x$. We can assign a function to \eqref{eq:to_resum} (that is, find a function whose asymptotic expansion is given by \eqref{eq:to_resum} for $\hbar\to0$) through the process of Borel resummation. For this, we will follow \cite{aniceto2019primer}.

Our wavefunctions can be written in terms of two functions, \eqref{eq:basis}. Those functions can be written as
\begin{align}
\psi_{\pm,x_0}(x,\hbar)=\exp\left(\mp\frac{1}{\ii\hbar}\int_{x_0}^xP_0(t,\hbar)\dd t\right)\Phi_{(\pm,x_0)}(x,\hbar),
\end{align}
where $\Phi_{(\pm,x_0)}$ is asymptotic to the formal series \eqref{eq:to_resum}, that we rewrite here for brevity as
\begin{align}
\Phi_{(\pm,x_0)}(x,\hbar)\simeq\sum_{g=0}^\infty\phi_{x_0,g}(x)(\pm \ii\hbar)^g.
\label{eq:as_exp}
\end{align}
The coefficients $\phi_{(x_0,g)}$ have a factorial growth due to \eqref{eq:fac_growth}, so \eqref{eq:as_exp} has zero radius of convergence. We can build a function that is asymptotic to \eqref{eq:as_exp} through the Borel-Padé resummation procedure: we first define
\begin{align}
\mathcal B[\Phi_{(\pm,x_0)}](s,\hbar)=\sum_{g=1}^\infty\frac{\phi_{(x_0,g)}}{(g-1)!}(\pm s)^{g-1}.
\end{align} 
This function has a non vanishing radius of convergence on the complex $s$ plane, so it can be analytically continued to define a function on the $s$ plane, up to isolated singularities. Suppose now that we want to find our solution at a certain phase $\theta$ of $\ii\hbar$. If the line $(0,\ee^{\ii\theta}\infty)$ does not meet a singularity, we can use the analytically extended Borel transform to define the \textit{Borel sum}
\begin{align}
\mathcal S_\theta\Phi_{(\pm,x_0)}(x,\hbar)=\phi_{(x_0,0)}+\int_0^{\ee^{\ii\theta}\infty}\ee^{\mp\frac{s}{\ii\hbar}}\mathcal B[\Phi_{(\pm,x_0)}](s,\hbar)\dd s.
\label{eq:bor_sum}
\end{align}
According to general theory of resurgence, for a large class of functions the Borel transform is well-behaved enough at infinity to make \eqref{eq:bor_sum} converge if no singularities are on the integration path. The function \eqref{eq:bor_sum} has the asymptotic behaviour \eqref{eq:as_exp}, so it can be used to build the solution $\psi_{\pm,x_0}$. We will see in appendix \ref{app:num} a numerical implementation of such computation.

When singularities are on the integration path, \eqref{eq:bor_sum} is ill-defined. In the general theory, the singularities appear at determinate values of $s$, that can be obtained in the following way: first consider the general solution to \eqref{eq:fin_dif_main}, to be written asymptotically as\footnote{We add the index $x_0$ to indicate that the actions are normalized at $x_0$. Furthermore, $q_{x_0}$ will indicate from now on
\begin{align}
q_{x_0}=\exp\left(-\frac{2\pi}{\hbar}\left(x-x_0\right)\right).\label{eq:q_def}
\end{align}}
\begin{align}
\begin{aligned}
\psi_{x_0}(x,\hbar)\simeq&\sum_{n=-\infty}^\infty a_n(\hbar)q_{x_0}^{n}\exp\left(-\frac{1}{\ii\hbar}\int_{x_0}^x P_0(t)\dd t\right)\Phi_{(+,x_0)}+\\
+&\sum_{n=-\infty}^\infty b_n(\hbar)q_{x_0}^{n}\exp\left(\frac{1}{\ii\hbar}\int_{x_0}^x P_0(t)\dd t\right)\Phi_{(-,x_0)}.
\label{eq:as_exp_big}
\end{aligned}
\end{align}
Expanding the $q_{x_0}$, we can rewrite
\begin{align}
\psi_{x_0}(x,\hbar)\simeq&\sum_{n=-\infty}^\infty a_n(\hbar)\exp\left(-\frac{1}{\ii\hbar}\left(\int_{x_0}^x P_0(t)\dd t+2\pi\ii n \left(x-x_0\right)\right)\right)\Phi_{(+,x_0)}+\\
+&\sum_{n=-\infty}^\infty b_n(\hbar)\exp\left(-\frac{1}{\ii\hbar}\left(-\int_{x_0}^x P_0(t)\dd t+2\pi\ii n \left(x-x_0\right)\right)\right)\Phi_{(-,x_0)}.
\label{eq:transseries}
\end{align}
In the terminology of \cite{aniceto2019primer}, \eqref{eq:transseries} is a \textit{transseries expansion} of $\psi_{x_0}(x,\hbar)$. The series $\Phi_{(\pm,x_0)}$ are denoted as \textit{sectors}, and the terms given by
\begin{align}
A_{x_0}^{(\pm,n)}=\pm\int_{x_0}^xP(t,\hbar)\dd t+2\pi\ii n\left(x-x_0\right),\label{eq:action}
\end{align}
are the \textit{actions} of the transseries\footnote{In a more general transseries, the sectors would also depend on $n$, as each action is allowed to have its own sector. In our expansion, once a quantum action $P_{\text T}(x,\hbar)$ is determined all possible quantum actions are of the form $\pm P_{\text T}(x,\hbar)+2\pi\ii n$. As the sectors $\Phi_{(\pm,x_0)}$ are determined by the coefficients $P_m$ with $m>0$, the sectors are independent from the choice of $n$, and only depend on the choice of sign. As a consequence, we have only two distinct sectors once the base point is fixed, $\Phi_{(+,x_0)}$ and $\Phi_{(-,x_0)}$.}.

\begin{figure}
\centering
\includegraphics{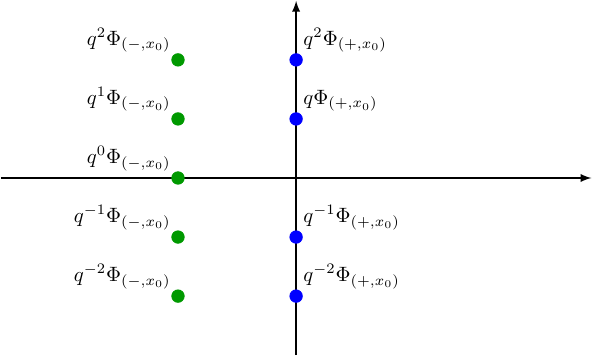}
\caption{Array of singularities for $\mathcal B[\Phi_{(+,x_0)}]$ on the $s$ plane when $x$ is in a classically allowed zone. All the singularities are at $A_{x_0}^{(n,\pm)}-A_{x_0}^{(0,+)}$, so the singularities of the $+$ sectors are on the imaginary axis, while the singularities of the $-$ sectors are on a vertical axis parallel to the imaginary axis, but distinct from it.}
\label{fig:borel_sings_big1}
\end{figure}

\begin{figure}
\centering
\includegraphics{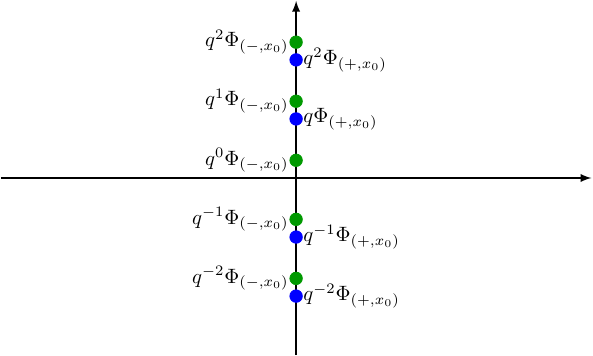}
\caption{Array of singularities for $\mathcal B[\Phi_{(+,x_0)}]$ on the $s$ plane when $x$ is in a classically forbidden zone. All the singularities are at $A_{x_0}^{(\pm,n)}-A_{x_0}^{(+,0)}$ and as $A_{x_0}^{(-,0)}-A_{x_0}^{(+,0)}$ is purely imaginary, they are all on the imaginary axis.}
\label{fig:borel_sings_big2}
\end{figure}

The actions dictate where the singularities for the Borel transform of the sector $\Phi_{(+,x_0)}$ will appear at the points $2\pi\ii n(x-x_0)$ (with the exception of $n=0$) and $A_{x_0}^{(-,n)}-A_{x_0}^{(+,0)}$, and likewise for the Borel transform of the sector $\Phi_{(-,x_0)}$ we will have singularities at $2\pi\ii n(x-x_0+1)$ (again, excepting $n=0$) and $A_{x_0}^{(+,n)}-A_{x_0}^{(-,0)}$. We picture the array of singularities for the Borel transform of $\Phi_{(+,x_0)}$ in figures \ref{fig:borel_sings_big1} and \ref{fig:borel_sings_big2}.

\begin{figure}
\centering
\includegraphics{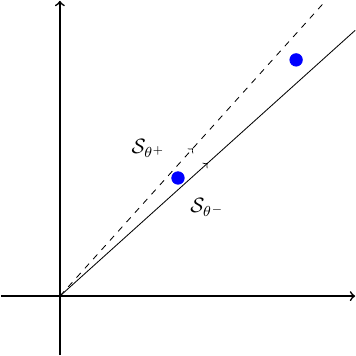}
\includegraphics{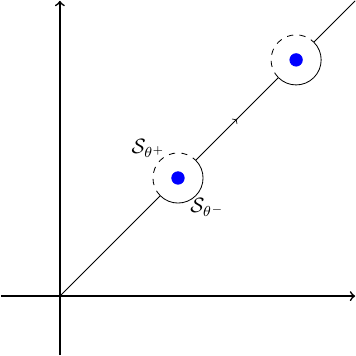}
\caption{Avoiding the singularities in the $\theta$ direction. On the left, we have two paths approaching the paths for the left resummation $\mathcal S_{\theta^+}$ and for the right resummation $\mathcal S_{\theta^-}$, avoiding the singularities on their left and right respectively. Those paths are equivalent to the paths on the right, in which the straight lines are common to both paths, the dashed semicircles belong to $\mathcal S_{\theta^+}$ and the continuous semicircles belong to $\mathcal S_{\theta^-}$.}
\label{fig:avoiding_sings}
\end{figure}

As stated, we cannot perform the integration on a line $(0,\ee^{\ii\theta}\infty)$ when a singularity is in the integration path. We can instead define a \textit{left resummation} $\mathcal S_{\theta^+}$ and a \textit{right resummation} $\mathcal S_{\theta^-}$ by deforming the contour in order to avoid the singularity, as in figure \ref{fig:avoiding_sings}. Due to the presence of singularities, the two resummations do not coincide, and the function $\mathcal S_\theta$ describing the resummation at angle $\theta$ has a discontinuity at values of the angles for which a singularity is met. This discontinuity is measured by the \textit{Stokes automorphism}, defined by
\begin{align}
\mathcal S_{\theta^+}=\mathcal S_{\theta^-}\circ\underline{\mathfrak S}_\theta.
\end{align}
The action of the Stokes automorphism is to reshuffle the various sectors in the asymptotic expansion, changing the constants $a_n$ and $b_n$ of the asymptotic expansion \eqref{eq:as_exp_big}. In particular, according to \cite{aniceto2019primer}, the Stokes automorphism $\underline{\mathfrak S}_{\theta}$ acts through a series of $\hbar$-dependent constants called the \textit{Borel residues}: an example of such an action is, for $x$ in the classically allowed region
\begin{align}
\exp\left(-\frac{1}{\ii\hbar}A_{x_0}^{(-.0)}\right)\underline{\mathfrak S}_0\Phi_{(-,0)}=-\exp\left(-\frac{1}{\ii\hbar}A_{x_0}^{(+,0)}\right)\mathrm S^{(x_0)}_{(-,0)\to(+,0)}\Phi_{(+,0)}.
\end{align}
Here $\mathrm S^{(x_0)}_{(-,0)\to(+,0)}$ is the Borel residue describing the transition from $(-,0)$ to $(+,0)$, and this is the only residue that appears due to the fact that $\Phi_{(+,0)}$ is the only singularity encountered in the $0$ direction from $(0,-)$. The Borel residues can depend on $x$.

The effect on the transseries can then be seen by applying the Stokes automorphism to \eqref{eq:as_exp_big}:
\begin{align}
\begin{aligned}
\underline{\mathfrak S}_0\psi{(x,\hbar)}=&\sum_{n=-\infty}^{\infty}(a_n-\mathrm S^{(x_0)}_{(-,n)\to(+,n)}b_n)\exp\left(-\frac{1}{\ii\hbar}A_{x_0}^{(+,n)}\right)\Phi_{(+,x_0)}+\\
+&\sum_{n=-\infty}^{\infty}b_n\exp\left(-\frac{1}{\ii\hbar}A^{(-,n)}_{x_0}\right)\Phi_{(-,x_0)}.
\end{aligned}
\end{align}

The effect of the Stokes automorphism then amounts to changing the boundary conditions of the solution, in order to obtain a different solution to the finite difference equation. Once an initial condition is established, the Stokes automorphism can then be used to define a function that is continuous on the $\hbar$ plane, by starting with a given resummation $\mathcal S_{\theta}\psi(x,\hbar)$ and applying the Stokes autmorphism to the asymptotic expansion $\psi_{x_0}(x,\hbar)$ whenever singularities are crossed. As an example, if there is a singularity at $\theta^*$, the solution can be specified by first choosing an asymptotic expansion $\psi_{x_0}(x,\hbar)$ that will hold at $\theta<\theta^*$ (until other singularities are encountered), and then the asymptotic expansion for $\theta>\theta^*$ will be given by $\underline{\mathfrak S}_{\theta^*}\psi_{x_0}(x,\hbar)$ (again, until other singularities are encountered). The resummed function will then be $\mathcal S_\theta\psi_{x_0}(x,\hbar)$ for $\theta<\theta^*$ and $\mathcal S_\theta\underline{\mathfrak S}_{\theta^*}\psi_{x_0}(x,\hbar)$ for $\theta>\theta^*$. The resulting function will be continuous in $\theta$, and will be defined at $\theta^*$ by continuity. We will study in great detail the Borel residues of a particular finite difference equation in chapter \ref{chapter3}.

We conclude this part by noting the fact that the quantum periods $\Pi^{(i,j)}(\hbar)$ also have non trivial resurgent properties. This is because the Borel residues in general depend on the base point chosen for the integration. The resurgent properties of the quantum periods of ordinary WKB are collected in the Dillinger-Delabaere-Pham formula \cite{delabaere1999resurgent}. While we did not manage to get a similar equation for our deformation of WKB, we will study the properties of the quantum periods for the harmonic oscillator in chapter \ref{chapter5}.

%% file: 3connformulae.tex
\chapter{The linear potential}
\label{chapter3}
In the WKB approach to the standard Schr\"odinger equation, the linear potential $V(x)=x$ is of great importance, due to the approach of \textit{deformation theory}. A general standard Schr\"odinger equation of the form
\begin{align}
-\frac{\hbar^2}2\psi''(x,\hbar)+V(x)\psi(x,\hbar)=E\psi(x,\hbar),
\end{align}
can be transformed into a standard Schr\"odinger equation with a linear potential
\begin{align}
-\frac{\hbar^2}2\psi''(\phi,\hbar)+\phi\psi(\phi,\hbar)=E\psi(\phi,\hbar),\label{eq:gen}
\end{align}
where $\varphi=\varphi(x,\hbar)$ is a deformation of the $x$ coordinate that can be computed through a non linear equation. We will detail this procedure and give the equation for $\varphi$ in chapter \ref{chapter4}. Equation \eqref{eq:gen} can be solved by
\begin{align}
\psi(x,\hbar)=A(\hbar)\Ai\left(-\frac{2^{\frac13}}{(\hbar)^\frac23}(E-\phi(x,\hbar))\right)+B(\hbar)\Bi\left(-\frac{2^{\frac13}}{(\hbar)^\frac23}(E-\phi(x,\hbar))\right),
\end{align}
with $\Ai$ and $\Bi$ the Airy functions. Airy functions have various definitions in literature: we can define them as the integrals
\begin{align}
&\Ai(z)=\frac1{2\pi\ii}\int_{\infty \ee^{-\frac{2\pi\ii}3}}^{\infty \ee^{\frac{2\pi\ii}3}}\exp\left(\frac13t^3-zt\right)\dd t,\\
&\Bi(z)=\frac1{2\pi}\int_{-\infty }^{\infty \ee^{\frac{2\pi\ii}3}}\exp\left(\frac13t^3-zt\right)\dd t+\frac1{2\pi}\int_{-\infty }^{\infty \ee^{-\frac{2\pi\ii}3}}\exp\left(\frac13t^3-zt\right)\dd t,
\end{align}
where $z$ is a complex variable. The reader can refer to \cite{NIST:DLMF} for additional information on their definitions, properties and asymptotic expansions. The asymptotic expansion of the Airy functions is very important to determine the asymptotic expansions of the solutions to a general standard Schr\"odinger equation. As our goal is to follow the WKB procedure for the standard Schr\"odinger equation, we will begin by examining the finite difference Schr\"odinger equation with linear potential $V(x)=gx$, where $g$ is a positive or negative coupling constant.

We start by rewriting \eqref{eq:fin_dif_main} with the linear potential as
\begin{align}
\psi(x+\ii\hbar,\hbar)+\psi(x-\ii\hbar,\hbar)=2(E-gx+1)\psi(x,\hbar).
\label{eq:lin_fin_dif}
\end{align}
The Bessel functions $J_\nu(z)$ and $Y_\nu(z)$ are defined in the following way: $J_\nu(z)$ is given by the series
\begin{align}
J_\nu(z)=\left(\frac12z\right)^\nu\sum_{k=0}^\infty(-1)^k\frac{\left(\frac14z\right)^\nu}{k!\Gamma(\nu+k+1)},
\end{align}
defining an analytic function of $z\in\mathbb C$, except for a branching point at $z=0$ present when $\nu$ is non integer (the principal branch of this function is defined by taking the principal branch of $z^\nu$). $Y_\nu(z)$ is defined by
\begin{align}
Y_\nu(z)=\frac{J_\nu(z)\cos(\nu\pi)-J_{-\nu}(z)}{\sin(\nu\pi)},
\end{align}
where the expression on the RHS is replaced by a limit procedure when $\nu$ is an integer $n$, giving
\begin{align}
Y_n(z)=\frac1\pi\left.\frac{\partial J_\nu(z)}{\partial\nu}\right|_{\nu=n}+\frac{(-1)^n}\pi\left.\frac{\partial J_\nu(z)}{\partial\nu}\right|_{\nu=-n}.
\end{align}
The main properties of the Bessel functions are described in \cite{NIST:DLMF}. In particular, we will use the following property:
\begin{align}
J_{\frac{y}{a}+1}\left(\frac1a\right)+J_{\frac{y}{a}-1}\left(\frac1a\right)=2yJ_{\frac ya}\left(\frac1a\right).
\label{eq:bessel_prop}
\end{align}
The same property holds for $Y$. It is easy to see that we can take the most general solution to \eqref{eq:lin_fin_dif} to be
\begin{align}
\psi(x,\hbar)=A(x,\hbar)J_{\frac{E-g x+1}{\ii g\hbar}}\left(\frac{1}{\ii g\hbar}\right)+B(x,\hbar)Y_{\frac{E-g x+1}{\ii g\hbar}}\left(\frac{1}{\ii g\hbar}\right),
\end{align}
where as always $A$ and $B$ are periodic in their $x$ dependency, with period $\ii\hbar$.

As the Bessel functions solve the finite difference Schr\"odinger equation with linear potential, we can see them as parallels of the Airy functions, and use their asymptotic expansions to write general asymptotic expansions for our WKB ansatz. Once the importance of the Bessel functions is established, the second step is obtaining their asymptotic expansions and their resurgence properties. For this, we will combine the ordinary Debye expansions for the Bessel functions and an useful integral representation given in \cite{howls1999resurgence}.

\section[Integral representation]{Integral representation and resurgence properties}
\label{sec:int_rep}
\begin{figure}
\centering
\includegraphics[scale=0.6]{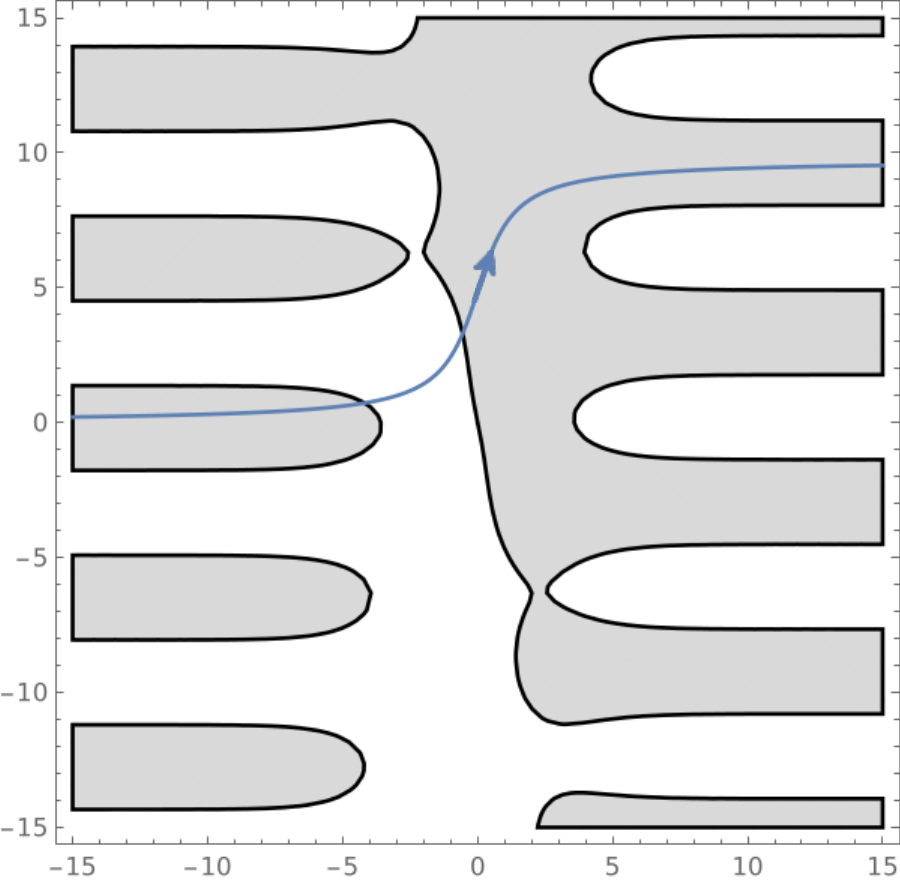}
\caption{A possible oriented integration contour for the integral $I_\gamma(y,a)$ in the $\tau$ plane. Here we have chosen $y=5$ and $\theta=0.2$. The integration path $\gamma$ (in blue) goes to infinity in regions in which \eqref{eq:exp_decay} holds: those regions are pictured in light grey in the plot.}
\label{fig:paths}
\end{figure}

The Bessel functions can be written in terms of the integral
\begin{align}
I_\gamma(y,a)=\int_{\gamma}\exp\left(-\frac1a\left(y\tau-\sinh\tau\right)\right)\dd\tau,
\label{eq:int_rep}
\end{align}
where the complex contour $\gamma$ starts and ends at infinity in such a way that the integral converges. Such an integral can be studied with the techniques developed in \cite{berry1991hyperasymptotics} where $a$ is the expansion parameter, together with \cite{howls2004higher} to deal with the case in which the integrated function depends on the expansion parameter and an additional parameter $y$. We will take $a$ to be a complex parameter of phase $\theta$, while $y$ will be a real parameter. In order to ensure convergence of the integral, the path $\gamma$ must go to infinity in regions in which
\begin{align}
\Re\left(\ee^{-\ii\theta}\left(y\tau-\sinh\tau\right)\right)>0:
\label{eq:exp_decay}
\end{align}
this ensures that the integral converges. We picture an example of this in figure \ref{fig:paths}.

We can see that \eqref{eq:int_rep} has the property \eqref{eq:bessel_prop}. To see that, we compute
\begin{align}
I_\gamma(y+a,a)+I_\gamma(y-a,a)=\int_\gamma\cosh\tau\exp\left(-\frac1a\left(y\tau-\sinh\tau\right)\right)\dd\tau.
\end{align}
We can perform an integration by part to obtain\footnote{The evaluation on the boundaries of $\gamma$ gives zero due to the convergence at infinity of the integral.}
\begin{align}
I_\gamma(y+a,a)+I_\gamma(y-a,a)=2y\int_\gamma\exp\left(-\frac1a(y\tau-\sinh\tau)\right)\dd\tau=2yI_\gamma(y,a).
\end{align}

\subsection{Saddle points and steepest descent contours}

Integrals such as \eqref{eq:int_rep} are usually written in terms of \textit{steepest descent contours}. As the integrand is an entire function of $\tau$, the integration path can be modified arbitrarily, provided that its endpoints are kept the same. We will exploit this property to write our integration path in a convenient way, as a composition of steepest descent contours.

First, we have to define \textit{saddle points} for the integration. Let us denote
\begin{align}
v(y,\tau)=y\tau-\sinh\tau,
\end{align}
that is the factor in the integrand. Saddle points of exponential integrals are defined as the zeroes in $\tau$ of the $\tau$ derivative of $v(x,\tau)$. Those can be trivially computed: all saddle points can be written as
\begin{align}
\tau^{(\pm,n)}=\pm\arccosh y +2\pi\ii n.
\end{align}
At those saddles, we can define
\begin{align}
v^{(\pm,n)}=v(\tau^{(\pm,n)},y)=y\arccosh y-\sqrt{y^2-1}+2\pi\ii n y.
\end{align}
While this representation is convenient for $y>1$, in the region $-1<y<1$ it is more convenient to use the analytic continuation
\begin{align}
v^{(\pm,n)}=v(\tau^{(\pm,n)},y)=\ii(y\arccos y-\sqrt{1-y^2}+2\pi n y).
\end{align}
Steepest descent contours are paths $\gamma^{(n,\pm)}$ defined by
\begin{align}
\gamma^{(\pm,n)}=\{\tau\in\mathbb C|\Im(\ee^{-\ii\theta}(v(\tau,y)-v^{(\pm,n)}))=0,\Re(\ee^{-\ii\theta}(v(\tau,y)-v^{(\pm,n)}))>0\}.
\label{eq:steep_cont}
\end{align}
Steepest descent contours can also be obtained as the flows of the vector field \cite{aniceto2019primer}
\begin{align}
X^{(\pm,n)}=-2\frac{\dd}{\dd\bar\tau}\Re\left(-\frac1a(v(\tau,y)-v^{(\pm,n)})\right).
\label{eq:or_vec}
\end{align}
Using the vectors $X^{(\pm,n)}$, we are also able to define an orientation on the steepest descent contours.

\begin{figure}
\centering
\includegraphics[scale=0.5]{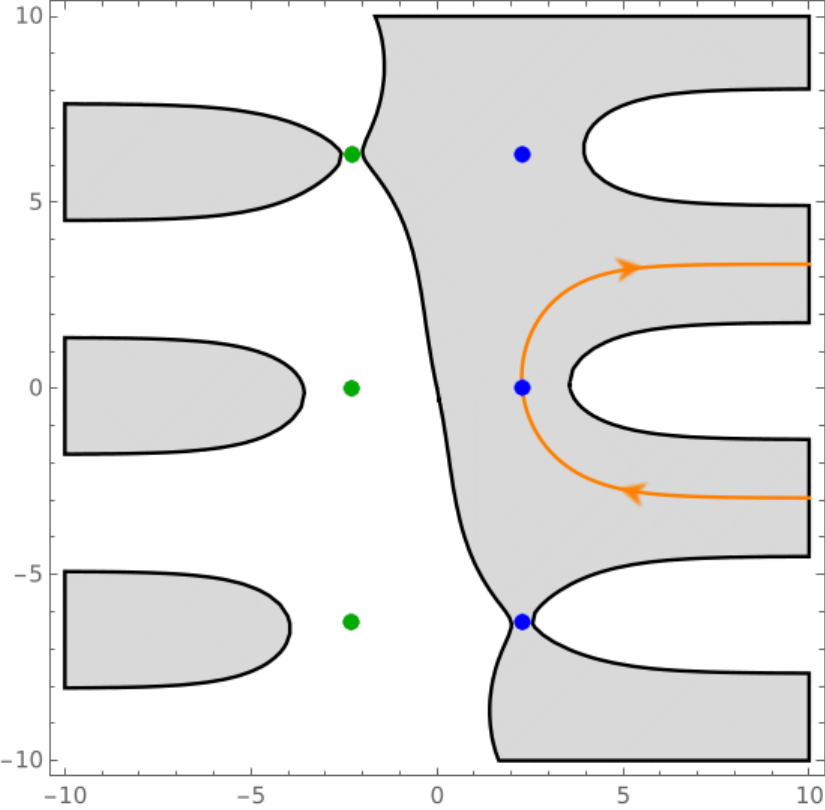}
\includegraphics[scale=0.5]{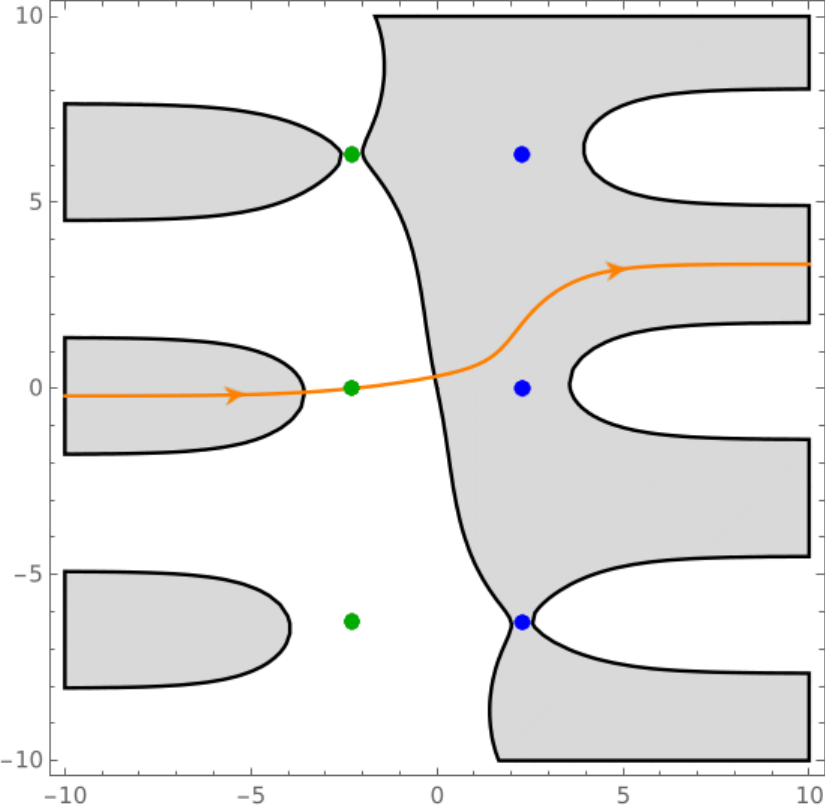}
\caption{$\tau$ plane with the saddle points $\tau^{(-,n)}$ (on the left) and $\tau^{(-,n)}$ (on the right) with $n=-1,0,1$ in blue. In the left picture, we plot steepest descent contour $\gamma^{(+,0)}$ in orange. In the same color, we plot in the right picture the steepest descent contour $\gamma^{(-,0)}$. We have chosen $y=5$ and $\theta=0.2$. The orientation of the arrows is determined by the vector field \eqref{eq:or_vec}.}
\label{fig:steep}
\end{figure}

We select an orientation on those paths by choosing the counter clockwise orientation as the default orientation. We plot examples of such steepest contours in figure \ref{fig:steep}. Any integration contour $\gamma$ can then be written as
\begin{align}
\gamma=\sum_{n=-\infty}^\infty c_n^{+}\gamma^{(+,n)}+\sum_{n=-\infty}^\infty c_n^{-}\gamma^{(-,n)},
\end{align}
where $c_{n}^{\pm}$ can only be $-1,1$ or $0$. As an example, the path of figure \ref{fig:paths} can be written as $\gamma^{(+,1)}+\gamma^{(-,0)}$.

\subsection{From steepest descents to asymptotic expansion}

Splitting the path $\gamma$ in steepest descent contours allows us to get the asymptotic behaviour of any function $I_\gamma(y,a)$ solely from the path decomposition. Following \cite{berry1991hyperasymptotics}, we can define the functions
\begin{align}
&T^{(\pm,n)}=\frac{1}{\sqrt a}\int_{\gamma^{(\pm,n)}}\exp\left(-\frac1a(v(\tau,y)-v^{(\pm,n)})\right)\dd\tau,\\
&I^{(\pm,n)}=\sqrt a\exp\left(-\frac1a v^{(\pm,n)}\right)T^{(\pm,n)}.\label{eq:int_def_bor}
\end{align}
As per the cited article, $T^{(\pm,n)}$ are single-valued functions of $a$, while $I^{(\pm,n)}$ are double-valued. The integral $I_\gamma$ is then decomposed as
\begin{align}
I_\gamma(y,a)=\sum_{n=-\infty}^\infty c_n^+ I^{(+,n)}(y,a)+\sum_{n=-\infty}^\infty c_n^- I^{(-,n)}(y,a).
\end{align}
The advantage of this decomposition is that there is a simple procedure for getting the asymptotic expansion of $T^{(\pm,n)}(y,a)$, given by
\begin{align}
T^{(\pm,n)}(y,a)\simeq\sum_{k=0}^\infty T_k^{(\pm,n)}(y)a^k,
\end{align}
where the coefficients $T_k^{(\pm,n)}$ are given by
\begin{align}
T_k^{(\pm,n)}=\frac{\Gamma\left(k+\frac12\right)}{2\pi\ii}\oint_{\gamma^{(\pm,n)}}\frac{1}{(v(\tau,y)-v^{(\pm,n)})^{k+\frac12}}\dd\tau.
\end{align}
The integrals in $T_k^{(\pm,n)}$ are loop integrals, and due to the fact that at saddle points the derivative of $v$ with respect to $\tau$ vanishes the term inside the brackets in the denominator of the integral is always of order $\tau^2$, so the square root gets cancelled and the integral can be evaluated through residue computation. It is easy to see that coefficients associated to the same sign are equal between them: this is because we can absorb the $2\pi\ii ny$ of $v^{(\pm,n)}$ inside $v(\tau,y)$ exploiting the periodicity of $\sinh\tau$, obtaining
\begin{align}
T_k^{(\pm,n)}=\frac{\Gamma\left(k+\frac12\right)}{2\pi\ii}\oint_{\gamma^{(\pm,n)}}\frac{1}{(v(\tau+2\pi\ii n,y)-v^{(\pm,0)})^{k+\frac12}}\dd\tau.
\end{align}
With a translation $\tau\to\tau-2\pi\ii n$, $\gamma^{(\pm,n)}$ becomes $\gamma^{(\pm,0)}$, so we obtain
\begin{align}
T^{(\pm,n)}_k=T^{(\pm,0)}_k
\end{align}
for every $n$: as a consequence, we can drop the $n$ label from the coefficients $T_k^{(\pm,n)}$ and the functions $T^{(\pm,n)}$. We will have only two sets of coefficients $T_k^{(\pm)}$ and two functions $T^{(\pm)}$, so the asymptotic expansion of $I_\gamma$ can be rewritten as
\begin{align}
\begin{aligned}
I_\gamma(y,a)\simeq&\sum_{n=-\infty}^\infty c_n^+\exp\left(-\frac{2\pi\ii n}ay\right)\exp\left(-\frac1a v^{(+,0)}\right)\sum_{g=0}^\infty T_g^{(+)}a^{g+\frac12}+\\
+&\sum_{n=-\infty}^\infty c_n^-\exp\left(-\frac{2\pi\ii n}ay\right)\exp\left(\frac1a v^{(+,0)}\right)\sum_{g=0}^\infty T_g^{(-)}a^{g+\frac12}
\end{aligned}
\end{align}
We have also used here $v^{(\pm,n)}=v^{(\pm,0)}+2\pi\ii ny$ and $v^{(+,0)}=-v^{(-,0)}$. This expansion is exactly of the form of the WKB expansion discussed in chapter \ref{chap:1}, as expected from the fact that $I_\gamma$ describes Bessel functions, that come out of a WKB problem with a linear potential.

We conclude this part by giving some coefficients $T_k^{(\pm)}$. Those can be computed algorithmically by residue computation, using the generalized binomial theorem to extract the residue. The first terms of the expansion are, simplified in the region $y>1$
\begin{align}
\begin{aligned}
&T_0^{(+)}=-\frac{\ii\sqrt{2\pi }}{(y^2-1)^\frac14}=-\ii T_0^{(-)},\\
&T_1^{(+)}=-\frac{\ii\sqrt{2\pi }( 2 y^2+3)}{24 \left(y^2-1\right)^{\frac{7}{4}}}=\ii T_1^{(-)},\\
&T_2^{(+)}=-\frac{\ii\sqrt{2\pi }4 \left(y^2+75\right)y^2+81}{1152\left(y^2-1\right)^{\frac{13}{4}}}=-\ii T_2^{(-)},\\
&T_3^{(+)}=-\frac{\ii\sqrt{2\pi }(1112 y^6-117684y^4-278478 y^2-30375)}{414720 \left(y^2-1\right)^{\frac{19}{4}}}=\ii T_3^{(-)}.
\end{aligned}
\label{eq:coef_steep}
\end{align}
Up to a multiplicative constant, the coefficients are the same of the Debye expansions of the Bessel functions. Those coefficients will allow us to rewrite the Debye expansions of the Bessel functions in terms of the asymptotic expansions of integrals $I_\gamma(y,a)$ by choosing an appropriate $\gamma$ path.

\subsection{The Stokes phenomenon}

The steepest descent contours depend on the phase of $a$, $\theta$. For almost any value of $\theta$, the steepest descent paths are as pictured in figure \ref{fig:steep}: they never present bifurcations, and only cross one saddle point. There is a discrete set of values of $\theta$ for which this does not happen. The description of the same path $\gamma$ in terms of steepest descent contours is modified when $\theta$ crosses those thresholds: this is an example of the \textit{Stokes phenomenon} in the context of steepest descent integrals. As we will see, this will change the asymptotic behaviour of the function $I_\gamma$.

\begin{figure}
\centering
\includegraphics[scale=0.6]{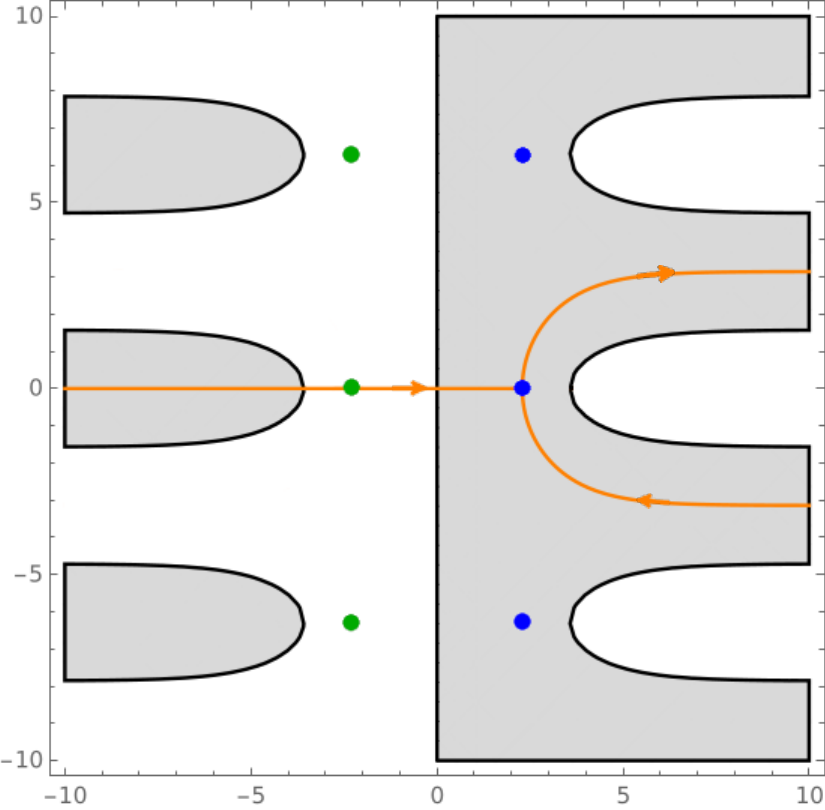}
\caption{Stokes phenomenon. At $\theta=0$, the steepest descent contour $\gamma^{(-,0)}$ passes through the saddle $\tau^{(+,0)}$. We have chosen $y=5$.}
\label{fig:stokes_phen_1}
\end{figure}

The Stokes phenomenon happens when two saddles are on the same steepest descent contour. This happens whenever
\begin{align}
\Im\left(\ee^{-\ii\theta}(v^{(s_1,n)}-v^{(s_2,m)})\right)=0,\quad \Re\left(\ee^{-\ii\theta}(v^{(s_1,n)}-v^{(s_2,m)})\right)>0,
\end{align}
where $s_1$ and $s_2$ are signs (variables that can only take value $+$ or $-$). In this case, the steepest descent contour from $\tau^{(s_2,m)}$ will flow into $\tau^{(s_1,n)}$ and bifurcate at that point. We picture such an example in figure \ref{fig:stokes_phen_1}.

\begin{figure}
\centering
\includegraphics[scale=0.6]{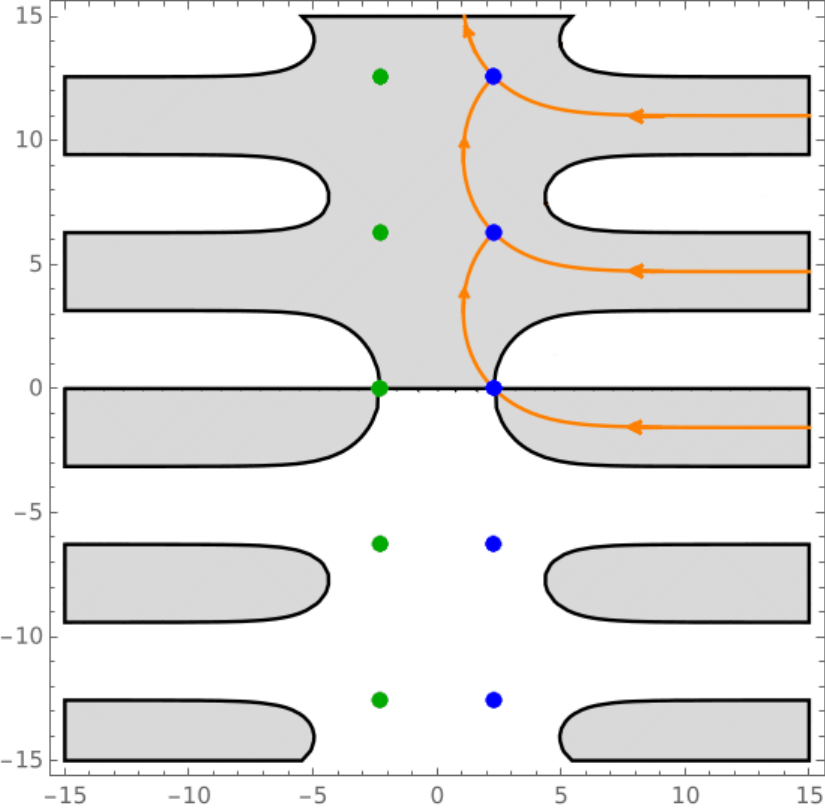}
\caption{Another example of Stokes phenomenon. At $\theta=\frac{\pi}2$, the steepest descent contour $\gamma^{(+,0)}$ passes through all the saddles saddle $\tau^{(+,n)}$ with $n$ strictly positive. We have chosen $y=5$.}
\label{fig:stokes_phen_2}
\end{figure}

\begin{figure}
\centering
\includegraphics[scale=0.6]{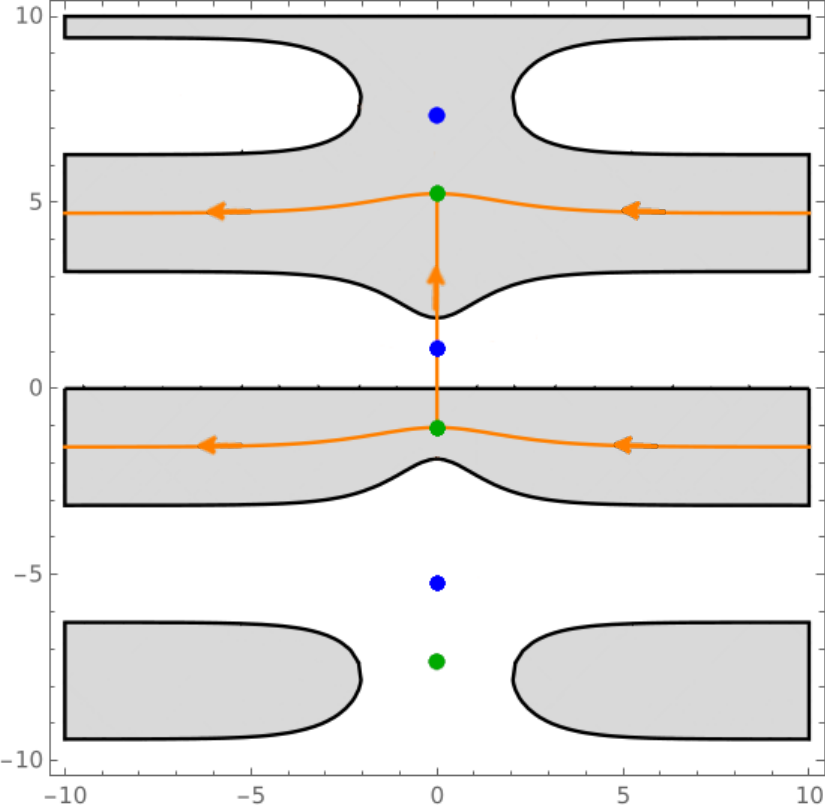}
\caption{Another example of Stokes phenomenon. At $\theta=\frac{\pi}2$, the steepest descent contour $\gamma^{(+,0)}$ passes through the saddles $\tau^{(-,0)}$ and $\tau^{(-,1)}$. We have chosen $y=\frac12$.}
\label{fig:stokes_phen_3}
\end{figure}

There are different values of $\theta$ at which we have a Stokes phenomenon. In general, a Stokes phenomenon for the path $\gamma^{(s_2,m)}$ will happen whenever $\theta=\arg(v^{(s_1,n)}-v^{(s_2,m)})$. If we take $y>1$, when $s_1$ and $s_2$ are different, the situation is similar to figure \ref{fig:stokes_phen_1}. When $s_1=s_2$, however, we have a different situation: in that case we can either have $\theta=\frac\pi2$ or $\theta=-\frac\pi2$, depending on the sign of $y$. In this case we have a Stokes phenomenon in which many steepest descent paths are conjoined, as pictured in figure \ref{fig:stokes_phen_2}. Lastly, when $0<y<1$ all the saddles are on the same vertical line, so the Stokes phenomenon only happens for $\theta=\pm\frac{\pi}2$. In this case, a saddle point is conjoined to two different saddles. We picture this situation in figure \ref{fig:stokes_phen_3}.

We now compute the effect of the Stokes phenomenon, by observing how the steepest descent paths get deformed when $\theta$ crosses a value at which the Stokes phenomenon appears. There are three possible cases, that we will examine separately.

\subsubsection{Case 1: $y>1$, $\theta\neq\pm\frac{\pi}2$}

\begin{figure}
\centering
\includegraphics[scale=0.6]{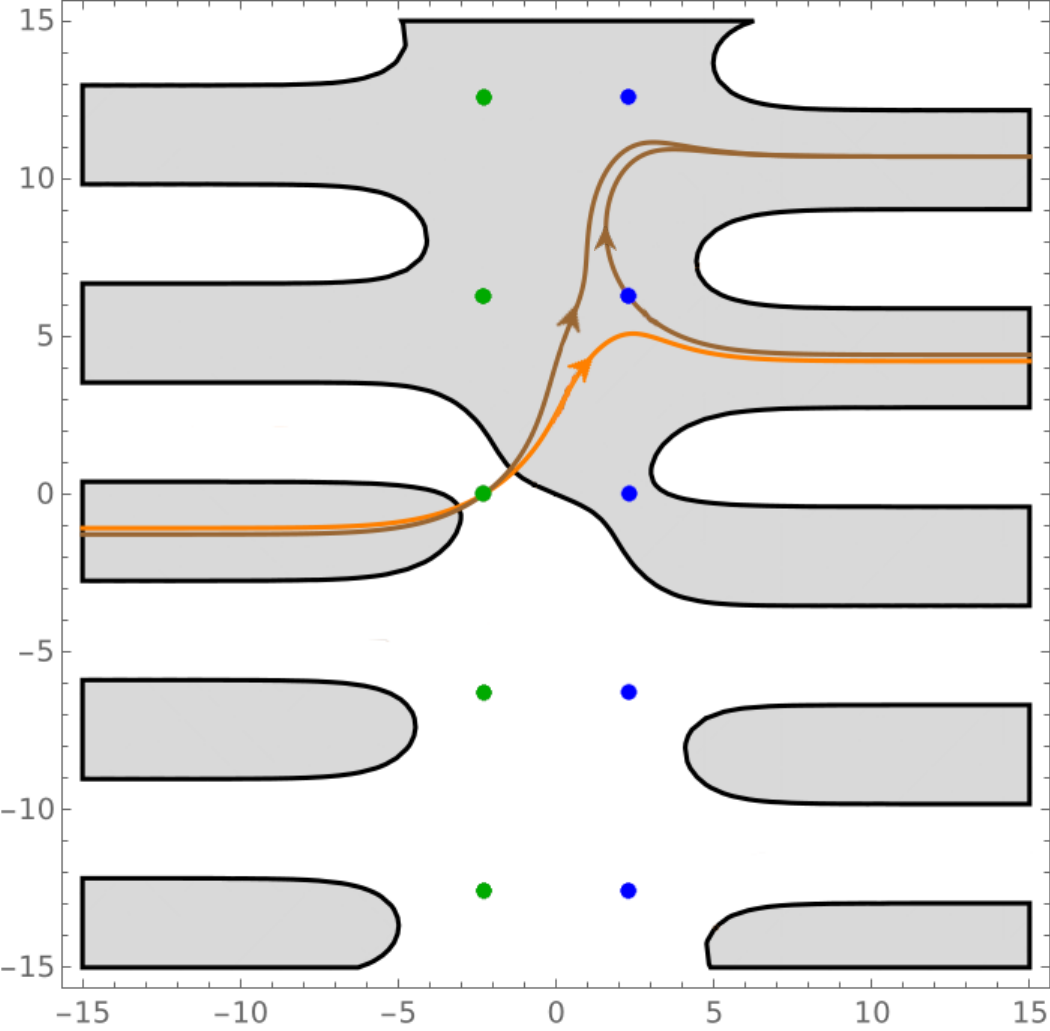}
\caption{Computation of the effect of the Stokes phenomenon between $\tau^{(-,0)}$ and $\tau^{(+,1)}$. The orange line is $\gamma^{(-,0)}$ for $\theta^*-\epsilon$ (here $\epsilon=0.1$), while the brown path passing through $\tau^{(-,0)}$ is the path $\gamma^{(-,0)}$ for $\theta^*+\epsilon$, and the brown path passing through $\tau^{(+,1)}$ is the path $\gamma^{(+,1)}$ (that undergoes no Stokes phenomenon, so we do not need to specify the deformation). We have chosen $y=5$.}
\label{fig:stokes_comp_1}
\end{figure}

This case has been pictured in figure \ref{fig:stokes_phen_3}, but we will choose a situation in which the saddles participating in the Stokes phenomenon are not horizontally aligned, to give a more generic picture (even if this does not change the final result). We start by picking
\begin{align}
\theta^*=\arg(v^{(+,n)}-v^{(-,0)}).
\end{align}
We compute the paths of steepest descent after deformation of $\theta^*$ by a small positive value. We picture the computation in figure \ref{fig:stokes_comp_1}. We can recover the path before the deformation by combining the two pictured paths. The Stokes phenomenon then amounts to
\begin{align}
\gamma^{(-,0)}\to\gamma^{(-,0)}-\gamma^{(+,n)}.
\label{eq:path_jump_1}
\end{align}
This happens for positive and negative values of $n$, and the same thing also happens for the paths $\gamma^{(+,0)}$: after crossing the angle $\arg(v^{(-,n)}-v^{(+,0)})$, we have
\begin{align}
\gamma^{(+,0)}\to\gamma^{(+,0)}-\gamma^{(-,n)}.
\end{align}

\subsubsection{Case 2: $y>1, \theta=\pm\frac\pi2$}

\begin{figure}
\centering
\includegraphics[width=0.49\textwidth]{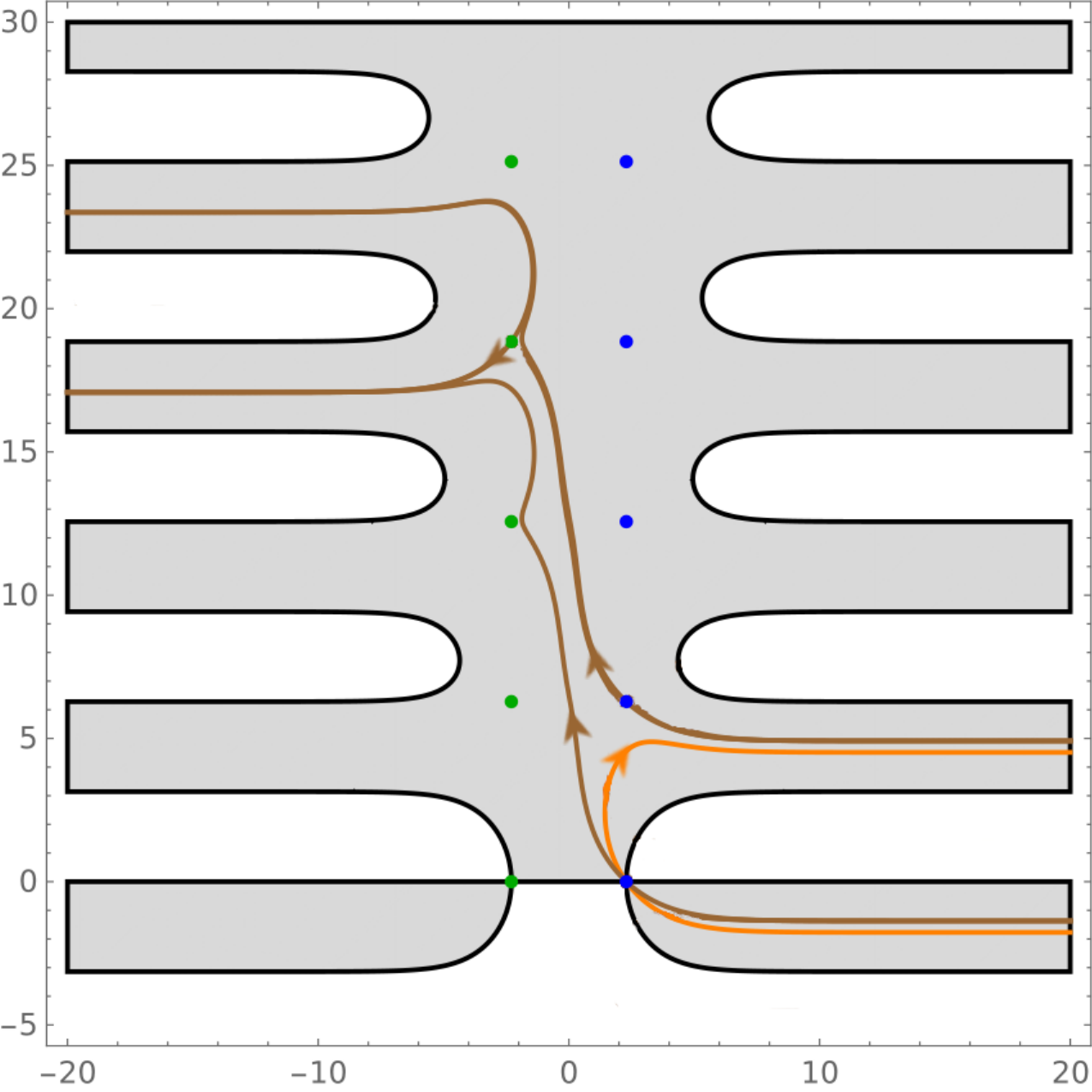}
\includegraphics[width=0.49\textwidth]{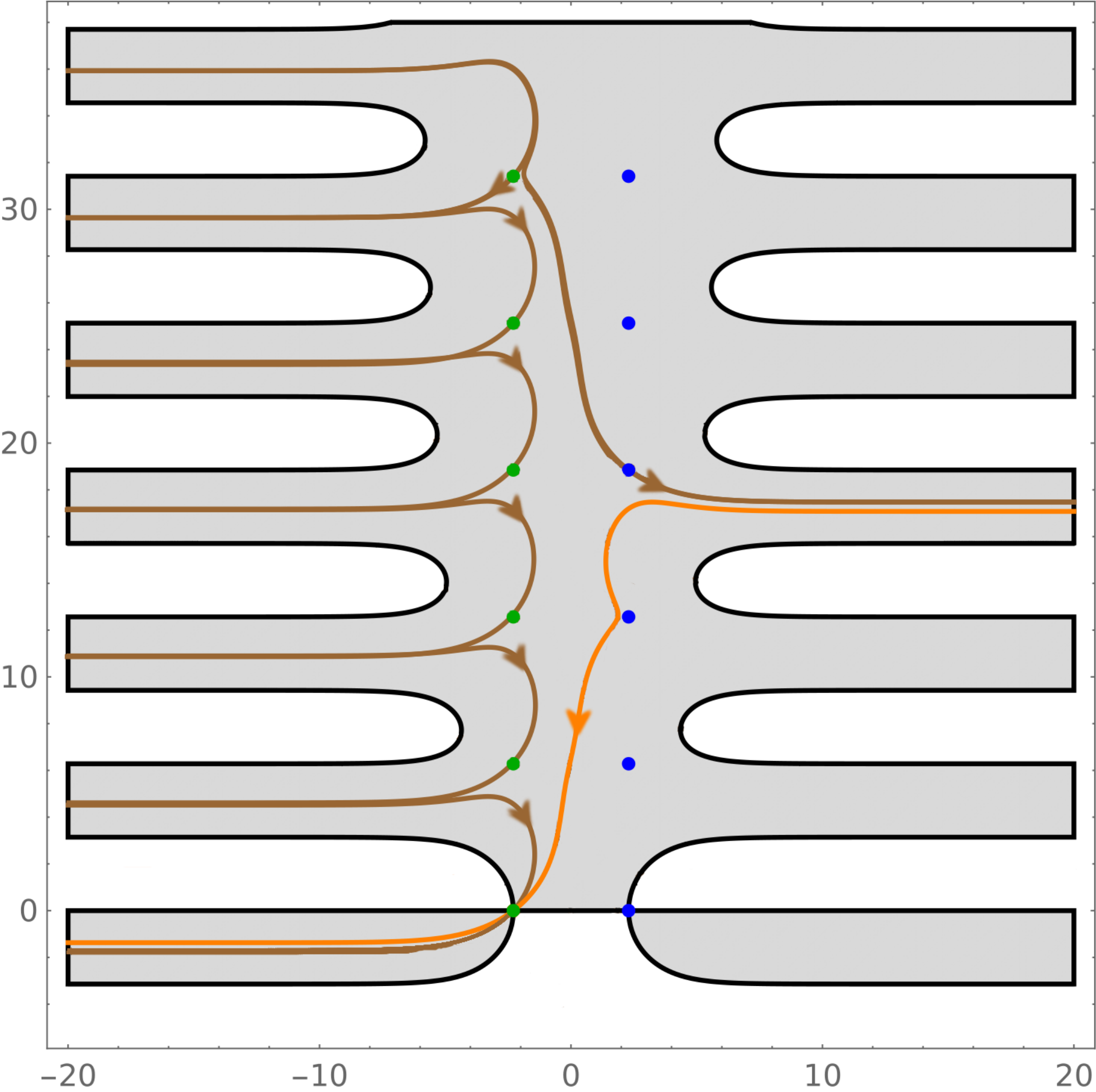}
\caption{Stokes phenomenon for $y=5$, $\theta=\frac\pi2$. On the left, we have the analysis of the positive saddle, while on the right we have the analysis of the negative saddle. We have chosen $\epsilon=0.1$. In orange we have the paths for $\frac\pi2-\epsilon$, in brown the paths for $\frac\pi2+\epsilon$.}
\label{fig:stokes_phen_4}
\end{figure}

This case is more complicated. Let us start with $\theta=\frac\pi2$: we will have to pick a deformation parameter $\epsilon$, and only in the limit $\epsilon\to0$ we will see the full description. We refer to figure \ref{fig:stokes_phen_4}.

Starting for the Stokes phenomenon related to the saddle $\tau_{(+,0)}$, we see that the orange path $\gamma_{(+,0)}$ gets deformed into a path that goes below the singularity $\tau_{(-,N)}$, with $N$ dependent from $\epsilon$. We don't need the exact dependency, we only need to know that $N\to\infty$ as $\epsilon\to0$. Then we can use the path $\gamma_{(-,N)}$ to reach the endpoint of $\gamma_{(+,1)}$, that we can use to reconnect the brown path to the endpoint of the orange path. Assigning the correct orientations, the effect of the Stokes phenomenon is then
\begin{align}
\gamma_{(+,0)}\to\gamma_{(+,0)}-\gamma_{(-,N)}-\gamma_{(+,1)}.
\label{eq:stok_pi2}
\end{align}
For the saddle $\tau_{(-,0)}$, we see that in order to reproduce the orange path, we first have to go up $M$ times, and then use the path from $\tau_{(+,m)}$ to cross over the other side. Here $M$ and $m$ are also functions of $\epsilon$, and also in this case they go to infinity with $\epsilon\to0$. The Stokes phenomenon is then
\begin{align}
\gamma_{(-,0)}\to\sum_{n=0}^M\gamma_{(-,n)}-\gamma_{(+,m)}.
\label{eq:stok_pi2_2}
\end{align}
Even if there is no path describing the limit for $\epsilon\to0$, we will see that the asymptotic description will be well-defined.

For $\theta=\frac{3\pi}2$, we have the same situation with the roles of $\tau_{(+,0)}$ and $\tau_{(-,0)}$ interchanged.

\subsubsection{Case 3: $0<y<1,\theta=\pm\frac\pi2$}

\begin{figure}
\centering
\includegraphics[width=0.49\textwidth]{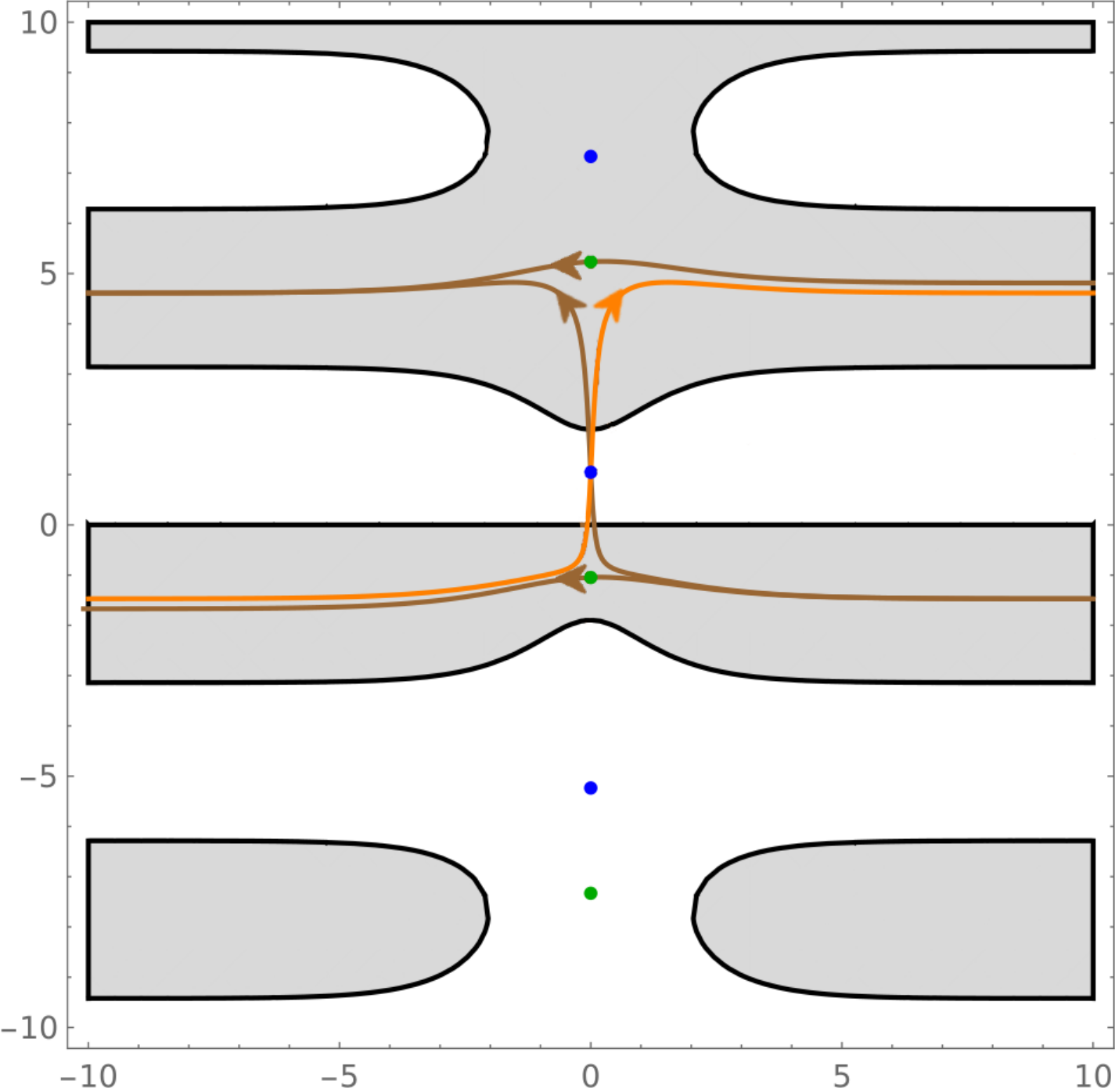}
\includegraphics[width=0.49\textwidth]{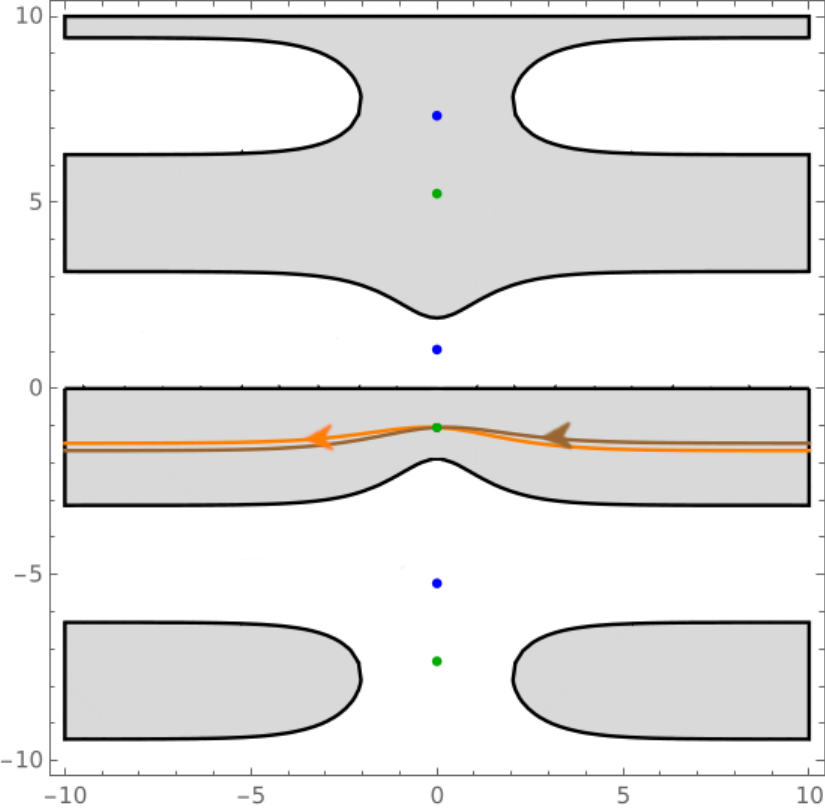}
\caption{Stokes phenomenon for $y=\frac12$, $\theta=\frac\pi2$. On the left, we have the analysis of the positive saddle, while on the right we have the analysis of the negative saddle. We have chosen $\epsilon=0.1$. In orange we have the paths for $\frac\pi2-\epsilon$, in brown the paths for $\frac\pi2+\epsilon$.}
\label{fig:stokes_phen_5}
\end{figure}

When $0<y<1$, all singularities are on the same vertical line. We picture the arrangement in figure \ref{fig:stokes_phen_5}.

As can be seen from the picture, the Stokes phenomenon is simpler than in the previous case. For the negative saddle $\tau_{(-,0)}$ there is no Stokes phenomenon at all, as the orange and brown paths have the same endpoints at infinity. For the saddle $\tau_{(+,0)}$, we have
\begin{align}
\gamma_{(+,0)}\to-\gamma_{(-,1)}+\gamma_{(+,0)}-\gamma_{(-,0)}.
\label{eq:stok_pi2_m1}
\end{align}
Note that in order to keep the continuity of the integration (that requires the same sign for $\gamma_{(+,0)}$ on both sides of \eqref{eq:stok_pi2_m1}) we have to flip the overall sign. This is analogous to what happens in \cite{aniceto2019primer}.

At $\theta=-\frac\pi2$ the opposite phenomenon appears. The positive saddle $\tau_{(+,0)}$ does not have a Stokes phenomenon, while the contour around the negative saddle has to be modified as
\begin{align}
\gamma_{(-,0)}\to-\gamma_{(+,-1)}+\gamma_{(-,0)}-\gamma_{(+,0)}.
\end{align}

\section{Stokes automorphism of the WKB function}

Now that we have established a relation between the integral representation \eqref{eq:int_rep}, steepest descent paths and the Stokes automorphism in that context, we turn to the next step towards our goal. We now link the description of section \ref{sec:int_rep} with the concepts of section \ref{sec:bor_an}. This will allow us to write Stokes automorphisms for the sectors appearing in our WKB solution.

\subsection{Integral representation and Borel transform}

Our first step is to obtain an asymptotic expansion for the functions $I^{(\pm,n)}$, in order to obtain our link to the WKB solutions.

\begin{figure}
\centering
\includegraphics[scale=1.5]{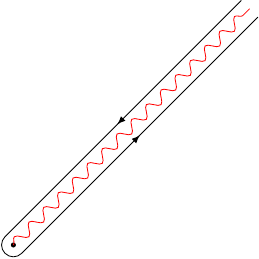}
\caption{Transformed path of steepest descent $\gamma_\theta^*$ after change of variable \eqref{eq:var_chan}. The path has to be deformed to go around the branch cut that can be put on the direction of integration $\ee^{\ii\theta}$. The integration goes around the origin (black dot).}
\label{fig:inttrans}
\end{figure}

We rewrite the integrals $I^{(\pm,n)}$ as
\begin{align}
I^{(\pm,n)}=\exp\left(-\frac1av^{(\pm,n)}\right)\int_{\gamma^{(\pm,n)}}\exp\left(-\frac1a(v(\tau,y)-v^{(\pm,n)})\right)\dd \tau.
\end{align}
We focus our attention on the integral, on which we perform the change of variables
\begin{align}
v(\tau,y)-v^{(\pm,n)}=s(\tau,y).
\label{eq:var_chan}
\end{align}
As a power series, $s(\tau,y)$ reads
\begin{align}
s(\tau,y)=\sum_{k=2}^\infty v_k(\tau-\tau^{(\pm,n)})^k,\label{eq:chan_var_1}
\end{align}
for some coefficients $v_k^{(\pm)}$ that can be computed by expansion\footnote{Explicit computation gives the coefficients as
\begin{align}
v_k^{(\pm)}=\frac{1}{k!}\begin{cases}
\mp\sqrt{y+1}\sqrt{y-1},&k\text{ even}\\
-y,&k\text{ odd}.
\end{cases}
\end{align}}. Due to the fact that the series starts at $k=2$, we can perform an inversion of the series as
\begin{align}
\tau-\tau^{(\pm,n)}=\sum_{k=1}^\infty c_k s^{\frac k2},\label{eq:chan_var_2}
\end{align}
for a series of coefficients $c_k$ determined by plugging \eqref{eq:chan_var_1} into \eqref{eq:chan_var_2} and expanding through the generalized binomial theorem to obtain the identity between LHS and RHS. The integral gets transformed into
\begin{align}
I^{(\pm,n)}=\exp\left(-\frac1av^{(\pm,n)}\right)\int_{\gamma^*_\theta}\ee^{-\frac{s}{a}}\frac{1}{y-\cosh\tau(s,y)}\dd s,
\label{eq:bridge_integral}
\end{align}
where $\tau(s,y)$ is either implicitly defined by \eqref{eq:var_chan} or explicitly by \eqref{eq:chan_var_2}, and we describe the transformed path $\gamma_\theta^*$ in the following way: as on the path of steepest descent $\Im s=\Im a$ and $\Re (s/a)>0$, the path $\gamma^{(\pm,n)}$ gets transformed in a path going from $\ee^{\ii\theta}\infty$ to $0$ and back. The integrand \eqref{eq:bridge_integral} has an expansion in $\sqrt s$, so there's a branch point at $0$ and a branch cut to consider, that we orient along the $\theta$ direction. We plot this path in figure \ref{fig:inttrans}.
As stated, due to \eqref{eq:chan_var_2}, the asymptotic expansion of $(y-\cosh\tau(s,y))^{-1}$ will contain square roots. We can write the expansion as
\begin{align}
\frac{1}{y-\cosh\tau(s,y)}\simeq\sum_{k=0}^\infty\hat B_k s^k+\frac{1}{\sqrt s}\sum_{k=0}^\infty B_k s^k.
\end{align}
This kind of expansion is present in \cite{aniceto2019primer}. The integration along $\gamma_\theta^*$ gets rid of the coefficients $\hat B_k$, due to the lack of a square root branch cut and the exponential decay on the path of steepest descent. We are only left with the coefficients $B_k$, and the integration can be expanded as
\begin{align}
I^{(\pm,n)}\simeq-2\exp\left(-\frac1av^{(\pm,n)}\right)\int_{0}^{\ee^{\ii\theta}\infty}\ee^{-\frac{s}{a}}\frac{1}{\sqrt s}\sum_{k=0}^\infty B_k s^k\dd s,
\end{align}
using the branch cut to transform the integral over the path $\gamma_\theta^*$ into the simple line integral, picking one of the two determinations. The integration of the formal series can now be carried out, obtaining
\begin{align}
I^{(\pm,n)}\simeq-2\exp\left(-\frac1av^{(\pm,n)}\right)\sqrt a\sum_{k=0}^\infty B_k\Gamma\left(\frac12+k\right)a^k.
\end{align}
We give some coefficients of this expansion for $I^{(+,0)}$ in the region $y>1$ as
\begin{align}
&-2\Gamma\left(\frac12\right)B_0=-\frac{\ii\sqrt{2\pi }}{(y^2-1)^{\frac14}},\\
&-2\Gamma\left(\frac32\right)B_1=-\frac{\ii\sqrt{2\pi }(2y^2+3)}{24(y^2-1)^{\frac74}},\\
&-2\Gamma\left(\frac52\right)B_2=\frac{\ii\sqrt{2\pi }(4y^2(y^2+75)+81)}{1152(y^2-1)^{\frac{11}4}}.
\end{align}
As expected, those coefficients coincide with the coefficients of \eqref{eq:coef_steep}, up to normalization. Let us now examine our WKB transseries \eqref{eq:transseries}. We will examine the WKB solution to the problem\footnote{The additional $g$ factor in \eqref{eq:transseries} has been set to $1$ for simplicity: it can be added back by a simple coordinate rescaling.}
\begin{align}
\psi(x+a,a)+\psi(x-a,a)=2(1+x-x_0)\psi(x,a),
\end{align}
so the turning point is located at $x=x_0$. In order to bridge between the two problems, we have to set $1+x-x_0= y$. The sector $\Phi_{(+,1)}$ is given by
\begin{align}
\Phi_{(+,1)}\simeq\frac{1}{(1+(x-x_0))^2-1)^{\frac14}}+\frac{(3+2(1+(x-x_0))^2)}{24((1+(x-x_0))^2-1)^{\frac74}}a+o(a^2).
\end{align}

%In order to compare the two problems, we have to set $y=x$, $a=\ii\hbar$ and $x_0=1$. The sector $\Phi_{(+,1)}$ gets expanded as (in the region $y>1$)

Furthermore, the action $A_1^{(\pm,n)}$ is given by $A_1^{(\pm,n)}=v^{(\pm,n)}+2\pi\ii n$. As the WKB solution and $I^{(\pm,n)}$ solve the same finite difference equation and they are asymptotic to each other, their asymptotic behaviours must be proportional:
\begin{align}
I^{(\pm,n)}\simeq C^{(\pm)}\exp\left(\frac{2\pi\ii n}{a}\right)\exp\left(-\frac{1}{a}A_1^{(\pm,n)}\right)\Phi_{(\pm,1)},
\end{align}
where the constant $C$ is determined by comparing the first coefficient in each asymptotic expansion, obtaining $C^{(-)}=\sqrt{2\pi\ a }$ and $C^{(+)}=-\ii\sqrt{2\pi } a$.

The asymptotic relation can be upgraded to a function equality, as $I^{(\pm,0)}$ is a function that can be computed at each value of $x$. Then, with $\theta=\arg a$, we have
\begin{align}
I^{(\pm,0)}=C^{(\pm)}\exp\left(-\frac1a A_1^{(\pm,0)}\right)\mathcal S_\theta\Phi_{(\pm,1)}.
\end{align}

\subsection{Computation of the Stokes automorphism}
\label{sec:stok_auto}
Using the results of section \ref{sec:int_rep}, we are now ready to describe explicit Stokes automorphisms of the WKB sectors. At a Stokes line, there is a difference between the resummation $\mathcal S_{\theta^+}$ and the resummation $\mathcal S_{\theta^-}$: this difference can be understood by using the discontinuous jumps of the integrals $I^{(\pm,n)}$. As always, we work case by case.

\subsubsection{Case 1: $y>1$}

Let us start with the sector $\Phi_{(-,0)}$, and choose $\theta_n=\arg(v_1^{(+,n)}-v_1^{(-,0)})$. At this value of $\theta$, the integral $I^{(-,0)}$ undergoes a transition, from the path deformation \eqref{eq:path_jump_1}. $I^{(-,0)}$ then changes as
\begin{align}
I^{(-,0)}\to I^{(-,0)}-I^{(+,n)}=I^{(-,0)}-\exp\left(-\frac{2\pi\ii n}ay\right) I^{(+,0)}.\label{eq:first_stokes}
\end{align}
The Stokes automorphism at $\theta$ is defined as $\mathcal S_{\theta^+}=\mathcal S_{\theta^-}\circ\underline{\mathfrak S}_\theta$. In order to reproduce \eqref{eq:first_stokes}, we must have (using $y-1=x-x_0$)
\begin{align}
\begin{aligned}
&\underline{\mathfrak S}_\theta\exp\left(-\frac1a A^{(-,0)}_1\right)\Phi_{(-,1)}=\\
&=\exp\left(-\frac1a A^{(-,0)}_1\right)\Phi_{(-,1)}+\ii \exp\left(-\frac{2\pi\ii n}a\left(y-1\right)\right)\exp\left(-\frac1a A^{(+,0)}_1\right)\Phi_{(+,1)}=\\
&=\exp\left(-\frac1a A^{(-,0)}_1\right)\Phi_{(-,1)}+\ii q_{x_0}^n\exp\left(-\frac1a A^{(+,0)}_1\right)\Phi_{(+,1)}=\\
&=\exp\left(-\frac1a A^{(-,0)}_1\right)\Phi_{(-,1)}+\ii\exp\left(-\frac1a A^{(+,n)}_1\right)\Phi_{(+,1)}.
\end{aligned}
\end{align}
This gives our first Borel residue for the linear problem: $\mathrm S_{(-,0)\to(+,n)}^{(1)}=-\ii$. Analogously, we also get $\mathrm S^{(1)}_{(+,0)\to(-,n)}=-\ii$ by reversing the roles of $\Phi_{(-,1)}$ and $\Phi_{(+,1)}$.

We now choose $\theta=\frac\pi2$. This automorphism has to be done through a limit procedure. We have to select an $\epsilon$ to define $\theta^\pm=\frac\pi2\pm\epsilon$ and then send $\epsilon\to0$. We start from the positive sector, where the path deformation is \eqref{eq:stok_pi2}. The automorphism in terms of the $I^{(\pm,0)}$ is
\begin{align}
I^{(+,0)}\to I^{(+,0)}-\exp\left(-\frac{2\pi\ii}aN\right)I^{(-,0)}-\exp\left(-\frac{2\pi\ii}ay\right)I^{(+,0)}.
\end{align}
In the limit $\epsilon\to0$ we have $N\to\infty$. In the exponential the phase of $a$ going to $\pi/2$ and $y>0$ the argument of the exponential is real negative, so $q^N\to0$ in the limit. We have
\begin{align}
I^{(+,0)}\to I^{(+,0)}-\exp\left(-\frac{2\pi\ii}ay\right)I^{(+,0)}.
\end{align}
The Stokes automorphism for $\Phi_{(+,1)}$ is then
\begin{align}
\begin{aligned}
\underline{\mathfrak S}_{\frac\pi2}\exp\left(-\frac1a A^{(+,0)}_1\right)\Phi_{(+,1)}&=\exp\left(-\frac1a A^{(+,0)}_1\right)\Phi_{(+,1)}-q_{x_0}\exp\left(-\frac1a A^{(+,0)}_1\right)\Phi_{(+,1)}=\\
&=\exp\left(-\frac1a A^{(+,0)}_1\right)\Phi_{(+,1)}-\exp\left(-\frac1a A^{(+,1)}_1\right)\Phi_{(+,1)}.
\end{aligned}
\end{align}
The Borel residues for this transition are then $\mathrm S^{(1)}_{(+,0)\to(+,n)}=\delta_{n,1}$ for $n>0$. For the other saddle, we have before $\epsilon\to0$ according to \eqref{eq:stok_pi2_2}
\begin{align}
I^{(-,0)}\to \sum_{n=0}^M\exp\left(-\frac{2\pi\ii n}ay\right)I^{(-,0)}+\exp\left(-\frac{2\pi\ii m}ay\right)I^{(+,0)}.
\end{align}
As discussed, in the limit $\epsilon\to0$ both $M$ and $m$ go to infinity. The positive saddle disappears and the contribution from negative saddles can be summed up to have
\begin{align}
I^{(-,0)}\to \frac{1}{1-\exp\left(-\frac{2\pi\ii}ay\right)}I^{(-,0)}.
\end{align}
For the sector, we have
\begin{align}
\begin{aligned}
\underline{\mathfrak S}_{\frac\pi2}\exp\left(-\frac1a A^{(-,0)}_1\right)\Phi_{(-,1)}&=\frac{1}{1-q_{x_0}}\exp\left(-\frac1a A^{(-,0)}_1\right)\Phi_{(-,1)}=\\
&=\sum_{n=0}^\infty\exp\left(-\frac1a A^{(-,n)}_1\right)\Phi_{(-,1)}.
\end{aligned}
\end{align}
The Borel residues for this transition are then $\mathrm S_{(-,0)\to(-,n)}^{(1)}=-1$ for $n>0$. For $\theta=-\frac{\pi}2$ we have the opposite situation: for $I^{(-,0)}$ we have
\begin{align}
I^{(-,0)}\to I^{(-,0)}-\exp\left(\frac{2\pi\ii N}ay\right)I^{(-,0)}-\exp\left(\frac{2\pi\ii}ay\right) I^{(+,0)}.
\end{align}
With the phase of $a$ being $-\pi/2$, the argument of the exponential is now real and negative, so $\exp\left(\frac{2\pi\ii N}ay\right)$ tends to $0$ as $N\to\infty$. We are left with
\begin{align}
I^{(-,0)}\to I^{(-,0)}-\exp\left(\frac{2\pi\ii}ay\right)I^{(-,0)},
\end{align}
so the Stokes automorphism for the sector $\Phi_{(-,1)}$ is
\begin{align}
\begin{aligned}
\underline{\mathfrak S}_{-\frac{\pi}2}\exp\left(-\frac1a A^{(-,0)}_1\right)\Phi_{(-,1)}&=\exp\left(-\frac1a A^{(-,0)}_1\right)\Phi_{(-,1)}-q_{x_0}\exp\left(-\frac1a A^{(-,0)}_1\right)\Phi_{(-,1)}=\\
&=\exp\left(-\frac1a A^{(-,0)}_1\right)\Phi_{(-,1)}-\exp\left(-\frac1a A^{(-,1)}_1\right)\Phi_{(-,1)}.
\end{aligned}
\end{align}
We see that $\mathrm S^{(1)}_{(-,0)\to(-,n)}=\delta_{n,-1}$ for $n<0$. For $I^{(+,0)}$ we have
\begin{align}
I^{(+,0)}\to \sum_{n=0}^M\exp\left(\frac{2\pi\ii n}ay\right)I^{(+,0)}+\exp\left(\frac{2\pi\ii m}ay\right)I^{(-,0)}.
\end{align}
As before, the limit removes the contribution from the negative saddle, so we get
\begin{align}
\begin{aligned}
\underline{\mathfrak S}_{-\frac{\pi}2}\exp\left(-\frac1a A^{(+,0)}_1\right)\Phi_{(+,1)}&=\frac{1}{1-q_{x_0}^{-1}}\exp\left(-\frac1a A^{(+,0)}_1\right)\Phi_{(+,1)}=\\
&=\sum_{n=0}^\infty\exp\left(-\frac1a A^{(+,n)}_1\right)\Phi_{(+,1)}.
\end{aligned}
\end{align}
The Borel residues are then $\mathrm S^{(1)}_{(+,0)\to(+,n)}=-1$ for $n<0$.

\subsubsection{Case 2: $0<y<1$}

In this case there are only two Stokes automorphisms to examine, $\underline{\mathfrak S}_{\frac\pi2}$ and $\underline{\mathfrak S}_{\frac{3\pi}2}$. For $\theta=\frac\pi2$, according to $\eqref{eq:stok_pi2_m1}$ we have
\begin{align}
I^{(+,0)}\to I^{(+,0)}-I^{(-,0)}-\exp\left(-\frac{2\pi\ii}ay\right)I^{(-,0)}.
\end{align}
$I^{(-,0)}$ has a trivial automorphism. The Stokes automorphism for the sector is then
\begin{align}
\begin{aligned}
&\underline{\mathfrak S}_{\frac{\pi}2}\exp\left(-\frac1a A^{(+,0)}_1\right)\Phi_{(+,1)}=\\
&=\exp\left(-\frac1a A^{(+,0)}_1\right)\Phi_{(+,1)}+\ii(1+q_{x_0})\exp\left(-\frac1a A^{(-,0)}_1\right)\Phi_{(-,1)}=\\
&=\exp\left(-\frac1a A^{(+,0)}_1\right)\Phi_{(+,1)}+\ii\exp\left(-\frac1a A^{(-,0)}_1\right)\Phi_{(-,1)}+\ii\exp\left(-\frac1a A^{(-,1)}_1\right)\Phi_{(-,1)}.
\end{aligned}
\end{align}
The non zero Borel residues are then $\mathrm S^{(1)}_{(+,0)\to(-,0)}=\mathrm S^{(1)}_{(+,0)\to(-,1)}=-\ii$. As we can see, the Borel residues jump discontinuously when $y$ goes across the turning point $y=1$.

For $\theta=-\frac{\pi}2$ we have that $I^{(+,0)}$ does not change, while
\begin{align}
I^{(-,0)}\to I^{(-,0)}-I^{(+,0)}-\exp\left(\frac{2\pi\ii}ay\right)I^{(+,0)}.
\end{align}
The Stokes automorphism for the sector is
\begin{align}
\begin{aligned}
&\underline{\mathfrak S}_{-\frac{\pi}2}\exp\left(-\frac1a A^{(-,0)}_1\right)\Phi_{(-,1)}=\\
&=\exp\left(-\frac1a A^{(-,0)}_1\right)\Phi_{(-,1)}+\ii(1+q_{x_0}^{-1})\exp\left(-\frac1a A^{(+,0)}_1\right)\Phi_{(+,1)}=\\
&=\exp\left(-\frac1a A^{(-,0)}_1\right)\Phi_{(-,1)}+\ii\exp\left(-\frac1a A^{(+,0)}_1\right)\Phi_{(+,1)}+\ii\exp\left(-\frac1a A^{(+,-1)}_1\right)\Phi_{(+,1)}.
\end{aligned}
\end{align}
The non zero Borel residues are then $\mathrm S^{(1)}_{(-,0)\to(+,0)}=\mathrm S^{(1)}_{(-,0)\to(+,-1)}=-\ii$.

\section{Expansion of the Bessel functions}

\subsection{Debye expansion}

In this section, we will use the $y$ variable. Furthermore, instead of $q_{x_0}$ we will use $q=\exp\left(-\frac{2\pi\ii}ay\right)$. As we have seen in the previous section, at the end of the computation we can always translate all the $q$ into $q_{x_0}$ for arbitrary turning points.

\begin{figure}
\centering
\includegraphics[scale=0.7]{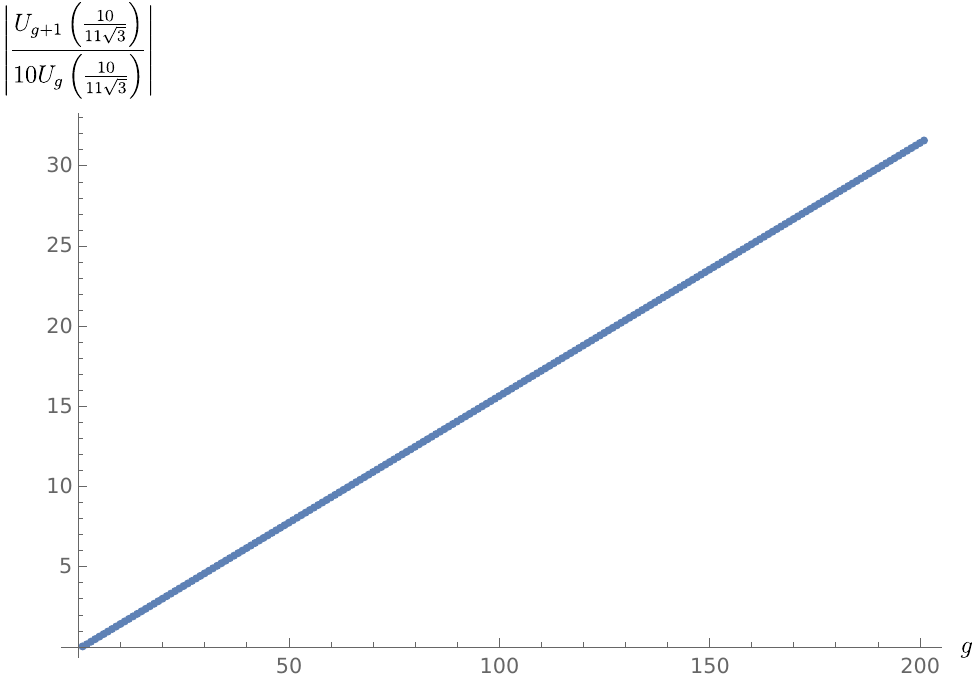}
\caption{Growth of the coefficients $y^{-g}U_g$. We have chosen $y=10$, so we have to evaluate the coefficients at $\frac{10}{11\sqrt3}$. The coefficients arrange on a line asymptotically, meaning that $y^{-g-1}U_{g+1}\simeq g x^{-g}U_g$: this is exactly a factorial growth. The Borel-Padé resummation is necessary to extract finite values from the asymptotic expansions in terms of the $y^{-g}U_g$ coefficients.}
\label{fig:fac_growth}
\end{figure}

The Bessel functions $J$ and $Y$ have an important expansion, called the \textit{Debye expansion}. Those expansions have been discussed very often in literature, and are available in various references (as an example, \cite{olver1997asymptotics}). Our exposition will use the nomenclatures established in \cite{NIST:DLMF}. We start by defining the polynomials $U_k(p)$, defined recursively as
\begin{align}
\begin{cases}
&U_0(p)=1,\\
&U_{k+1}(p)=\frac12p^2(1-p^2)U_k'(p)+\frac18\int_0^p(1-5t^2)U_k(t)\dd t.
\end{cases}
\end{align}
We give the starting polynomials of this succession:
\begin{align}
\begin{aligned}
&\frac{1}{\sqrt{2\pi}(y^2-1)^{\frac14}}U_0\left(\frac{y}{\sqrt{y^2-1}}\right)=\frac{1}{\sqrt{2\pi}(y^2-1)^{\frac14}},\\
&\frac{1}{\sqrt{2\pi}(y^2-1)^{\frac14}}y^{-1}U_1\left(\frac{y}{\sqrt{y^2-1}}\right)=-\frac{ 2 y^2+3}{24\sqrt{2\pi} \left(y^2-1\right)^{\frac{7}{4}}},\\
&\frac{1}{\sqrt{2\pi}(y^2-1)^{\frac14}}y^{-2}U_2\left(\frac{y}{\sqrt{y^2-1}}\right)=\frac{4 \left(y^2+75\right) y^2+81}{1152 \sqrt{2\pi}\left(y^2-1\right)^{\frac{13}{4}}},\\
&\frac{1}{\sqrt{2\pi}(y^2-1)^{\frac14}}y^{-3}U_3\left(\frac{y}{\sqrt{y^2-1}}\right)=\frac{1112 y^6-117684 y^4-278478 y^2-30375}{414720\sqrt{2\pi} \left(y^2-1\right)^{\frac{19}{4}}}.
\end{aligned}
\end{align}
We can see that those coefficients coincide with the coefficients \ref{eq:coef_steep}. The recursive definition is very efficient at generating coefficients, so we can use it to test their growth. As expected, the growth is factorial: we plot an example in figure \ref{fig:fac_growth}.

Getting the right expansion is now only a matter of normalization, provided by \cite{NIST:DLMF}. The expansions provided in the reference are for the functions $J_{\nu}(\nu z)$ and $Y_\nu(\nu z)$, with $\nu$ going to infinity on the real positive direction. For this expansion, it is more convenient to work with variables $y,a$ of \eqref{eq:bessel_prop}: we have to identify $y=\frac1z$ and $a=\frac{z}\nu$ to recover those variables. For $z$ and $\nu$ real and positive, $a$ will be a real variable, so due to the analysis of section \ref{sec:int_rep} we will have to take care of the Stokes phenomenon. Furthermore, the expansions depend on the value of $y$, and change between $y>1$ and $0<y<1$.

By comparison, we have
\begin{align}
\begin{aligned}
\frac{1}{(y^2-1)^{\frac14}}&\ee^{\mp\frac1a(y\arccosh y-\sqrt{y^2-1})}\sum_{g=0}^\infty (\pm y)^{-g}U_g\left(\frac{y}{\sqrt{y^2-1}}\right)a^g=\\
&=\exp\left(-\frac1a A^{(\pm,0)}_1\right)\Phi_{(\pm,1)}
\end{aligned}
\end{align}
in the region $y>1$, and
\begin{align}
\begin{aligned}
\frac{1}{(1-y^2)^{\frac14}}&\ee^{\mp\frac1a(y\arccos y-\sqrt{1-y^2})}\sum_{g=0}^\infty (\pm y)^{-g}U_g\left(\frac{\ii y}{\sqrt{y^2-1}}\right)a^g=\\&=\exp\left(-\frac1a A^{(\pm,0)}_1\right)\Phi_{(\pm,1)}
\end{aligned}
\end{align}
in the region $y<1$. As those will be the series entering the Debye expansions, we will be able to write them in terms of our sectors from the WKB algorithm. Following a common convention \cite{marino2021advanced}, we rename our sectors to
\begin{align}
\frac{1}{(y^2-1)^{\frac14}}\beta_{\pm}(y,a)=\exp\left(-\frac1a A_1^{(\pm,0)}\right)\Phi_{(\pm,1)}\label{eq:beta_al}
\end{align}
for $y>1$, and
\begin{align}
\frac{1}{(1-y^2)^{\frac14}}\beta_{\pm}(y,a)=\exp\left(-\frac1a A_1^{(\pm,0)}\right)\Phi_{(\pm,1)}\label{eq:beta_for}
\end{align}
for $0<y<1$. We will write the Debye expansions in terms of the asymptotic series $\beta_{\pm}$.

\subsubsection{Case 1: $y>1$}
The expansion for $J$ is
\begin{align}
J_{\frac ya}\frac{1}{(y^2-1)^{\frac14}}\left(\frac1a\right)\simeq\sqrt{\frac{a}{2\pi}}\beta_+(y,a).
\end{align}
The expansion for $Y$ as listed on \cite{NIST:DLMF} is
\begin{align}
Y_{\frac ya}\left(\frac1a\right)\simeq-2\frac{1}{(y^2-1)^{\frac14}}\sqrt{\frac{a}{2\pi}}\beta_-(y,a).
\end{align}
This naive expansion would ignore subleading contributions. As we have seen, $\beta_-(y,a)$ cannot be resummed without ambiguity for $\arg a=0$, due to the presence of the Stokes phenomenon. A way to solve this is to deform $a$ with a slight imaginary part as $a\pm\ii\epsilon$, with $\epsilon>0$. Numerical analysis (as detailed in appendix \ref{app:num}) shows that the correct expansion is\footnote{We will still write $a$ instead of $a\pm\ii\epsilon$ in the expansion for brevity: it is intended that $a$ has to be deformed in expansion \eqref{eq:y_exp}}.
\begin{align}
\begin{aligned}
Y_{\frac ya}\left(\frac1a\right)\simeq&\sqrt{\frac{a}{2\pi}}\frac{1}{(y^2-1)^{\frac14}}\left(-2\beta_{-}(y,a)\pm\ii\beta_+(y,a)\right).
\end{aligned}
\label{eq:y_exp}
\end{align}
In the examined region, the additional term in \eqref{eq:y_exp} is a subleading term in the expansion of $y$, due to the fact that the exponential term is suppressed. This subleading term is important in getting all-orders connection formulae. Historically, in ordinary WKB the role of the Bessel functions is played by the Airy functions, and at first the connection formulae were derived neglecting the subleading term in the expansion of the Airy function $\Bi(z)$ \cite{langer1937connection}. It was only in \cite{silverstone1985jwkb} that the subleading term was considered, and the correct connection formulae were obtained.

In matrix form, those expansions can be summarized as
\begin{align}
\left(
\begin{array}{c}
J_\frac{y}{a}\left(\frac1a\right)\\
Y_\frac{y}{a}\left(\frac1a\right)
\end{array}
\right)\simeq
\sqrt{\frac a{2\pi}}\frac{1}{(y^2-1)^{\frac14}}\left(
\begin{array}{cc}
1&0\\
\pm\ii&-2
\end{array}
\right)\cdot
\left(
\begin{array}{c}
\beta_+(y,a)\\
\beta_-(y,a)
\end{array}
\right).
\end{align}

\subsubsection{Case 2: $0<y<1$}

This case has a more straightforward analysis, not presenting a Stokes phenomenon on $\arg a=0$. The expansion for $J$ is given by
\begin{align}
\begin{aligned}
J_{\frac{x}{a}}\left(\frac{1}{a}\right)\simeq&\sqrt{\frac{a}{2\pi}}\frac{1}{(1-y^2)^{\frac14}}\left(\ee^{-\ii\frac\pi4}\beta_+(y,a)\ee^{\ii\frac\pi4}\beta_-(y,a)\right).
\end{aligned}
\end{align}
For $Y$ the expansion is
\begin{align}
\begin{aligned}
Y_{\frac{x}{a}}\left(\frac{1}{a}\right)\simeq&\sqrt{\frac{a}{2\pi}}\frac{1}{(1-y^2)^{\frac14}}(-\ee^{\ii\frac\pi4}\beta_+(y,a)-\ee^{-\ii\frac\pi4}\beta_-(y,a)).
\end{aligned}
\end{align}
In matrix form, those expansions are given by
\begin{align}
\left(
\begin{array}{c}
J_\frac{y}{a}\left(\frac1a\right)\\
Y_\frac{y}{a}\left(\frac1a\right)
\end{array}
\right)\simeq
\frac{1}{(1-y^2)^{\frac14}}\sqrt{\frac a{2\pi}}\left(
\begin{array}{cc}
\ee^{-\ii\frac\pi4}&\ee^{\ii\frac\pi4}\\
-\ee^{\ii\frac\pi4}&-\ee^{-\ii\frac\pi4}
\end{array}
\right)\cdot
\left(
\begin{array}{c}
\beta_+(y,a)\\
\beta_-(y,a)
\end{array}
\right).
\end{align}

We stress that the asymptotic expansions that we have written are valid in the region $a>0$, up to small deformations to avoid the Stokes phenomenon. As in our WKB problem we have $a=\ii g\hbar$ (with $g$ the linear coupling), we will need to find a way to write an expansion for $a$ being purely imaginary, with its phase being $\pm\frac\pi2$. For this, we will use the concept of Stokes automorphism introduced in subsection \ref{sec:bor_an}. In the next subsection, we explain how to do so.

\subsection{The physical region}
\label{sec:phys_reg}
We now get to the main result of this chapter, the asymptotic expansion of the Bessel functions in the physical region. As stated, in the WKB problem we will have $a=\ii\hbar g$, with $g$ the strength of the linear coupling. In physical applications, $\hbar>0$ and $g$ is a real number that can be positive or negative. We then need an expansion for the Bessel functions for $a$ purely imaginary, with arbitrary sign of the imaginary part. We have seen that at $\arg a=\pm\frac\pi2$ we have a Stokes phenomenon, so we will need to make small deformations of $\hbar$ with an imaginary part, and take the limit for the imaginary part going to $0$.

Non trivial Stokes automorphisms happen at $\theta_{(\pm,n)}=\arg(A_1^{(\mp,n)}-A_1^{(\pm,0)})$ and at $\pm\frac\pi2$. It is very convenient to write those automorphisms in a matrix form. As we have computed in subsection \ref{sec:stok_auto}, the Stokes automorphisms in the region $y>1$ can be written as
\begin{align}
&\underline{\mathfrak S}_{\theta_{(+,n)}}=\left(\begin{array}{cc}
1&\ii q^n\\
0&1
\end{array}\right),
\quad
\underline{\mathfrak S}_{\theta_{(-,n)}}=\left(\begin{array}{cc}
1&0\\
\ii q^n&1
\end{array}\right),\\
&\underline{\mathfrak S}_{\frac\pi2}=\left(\begin{array}{cc}
1-q&0\\
0&\frac1{1-q}
\end{array}\right),
\quad
\underline{\mathfrak S}_{-\frac\pi2}=\left(\begin{array}{cc}
\frac{1}{1-q^{-1}}&0\\
0&1-q^{-1}
\end{array}\right),
\end{align}
while in the region $0<y<1$ we have the automorphisms
\begin{align}
&\underline{\mathfrak S}_{\frac\pi2}=\left(\begin{array}{cc}
1&\ii(1+q)\\
0&1
\end{array}\right),
\quad
\underline{\mathfrak S}_{-\frac\pi2}=\left(\begin{array}{cc}
1&0\\
\ii(1+q^{-1})&1
\end{array}\right).
\end{align}
Getting the right asymptotic expansions is now only a matter of multiplying the correct matrices. As always, we distinguish between $y>1$ and $y<1$.

\subsubsection{Case 1: $y>1$}

We start with $\arg a=\frac\pi2^-$, meaning that $a$ is deformed with a small positive real part. We start our Stokes automorphisms from the region in which $\arg a$ is a small positive number, where the expansion is given by
\begin{align}
\left(
\begin{array}{c}
J_\frac{y}{a}\left(\frac1a\right)\\
Y_\frac{y}{a}\left(\frac1a\right)
\end{array}
\right)\simeq
\sqrt{\frac a{2\pi}}\frac{1}{(y^2-1)^{\frac14}}\left(
\begin{array}{cc}
1&0\\
\ii&-2
\end{array}
\right)\cdot
\left(
\begin{array}{c}
\beta_+(y,a)\\
\beta_-(y,a)
\end{array}
\right)
\end{align}
For the argument of $a$ going to $\frac\pi2^-$, we encounter singularities at $\theta_{(-,n)}$ with $n>0$. We have to multiply the Stokes automorphism matrices to the expansion matrix in the following way:
\begin{align}
\left(
\begin{array}{c}
J_\frac{y}{a}\left(\frac1a\right)\\
Y_\frac{y}{a}\left(\frac1a\right)
\end{array}
\right)\to
\sqrt{\frac a{2\pi}}\frac{1}{(y^2-1)^{\frac14}}
\left(
\begin{array}{cc}
1&0\\
\ii&-2
\end{array}
\right)\cdot
\left(
\prod_{n=1}^\infty\underline{\mathfrak S}_{\theta_{(n,-)}}
\right)^{-1}\cdot
\left(
\begin{array}{c}
\beta_+(y,a)\\
\beta_-(y,a)
\end{array}
\right).
\end{align}
The product of Stokes automorphisms is given by
\begin{align}
\prod_{n=1}^\infty\underline{\mathfrak S}_{\theta_{(n,-)}}=\left(\begin{array}{cc}
1&0\\
\frac{\ii q}{q-1}&1,
\end{array}
\right)
\end{align}
so the final expansion for $\arg a=\frac\pi2^-$ is
\begin{align}
\left(
\begin{array}{c}
J_\frac{y}{a}\left(\frac1a\right)\\
Y_\frac{y}{a}\left(\frac1a\right)
\end{array}
\right)\simeq
\sqrt{\frac a{2\pi}}\frac{1}{(y^2-1)^{\frac14}}
\left(
\begin{array}{cc}
1&0\\
-\ii\frac{q+1}{q-1}&-2
\end{array}
\right)\cdot
\left(
\begin{array}{c}
\beta_+(y,a)\\
\beta_-(y,a)
\end{array}
\right).
\label{eq:first_exp}
\end{align}
For $\arg a=\frac\pi2^+$, we just have to cross the line at $\frac\pi2$: to do so, it is sufficient to multiply the matrix that we just obtained by $\underline{\mathfrak S}_{\frac\pi2}^{-1}$ on the right, obtaining for $\arg a=\frac\pi2^+$
\begin{align}
\left(
\begin{array}{c}
J_\frac{y}{a}\left(\frac1a\right)\\
Y_\frac{y}{a}\left(\frac1a\right)
\end{array}
\right)\simeq
\sqrt{\frac a{2\pi}}\frac{1}{(y^2-1)^{\frac14}}
\left(
\begin{array}{cc}
-q+1&0\\
\ii(q+1)&\frac{2}{q-1}
\end{array}
\right)\cdot
\left(
\begin{array}{c}
\beta_+(y,a)\\
\beta_-(y,a)
\end{array}
\right).
\label{eq:sec_exp}
\end{align}
For $\arg a=-\frac\pi2^+$, we start with $\arg a$ small and negative and cross the singularities $\theta_{(n,-)}$ with $n<0$. As we are crossing them clockwise, we have to introduce an additional $-1$, obtaining
\begin{align}
\left(
\begin{array}{c}
J_\frac{y}{a}\left(\frac1a\right)\\
Y_\frac{y}{a}\left(\frac1a\right)
\end{array}
\right)\to
\sqrt{\frac a{2\pi}}\frac{1}{(y^2-1)^{\frac14}}
\left(
\begin{array}{cc}
1&0\\
-\ii&-2
\end{array}
\right)\cdot
\prod_{n=1}^\infty\underline{\mathfrak S}_{\theta_{(-n,-)}}
\cdot
\left(
\begin{array}{c}
\beta_+(y,a)\\
\beta_-(y,a)
\end{array}
\right).
\end{align}
The expansion for $\arg a=\left(-\frac\pi2\right)^+$ is then
\begin{align}
\left(
\begin{array}{c}
J_\frac{y}{a}\left(\frac1a\right)\\
Y_\frac{y}{a}\left(\frac1a\right)
\end{array}
\right)\simeq
\sqrt{\frac a{2\pi}}\frac{1}{(y^2-1)^{\frac14}}
\left(
\begin{array}{cc}
1&0\\
\ii\frac{q^{-1}+1}{q^{-1}-1}&-2
\end{array}
\right)\cdot
\left(
\begin{array}{c}
\beta_+(y,a)\\
\beta_-(y,a)
\end{array}
\right).\label{eq:thir_exp}
\end{align},
For $\arg a=\left(-\frac\pi2\right)^-$, we also cross the Stokes line at $-\frac\pi2$ via multiplication by $\underline{\mathfrak S}_{-\frac\pi2}^{-1}$ on the right, obtaining
\begin{align}
\left(
\begin{array}{c}
J_\frac{y}{a}\left(\frac1a\right)\\
Y_\frac{y}{a}\left(\frac1a\right)
\end{array}
\right)\simeq
\sqrt{\frac a{2\pi}}\frac{1}{(y^2-1)^{\frac14}}
\left(
\begin{array}{cc}
-q^{-1}+1&0\\
-\ii(1+q^{-1})&\frac{2}{q^{-1}-1}
\end{array}
\right)\cdot
\left(
\begin{array}{c}
\beta_+(y,a)\\
\beta_-(y,a)
\end{array}
\right).
\label{eq:four_exp}
\end{align}
We note that \eqref{eq:thir_exp} and \eqref{eq:four_exp} can be obtained from \eqref{eq:first_exp} and \eqref{eq:sec_exp} respectively by changing $q\to q^{-1}$ and inverting the sign of the off diagonal element.

\subsubsection{Case 2: $0<y<1$}

In this case we only have two non trivial Stokes automorphisms. The matrix multiplications work the same way as in the case $y>1$, so we will directly provide results.

For $\arg a=\frac\pi2^-$ there is no Stokes automorphism when starting from $\arg a=0$. The expansion is then
\begin{align}
\left(
\begin{array}{c}
J_\frac{y}{a}\left(\frac1a\right)\\
Y_\frac{y}{a}\left(\frac1a\right)
\end{array}
\right)\simeq
\sqrt{\frac a{2\pi}}\frac{1}{(1-y^2)^{\frac14}}\left(
\begin{array}{cc}
\ee^{-\ii\frac\pi4}&\ee^{\ii\frac\pi4}\\
-\ee^{\ii\frac\pi4}&-\ee^{-\ii\frac\pi4}
\end{array}
\right)\cdot
\left(
\begin{array}{c}
\beta_+(y,a)\\
\beta_-(y,a)
\end{array}
\right).
\label{eq:for_1}
\end{align}
For $\arg a=\frac\pi2^+$, we multiply by the appropriate $\underline{\mathfrak S}_{\frac\pi2}$, obtaining
\begin{align}
\left(
\begin{array}{c}
J_\frac{y}{a}\left(\frac1a\right)\\
Y_\frac{y}{a}\left(\frac1a\right)
\end{array}
\right)\simeq
\sqrt{\frac a{2\pi}}\frac{1}{(1-y^2)^{\frac14}}\left(
\begin{array}{cc}
\ee^{-\ii\frac\pi4}&-\ee^{\ii\frac\pi4}q\\
-\ee^{\ii\frac\pi4}&-\ee^{-\ii\frac\pi4}(2+q)
\end{array}
\right)\cdot
\left(
\begin{array}{c}
\beta_+(y,a)\\
\beta_-(y,a)
\end{array}
\right).
\label{eq:for_2}
\end{align}
For $\arg a=\left(-\frac\pi2\right)^+$ there is no Stokes automorphism starting from $\arg a=0$, so we have again
\begin{align}
\left(
\begin{array}{c}
J_\frac{y}{a}\left(\frac1a\right)\\
Y_\frac{y}{a}\left(\frac1a\right)
\end{array}
\right)\simeq
\sqrt{\frac a{2\pi}}\frac{1}{(1-y^2)^{\frac14}}\left(
\begin{array}{cc}
\ee^{-\ii\frac\pi4}&\ee^{\ii\frac\pi4}\\
-\ee^{\ii\frac\pi4}&-\ee^{-\ii\frac\pi4}
\end{array}
\right)\cdot
\left(
\begin{array}{c}
\beta_+(y,a)\\
\beta_-(y,a)
\end{array}
\right).
\label{eq:for_3}
\end{align}
Lastly, for $\arg a=\left(-\frac\pi2\right)^-$ we have
\begin{align}
\left(
\begin{array}{c}
J_\frac{y}{a}\left(\frac1a\right)\\
Y_\frac{y}{a}\left(\frac1a\right)
\end{array}
\right)\simeq
\sqrt{\frac a{2\pi}}\frac{1}{(1-y^2)^{\frac14}}\left(
\begin{array}{cc}
-\ee^{-\ii\frac\pi4}q^{-1}&\ee^{\ii\frac\pi4}\\
-\ee^{\ii\frac\pi4}(2+q^{-1})&-\ee^{-\ii\frac\pi4}
\end{array}
\right)\cdot
\left(
\begin{array}{c}
\beta_+(y,a)\\
\beta_-(y,a)
\end{array}
\right).
\label{eq:for_4}
\end{align}

%% file: 4connformulaecomp.tex
\chapter{The connection problem}
\label{chapter4}

We now come to the main result of this thesis, the computation of connection formulae for the deformed Schr\"odinger equation. In the resummation of the WKB wavefunctions from the asymptotic series generated by the algorithm of chapter \ref{chap:1}, we have to impose continuity when crossing the turning points. The connection formulae are a tool to impose this continuity and obtain a smooth wavefunction on $\mathbb R$, and will allow us to obtain the normalizable spectrum for a large class of potentials, exactly or up to exponentially decaying corrections.

In this chapter, we present a review of the connection formulae of quantum mechanics and their derivation, as we will work by making parallels with the derivation in the standard case. We follow this review with a section dedicated to getting those formulae, and then conclude with a comparison between the two sets of formulae, showing that the formulae of standard quantum mechanics can be recovered in the $\Lambda\to\infty$ limit of deformed quantum mechanics. In the review we mainly follow \cite{silverstone1985jwkb}, \cite{marino2021advanced} and \cite{kawai2005algebraic}.

\section[Standard quantum mechanics]{Review: connection formulae of standard quantum mechanics}

\subsection{The linear problem: Airy functions and asymptotic expansion}

The derivation of the connection formulae starts with the analysis of the linear problem. The standard quantum mechanics eigenvalue problem for the linear potential is
\begin{align}
-\hbar^2\psi''(x,\hbar)=2x\psi(x),
\end{align}
and the linear problem is solved by the Airy functions $\Ai$ and $\Bi$ as
\begin{align}
\psi(x,\hbar)=A \Ai\left(-\hbar^{-\frac23}2^{\frac13}x\right)+B \Bi\left(-\hbar^{-\frac23}2^{\frac13}x\right),
\end{align}
with $A$ and $B$ constants. The Airy functions play the same role that the Bessel functions play in the deformed Schr\"odinger equation. Their asymptotic expansions can be written in terms of the series
\begin{align}
\beta(\nu)=\ee^\nu\sum_{k=0}^\infty c_k\nu^k,\quad c_k=\frac1{2\pi}\frac{\Gamma\left(k+\frac56\right)\Gamma\left(k+\frac16\right)}{2^kk!}:
\end{align}
by defining the function
\begin{align}
\zeta(z)=\frac23z^{\frac 32},
\end{align}
the asymptotic expansions of the Airy functions can be written in the following way: in the region $z>0$ they have an exponential behaviour, given by
\begin{align}
&\Ai(z)\simeq\frac{1}{2\sqrt\pi z^{\frac14}}\beta(-\zeta(z)),\\
&\Bi(z)\simeq\frac{1}{2\sqrt\pi z^{\frac14}}(2\beta(\zeta(z))\pm\ii\beta(-\zeta(z))),\label{eq:bi_exp_amb}
\end{align} 
while in the region $z<0$ they have an oscillatory behaviour, given by
\begin{align}
&\Ai(z)\simeq\frac{1}{2\sqrt\pi (-z)^{\frac14}}(\ee^{-\ii\frac\pi4}\beta(\ii\zeta(-z))+\ee^{\ii\frac\pi4}\beta(-\ii\zeta(-z))),\\
&\Bi(z)\simeq\frac{1}{2\sqrt\pi (-z)^{\frac14}}(\ee^{\ii\frac\pi4}\beta(\ii\zeta(-z))+\ee^{-\ii\frac\pi4}\beta(-\ii\zeta(-z))).
\end{align}
As in the Bessel case, there is a formula with a sign ambiguity: as at $\arg z=0$ we have a Stokes phenomenon, we have to deform $z$ with a small imaginary part to have a well defined expansion. The $+$ sign is chosen for a deformation $z\to z-\ii\epsilon$ with $\epsilon>0$, the $-$ sign is used for the other deformation.

\subsection{Uniform WKB}

The procedure to obtain connection formulae is based on \textit{uniform WKB}, that makes a variable transformation in the WKB problem, deforming the non linear potential to the linear one.

The standard WKB ansatz is given by
\begin{align}
\psi(x,\hbar)=\frac{1}{\sqrt{S'(x,\hbar)}}\exp\left(\frac{1}{\ii\hbar}S(x,\hbar)\right).
\end{align}
where $S(x,\hbar)$ is a series in $\hbar^2$ that obeys the Riccati equation
\begin{align}
(S'(x,\hbar))^2+\frac{\hbar^2}{2}\{S(x),x\}=2(E-V(x)),
\end{align}
with
\begin{align}
\{S(x),x\}=-2\sqrt{S'(x,\hbar)}\frac{\dd}{\dd x}\frac{1}{\sqrt{S'(x,\hbar)}}
\end{align}
the Schwarzian derivative. The choice of the exponential function is a convenient one, but this is not the only possibility. In the uniform WKB method, the exponential is substituted by an arbitrary function $f$, and the WKB ansatz is given by
\begin{align}
\psi(x,\hbar)=\frac{1}{\sqrt{\phi'(x,\hbar)}}f(\phi(x,\hbar)).
\end{align}
The $\phi$ function plays the role of the deformed quantum action in this ansatz. We impose that $f$ solves the equation
\begin{align}
f''(\phi)+\frac{1}{\hbar^2}\Pi(\phi)^2f(\phi)=0,\label{eq:uniform1}
\end{align}
for a function $\Pi(\phi)^2$ that we can choose arbitrarily. The reason for this is that \eqref{eq:uniform1} is a Schr\"odinger equation with an arbitrary potential, that can be fixed to suit our needs. The cost for this freedom is a rather complicated equation for $\phi(x,\hbar)$, that is also influenced by the choice of $\Pi$. The function $\phi(x,\hbar)$ will solve
\begin{align}
\Pi(\phi)^2(\phi'(x,\hbar))^2+\frac{\hbar^2}2\{\phi(x,\hbar),x\}=2(E-V(x)).\label{eq:uniform2}
\end{align}
We can recover the standard WKB ansatz by choosing $\Pi(\phi)^2=1$, but we will make a different choice. We choose $\Pi(\phi)^2=\phi$, so $f$ is given by
\begin{align}
f(\phi)=A \Ai(-2^{\frac13}\hbar^{-\frac23}\phi)+B \Bi(-2^{\frac13}\hbar^{-\frac23}\phi).
\end{align}
By doing so, we have effectively transformed the WKB problem with arbitrary potential in the linear WKB problem, at the cost of obtaining a non trivial differential equation for $\phi$:
\begin{align}
\phi(x,\hbar)(\phi'(x,\hbar))^2+\frac{\hbar^2}2\{\phi(x,\hbar),x\}=2(E-V(x)).
\end{align}
We now assume that $x_0$ is a point for which $E=V(x_0)$ and perform a series expansion of $\phi(x,\hbar)$ in even powers of $\hbar$ as
\begin{align}
\phi(x,\hbar)=\sum_{n=0}^\infty \phi_n(x)\hbar^{2n}.
\end{align}
The zero order satisfies
\begin{align}
\phi_0(x)=\left(\frac32\int_{x_0}^x\sqrt{2(E-V(t))}\dd t\right)^{\frac23},\label{eq:class}
\end{align}
and the generic equation for $\phi_n(x)$ will be of the form
\begin{align}
(\phi_0'(x))^2\phi_n(x)+2\phi_0(x)\phi_0'(x)\phi_n'(x)+\dots=0,
\end{align}
with $\dots$ indicating terms involving all the $\phi_{m}(x)$ with $m<n$. The equations for the $\phi_n$ are then linear non homogeneous equations, and we can choose the integration constants in such a way that $\phi_n(x_0)=0$ for $n>0$. Furthermore, we will have $\phi(x,\hbar)>0$ for $x>x_0$ and $\phi(x,\hbar)<0$ for $x<x_0$. $\phi(x,\hbar)$ can then be thought as the necessary coordinate modification to express the solution to the standard WKB Schr\"odinger equation in terms of the Airy functions.

\subsection{Uniform WKB and standard WKB}

The next step is to connect the standard WKB ansatz
\begin{align}
\psi(x,\hbar)=\frac{1}{\sqrt{S'(x,\hbar)}}\exp\left(\frac{1}{\ii\hbar}S(x,\hbar)\right)
\end{align}
to the $\beta$ functions that appear in the expansion of the Airy functions. To do so, we first define
\begin{align}
\beta_{\pm}=\beta\left(\pm\frac{2\ii}{3\hbar}\phi(x,\hbar)^{\frac32}\right)
\end{align}
and then take a different approach to the linkage: we \textit{define}
\begin{align}
S(x,\hbar)=-\frac{\ii\hbar}2\log\left(\frac{\beta_+}{\beta_-}\right)
\end{align}
and then show that the $S$ defined in this way can be used as quantum actions to solve the Schr\"odinger equation. In order to do so, we first compute
\begin{align}
S'(x,\hbar)=\frac{\phi'(x,\hbar)\sqrt{\phi(x,\hbar)}}2\frac{\beta_+'\beta_-+\beta_+\beta_-'}{\beta_+\beta_-}.
\end{align}
The numerator can be evaluated using Wronskian identities, giving
\begin{align}
\beta_+'\beta_-+\beta_+\beta_-'=2,
\end{align}
so we conclude
\begin{align}
S'(x,\hbar)=\phi'(x,\hbar)\sqrt{\phi(x,\hbar)}\frac{2}{\beta_+\beta_-}.
\end{align}
We immediately have
\begin{align}
\frac{1}{\sqrt{S'(x,\hbar)}}\exp\left(\pm\frac{1}{\hbar}S(x,\hbar)\right)=\frac{1}{\sqrt{\phi'(x,\hbar)}\phi(x,\hbar)^{\frac14}}\beta_{\pm}.\label{eq:clas_id}
\end{align}
As the RHS is a linear combination of $\Ai$ and $\Bi$ solving the Schr\"odinger equation, the LHS also does, so $S$ is a quantum action. Furthermore, when the expansion is taken around a turning point at $x=x_0$, $S$ will be normalized at $x_0$, fixing the normalization.

\subsection{Connection formulae}

\begin{figure}
\centering
\includegraphics[scale=1.2]{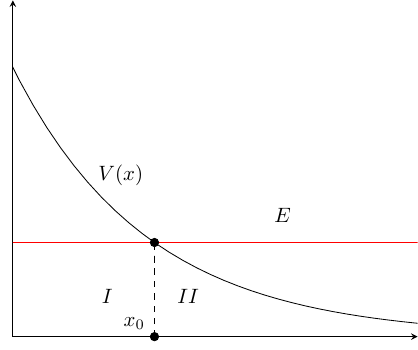}
\caption{Classically forbidden (region $I$) and allowed (region $II$) zones for the standard WKB connection formulae.}
\label{fig:clasall}
\end{figure}

We are now ready to obtain connection formulae for the standard Schr\"odinger equation. We will assume that $x_0$ is such as $E=V(x_0)$, for $x>x_0$ we have $E>V(x)$ and for $x<x_0$ we have $E<V(x)$: the classically allowed region is on the right of $x_0$, while the classically forbidden region is on the left of $x_0$. We will call the classically allowed region $II$ and the classically forbidden region $I$, as illustrated in \ref{fig:clasall}.

We use the uniform WKB ansatz with $\Pi(\phi)^2=\phi$ to write the wavefunction as
\begin{align}
\psi(x,\hbar)=\frac{1}{\sqrt{\phi(x,\hbar)}}(A \Ai(-2^{\frac13}\hbar^{-\frac23}\phi)+B \Bi(-2^{\frac13}\hbar^{-\frac23}\phi)).
\end{align}
We start from the classically allowed region $II$, where $\phi(x,\hbar)>0$. Using the asymptotic formulae, the expansion is given by
\begin{align}
\psi(x,\hbar)\simeq\frac{1}{\sqrt{\phi'}\phi^{\frac14}}\left[(B-\ii A)\ee^{\ii\frac\pi4}\beta\left(\frac{2\ii}{3\hbar}\phi^{\frac32}\right)+(B+\ii A)\ee^{-\ii\frac\pi4}\beta\left(-\frac{2\ii}{3\hbar}\phi^{\frac32}\right)\right].
\end{align}
Using \eqref{eq:clas_id}, we have
\begin{align}
\psi(x,\hbar)\simeq\frac{1}{\sqrt{S'}}\left[\left(B-\ii A\right)\ee^{\frac{\ii}\hbar S+\ii\frac\pi4}+\left(B+\ii A\right)\ee^{-\frac{\ii}\hbar S-\ii\frac\pi4}\right].\label{eq:clas}
\end{align}
In the classically forbidden region we have
\begin{align}
\psi(x,\hbar)\simeq\frac{1}{\sqrt{\phi'}(-\phi)^{\frac14}}\left[2B\beta\left(\frac{2}{3\hbar}(-\phi)^{\frac32}\right)+(A\pm\ii B)\beta\left(-\frac{2}{3\hbar}(-\phi)^{\frac32}\right)\right],\label{eq:asy_for}
\end{align}
according to the deformation of $\hbar$ ($+$ sign for $\hbar\to\hbar-\ii\epsilon$, $-$ sign otherwise). In order to deal with the signs coming from the roots of negative quantities, we define in the forbidden region
\begin{align}
T(x,\hbar)=\ii S(x,\hbar),
\end{align}
that is a real function, for which we have
\begin{align}
\frac{1}{\sqrt{S'(x,\hbar)}}\exp\left(\pm\frac{\ii}{\hbar}S(x,\hbar)\right)=(-\ii T'(x,\hbar))^{-\frac12}\ee^{\pm \frac1\hbar T(x,\hbar)}.
\end{align}
In this region, we have
\begin{align}
\frac1{\sqrt{\phi'}(-\phi)^\frac14}\beta\left(\pm\frac{2}{3\hbar}(-\phi)^{\frac32}\right)=\frac{1}{(-T')^{\frac12}}\ee^{\pm\frac 1\hbar T}.
\end{align}
Expansion \eqref{eq:asy_for} then reads
\begin{align}
\psi(x,\hbar)\simeq\frac{1}{(-T')^{\frac12}}\left[2 B\exp\left(\frac1\hbar T(x,\hbar)\right)+(A\pm \ii B)\exp\left(-\frac1\hbar T(x,\hbar)\right)\right].\label{eq:for}
\end{align}
\eqref{eq:clas} and \eqref{eq:for} allow us to link the two regions, dictating the changes in the multiplicative constants that are necessary to get a continuous asymptotic expansion. Getting the transition formulae is now only a matter of choosing appropriate $A$ and $B$ to isolate the various exponentials: for the $+$ deformation, we have
\begin{align}
&\frac{1}{\sqrt{S'}}\exp\left(\frac\ii\hbar S\right)\leftrightarrow \frac{1}{\sqrt{-T'}}\left(\ee^{-\ii\frac\pi4}\exp\left(\frac1\hbar T\right)+\ee^{\ii\frac\pi4}\exp\left(-\frac1\hbar T\right)\right),\label{eq:stand_1}\\
&\frac{1}{\sqrt{S'}}\exp\left(-\frac\ii\hbar S\right)\leftrightarrow \frac{1}{\sqrt{-T'}}\ee^{\ii\frac\pi4}\exp\left(\frac1\hbar T\right),\label{eq:stand_2}
\end{align}
while for the $-$ deformation we have
\begin{align}
&\frac{1}{\sqrt{S'}}\exp\left(\frac\ii\hbar S\right)\leftrightarrow \frac{1}{\sqrt{-T'}}\ee^{-\ii\frac\pi4}\exp\left(\frac1\hbar T\right),\label{eq:stand_5}\\
&\frac{1}{\sqrt{S'}}\exp\left(-\frac\ii\hbar S\right)\leftrightarrow \frac{1}{\sqrt{-T'}}\left(\ee^{\ii\frac\pi4}\exp\left(\frac1\hbar T\right)+\ee^{-\ii\frac\pi4}\exp\left(-\frac1\hbar T\right)\right)\label{eq:stand_6}.
\end{align}
This concludes the derivation for the formulae when $x>x_0$ is a classically allowed region. We can obtain the other case easily by repeating the same procedure, obtaining the other set of formulae: for the $+$ determination, we obtain
\begin{align}
&\frac{1}{\sqrt{S'}}\exp\left(\frac\ii\hbar S\right)\leftrightarrow \frac{1}{\sqrt{-T'}}\ee^{\ii\frac\pi4}\exp\left(-\frac1\hbar T\right),\label{eq:stand_3}\\
&\frac{1}{\sqrt{S'}}\exp\left(-\frac\ii\hbar S\right)\leftrightarrow \frac{1}{\sqrt{-T'}}\left(\ee^{\ii\frac\pi4}\exp\left(\frac1\hbar T\right)+\ee^{-\ii\frac\pi4}\exp\left(-\frac1\hbar T\right)\right),\label{eq:stand_4}
\end{align}
while for the $-$ determination we have
\begin{align}
&\frac{1}{\sqrt{S'}}\exp\left(\frac\ii\hbar S\right)\leftrightarrow \frac{1}{\sqrt{-T'}}\left(\ee^{-\ii\frac\pi4}\exp\left(\frac1\hbar T\right)+\ee^{\ii\frac\pi4}\exp\left(-\frac1\hbar T\right)\right)\label{eq:stand_7},\\
&\frac{1}{\sqrt{S'}}\exp\left(-\frac\ii\hbar S\right)\leftrightarrow \frac{1}{\sqrt{-T'}}\ee^{-\ii\frac\pi4}\exp\left(-\frac1\hbar T\right).\label{eq:stand_8}
\end{align}

This concludes the derivation of the WKB connection formulae for standard quantum mechanics. We will follow this procedure closely in deriving the formulae for deformed quantum mechanics, exploiting the asymptotic expansion of the Bessel functions that was obtained in chapter \ref{chapter3}.

\section[Deformed quantum mechanics]{Connection formulae of deformed quantum mechanics}
\label{sec:conn_form_def}
\subsection{Turning points}

The first main difference between the standard WKB ansatz and the deformed WKB ansatz is the doubling of turning points. While in standard WKB we have turning points for all $x_0$ such as $E=V(x_0)$, the turning points of the deformed WKB ansatz are given by $E=V(x_0)$ and $E=V(x_0)-2$, as seen in chapter \ref{chap:1}. We call the turning points $x_0$ such as $E=V(x_0)$ \textit{turning points at $1$}, while the turning points at $x_0$ will be called \textit{turning points at $-1$}. The reason for this nomenclature comes by examining the finite difference Schr\"odinger equation
\begin{align}
\psi(x+\ii\hbar,\hbar)+\psi(x-\ii\hbar,\hbar)=2(E-V(x)+1)\psi(x,\hbar):
\end{align}
at the turning points at $1$, the bracket on the RHS has value $1$, while at the turning points at $-1$ it has value $-1$.

In chapter \ref{chapter3}, we have only expanded the Bessel functions around a turning point at $1$. No other expansion is needed, as the expansions for the turning points at $-1$ can be obtained from the expansions at the turning points at $-1$. In fact, assume that $x_0$ is a turning point at $-1$: the function $\varphi(x,\hbar)=\exp\left(-\frac{2\pi}\hbar(x-x_0+1)\right)\psi(x,\hbar)$ solves
\begin{align}
\varphi(x+\ii\hbar,\hbar)+\varphi(x-\ii\hbar,\hbar)=-2(E+V(x)-1)\varphi(x,\hbar).\label{eq:general_problem}
\end{align}
For the function $\varphi$, $x_0$ is now a turning point at $1$. This trick has been vastly used in the works on finite difference WKB (see \cite{pasquier1992periodic} and \cite{fedotov2021wkb} as examples). We can focus on the turning points at $1$, and then obtain the connection formulae for the turning points at $-1$ by using this technique.

From now on, we will assume that $x_0$ is a turning point at $1$ for the WKB ansatz, with the added imposition of the turning point being \textit{simple}, meaning $V'(x_0)\neq0$ at the turning point.

\subsection{From the linear problem to a uniform WKB ansatz}

As we have seen, the most generic linear problem can be written as
\begin{align}
\psi(x+\ii\hbar,\hbar)+\psi(x-\ii\hbar,\hbar)=2(E-gx+1)\psi(x,\hbar),
\end{align}
where $g$ is a coupling constant. It is convenient to rewrite this as
\begin{align}
\psi(x+\ii\hbar,\hbar)+\psi(x-\ii\hbar,\hbar)=2(1-g(x-x_0))\psi(x,\hbar),
\end{align}
with $x_0=E/g$. The solution to this difference equation is
\begin{align}
\psi(x,\hbar)=A(q_{x_0})J_{\frac{1-g(x-x_0)}{\ii g\hbar}}\left(\frac{1}{\ii g \hbar}\right)+B(q_{x_0})Y_{\frac{1-g(x-x_0)}{\ii g\hbar}}\left(\frac{1}{\ii g \hbar}\right),
\label{eq:bess_ans}
\end{align}
where we recall that $q_{x_0}$ is defined as
\begin{align}
q_{x_0}=\exp\left(-\frac{2\pi}{\hbar}(x-x_0)\right),
\end{align}
and $A$ and $B$ are Laurent series in $q_{x_0}$. This solution can be used to build an ansatz for the general problem.

For the general problem \eqref{eq:general_problem}, we would have to follow the procedure of standard WKB, obtaining formulae corresponding to \eqref{eq:uniform1} and \eqref{eq:uniform2}. Unfortunately, getting those equations require usage of the chain rule for ordinary derivatives, and we do not have such a tool at our disposal in the deformed case. We can skip this step, at the cost of not having a well-defined equation for the deformation $\phi(x,\hbar)$. While this is surely problematic from a theoretical point of view (as an equation would help establish existence of $\varphi$), the final result will be independent of the particular deformation. Here lies the principal conjecture of this thesis: we will assume existence of this deformation, and test our results numerically to give a motivation for our conjecture.

We start from the classically allowed region, assuming that $x>x_0$ is the allowed region. The deformed WKB ansatz with the even-odd relation implemented is
\begin{align}
\psi_\pm(x,\hbar)=\exp\left(-\frac12\cosh\left(\frac{\ii\hbar\partial_x}2\right)^{-1}\log\sinh D_{\ii\hbar}S(x,\hbar)\right)\exp\left(\mp\frac{1}{\ii\hbar}S(x,\hbar)\right).\label{eq:def_ans}
\end{align}
Here we have denoted\footnote{The operator $D_{\ii\hbar}$ is known as a \textit{finite difference derivative}. We won't go into details in finite difference derivatives, but the reader can find details on \cite{levy1992finite} and \cite{kac2002quantum}.}
\begin{align}
D_{\ii\hbar}S(x,\hbar)=\frac{S\left(x+\frac{\ii\hbar}2,\hbar\right)-S\left(x-\frac{\ii\hbar}2,\hbar\right)}{\ii\hbar}.
\end{align}
In order to reproduce the uniform WKB ansatz in the deformed case, we define
\begin{align}
\begin{aligned}
\varphi_{\pm}(x,\hbar)=&\exp\left(-\frac12\cosh\left(\frac{\ii\hbar\partial_x}2\right)^{-1}\log\sinh D_{\ii\hbar}\phi(x,\hbar)\right)\times\\
&\times\frac{1}{(\phi(x,\hbar)^2-1)^{\frac 14}}\beta_{\pm}(\phi(x,\hbar),-\ii\hbar V'(x)).
\end{aligned}
\end{align}
$\beta_\pm$ are the series defined by \eqref{eq:beta_al} and \eqref{eq:beta_for}, used for the asymptotic expansion of the Bessel functions. The factor $V'(x)$ in the second argument of $\beta$ is necessary, as in \eqref{eq:bess_ans} we see that $\hbar$ is always multiplied by minus the slope of the potential at $x=x_0$. As in the ordinary WKB ansatz, $\phi(x,\hbar)$ is a deformation needed to have $\varphi_{\pm}(x,\hbar)$ as a solution to \eqref{eq:general_problem}, and $\phi(x,\hbar)$ is a series of the form
\begin{align}
\phi(x,\hbar)=\sum_{n=0}^\infty\hbar^{2n}\phi_n(x).
\end{align}
$\phi_0(x)$ is implicitly defined by
\begin{align}
-\frac{1}{V'(x)}\int_1^{\phi_0(x)}\arccosh t\dd t=\int_{x_0}^x\arccosh(E-V(t)+1)\dd t,\label{eq:first_term_eq}
\end{align}
in order to make the exponential terms match. From this equation, we have that $\phi_0(x)>1$ in the allowed region, $0<\phi_0(x)<1$ in the forbidden region. The equation for the higher terms is as before a linear non homogeneous differential equation of first order, so we can always set $\phi_n(x_0)=0$ for $n>0$. The difference with standard WKB is that the only way that we have found to compute the $\phi_n(x)$ for $n>0$ is to take a series expansion in $\hbar$ of the coefficients of $\beta_{\pm}$ and then match coefficients of the expansion. As stated, the existence of this function is the main conjecture of this thesis, but no details of $\phi$ will be needed in the practical computations.

As before, the linkage is made by forgetting for a moment the definition of $S(x,\hbar)$ in terms of chapter \ref{chap:1}, and redefining
\begin{align}
S(x,\hbar)=-\frac{\ii\hbar}2\log\frac{\varphi_+(x,\hbar)}{\varphi_-(x,\hbar)}.\label{eq:bridge}
\end{align}
We have to simplify the even-odd term in \eqref{eq:def_ans}. We now compute
\begin{align}
D_{\ii\hbar}S(x,\hbar)=-\frac12\log\frac{\varphi_+\left(x+\frac{\ii\hbar}2,\hbar\right)}{\varphi_-\left(x+\frac{\ii\hbar}2,\hbar\right)}\frac{\varphi_-\left(x-\frac{\ii\hbar}2,\hbar\right)}{\varphi_+\left(x-\frac{\ii\hbar}2,\hbar\right)}
\end{align}
The application of $\sinh$ gives
\begin{align}
\begin{aligned}
&\sinh D_{\ii\hbar}S(x,\hbar)=\\
=&\frac12\left(\frac{\varphi_-\left(x+\frac{\ii\hbar}2,\hbar\right)\varphi_+\left(x-\frac{\ii\hbar}2,\hbar\right)-\varphi_-\left(x-\frac{\ii\hbar}2,\hbar\right)\varphi_+\left(x+\frac{\ii\hbar}2,\hbar\right)}{\sqrt{\varphi_-\left(x+\frac{\ii\hbar}2,\hbar\right)\varphi_-\left(x-\frac{\ii\hbar}2,\hbar\right)\varphi_+\left(x+\frac{\ii\hbar}2,\hbar\right)\varphi_+\left(x-\frac{\ii\hbar}2,\hbar\right)}}\right).
\end{aligned}
\end{align}
The term in the numerator is the \textit{Casoratian} of \eqref{eq:general_problem}, as $\varphi_\pm(x,\hbar)$ solve the general problem. The Casoratian is the equivalent of the Wronskian for difference equations \cite{zwillinger2018crc}. The Casoratian is zero if and only if $\varphi_-$ and $\varphi_+$ can be obtained from one another by multiplication by a periodic function, and is a non zero periodic function otherwise. As $\varphi_-$ and $\varphi_+$ are independent solutions (as evident from their asymptotics), we conclude that
\begin{align}
\begin{aligned}
\varphi_-\left(x+\frac{\ii\hbar}2,\hbar\right)\varphi_+\left(x-\frac{\ii\hbar}2,\hbar\right)-&\varphi_-\left(x-\frac{\ii\hbar}2,\hbar\right)\varphi_+\left(x+\frac{\ii\hbar}2,\hbar\right)=\\&=2P\left(x+\frac{\ii\hbar}2,\hbar\right)^{-2},
\end{aligned}
\end{align}
where $P$ is a periodic function in $x$ of period $\ii\hbar$ and the additional factors are chosen for convenience. Continuing the computation, we get
\begin{align}
\begin{aligned}
&\log\sinh D_{\ii\hbar}S(x,\hbar)=-2\log P\left(x+\frac{\ii\hbar}2,\hbar\right)-\\
-&\frac12\log\left(\varphi_-\left(x+\frac{\ii\hbar}2,\hbar\right)\varphi_-\left(x-\frac{\ii\hbar}2,\hbar\right)\varphi_+\left(x+\frac{\ii\hbar}2,\hbar\right)\varphi_+\left(x-\frac{\ii\hbar}2,\hbar\right)\right).\label{eq:before_inv}
\end{aligned}
\end{align}
We now have to apply the operator $\cosh\left(\frac{\ii\hbar}2\partial_x\right)^{-1}$. To do so, we check the action of its inverse on test functions $f(x,\hbar)$:
\begin{align}
\cosh\left(\frac{\ii\hbar}2\partial_x\right)f(x,\hbar)=\frac12\left(f\left(x+\frac{\ii\hbar}2,\hbar\right)+f\left(x-\frac{\ii\hbar}2,\hbar\right)\right).\label{eq:straight_ap}
\end{align}
If $f$ has period $\ii\hbar$ in $x$, the operator is a simple translation, so the action of its inverse on $P(x,\hbar)$ is trivial. Due to the properties of $\log$, the second row of \eqref{eq:before_inv} is exactly of the form of the RHS of \eqref{eq:straight_ap}: we can then conclude that
\begin{align}
-\frac12\cosh\left(\frac{\ii\hbar}2\partial_x\right)^{-1}\log\sinh D_{\ii\hbar}S=\log P\left(x,\hbar\right)+\frac12\log\varphi_+(x,\hbar)\varphi_-(x,\hbar).
\end{align}
From this, we can conclude
\begin{align}
\begin{aligned}
&\exp\left(-\frac12\cosh\left(\frac{\ii\hbar}2\partial_x\right)^{-1}\log\sinh D_{\ii\hbar}S\right)\exp\left(\mp\frac{1}{\ii\hbar}S(x,\hbar)\right)=\\
=&P(x,\hbar)\varphi_+(x,\hbar)\varphi_-(x,\hbar)\left(\frac{\varphi_+(x,\hbar)}{\varphi_-(x,\hbar)}\right)^{\pm\frac12}=P(x,\hbar)\varphi_{\pm}(x,\hbar).
\end{aligned}
\end{align}
We can conclude that $\psi_\pm$ defined in \eqref{eq:def_ans} with the redefinition \eqref{eq:bridge} is a solution of \eqref{eq:general_problem}, as $\varphi_{\pm}(x,\hbar)$ are solutions to \eqref{eq:general_problem}. We can expand $P$ as
\begin{align}
P(x,\hbar)=\sum_{n=-\infty}^\infty p_n(\hbar)\exp\left(-\frac{2\pi}{\hbar}x\right):
\end{align}
in order to have \eqref{eq:first_term_eq}, due to the fact that the imaginary part of $\arccosh$ in the chosen determination can only go from $-\ii\pi$ to $\ii\pi$, we can assume that $p_n(x)=0$ for all $n\neq0$, so $P(x,\hbar)$ will reduce to a constant that is not important in our work.

We now have to consider other possible solutions, the ones obtained by multiplying the standard solution \eqref{eq:def_ans} by $q_{x_0}$. This can be done by translating the $\arccosh$ functions in \eqref{eq:first_term_eq} by $2\pi\ii$, obtaining
\begin{align}
\begin{aligned}
&-\frac{1}{V'(x)}\int_1^{\phi_0(x)}\arccosh t\dd t-\frac{2\pi\ii n}{V'(x)}(\phi_0(x)-1)=\\
=&\int_{x_0}^x\arccosh(E-V(t)+1)\dd t+2\pi\ii n(x-x_0),
\end{aligned}
\end{align}
with $n$ an integer. Expanding $\phi_0(x)$ to first order in $x-x_0$ shows that at least at this order $\phi_0(x)$ satisfies the updated equation. We naturally come to the definition of the equivalents of the $q_{x_0}$ factors in the uniform deformed WKB ansatz:
\begin{align}
q^{(\phi)}=\exp\left(\frac{2\pi n}{\hbar V'(x)}(\phi_0(x)-1)\right).
\end{align}
Our conclusion is
\begin{align}
\begin{aligned}
&\exp\left(-\frac12\cosh\left(\frac{\ii\hbar}2\partial_x\right)^{-1}\log\sinh D_{\ii\hbar}S\right)\exp\left(\mp\frac{1}{\ii\hbar}S(x,\hbar)\right)q_{x_0}^{\pm n}=\\
=&(q^{(\phi)})^{\pm n}\exp\left(-\frac12\cosh\left(\frac{\ii\hbar\partial_x}2\right)^{-1}\log\sinh D_{\ii\hbar}\phi(x,\hbar)\right)\times\\
&\times\frac{1}{(\phi(x,\hbar)^2-1)^{\frac 14}}\beta_{\pm}(\phi(x,\hbar),-\ii\hbar V'(x)).
\end{aligned}
\label{eq:bridge_1}
\end{align}
This is the conclusion in the classically allowed region to bridge between the deformed WKB ansatz and the uniform WKB ansatz adapted to the deformed case.

In the forbidden region (that we assume to be at $x<x_0$), $S(x,\hbar)$ is imaginary, so it is more convenient to work with the real function
\begin{align}
T(x,\hbar)=\ii S(x,\hbar).
\end{align}
$\psi_{\pm}(x,\hbar)$ becomes
\begin{align}
\psi_{\pm}(x,\hbar)=\exp\left(-\frac12\cosh\left(\frac{\ii\hbar}2\partial_x\right)^{-1}\log(-\ii \sin D_{\ii\hbar} T(x,\hbar))\right)\exp\left(\pm\frac{1}{\hbar}T(x,\hbar)\right). \label{eq:psi_for}
\end{align}
$\phi_0$ is still defined by \eqref{eq:first_term_eq}, and is smaller than $1$ in the forbidden region. We rewrite that equation as
\begin{align}
-\frac{1}{V'(x)}\int_{\phi_0(x)}^{1}\arccos t\dd t=\int_{x}^{x_0}\arccos(E-V(t)+1)\dd t,
\label{eq:for_eq}
\end{align}
As in the forbidden region the argument of the $\arccos$ in the RHS is between $-1$ and $1$, this is a real valued equality, and $\phi_0(x)<1$. We then rewrite $\varphi_{\pm}$ as
\begin{align}
\begin{aligned}
\varphi_{\pm}(x,\hbar)=&\exp\left(-\frac12\cosh\left(\frac{\ii\hbar\partial_x}2\right)^{-1}\log\sinh D_{\ii\hbar}\phi(x,\hbar)\right)\times\\
&\times\frac{\ee^{-\ii\frac\pi4}}{(1-\phi(x,\hbar)^2)^{\frac 14}}\beta_{\pm}(\phi(x,\hbar),-\ii\hbar V'(x)).
\end{aligned}
\end{align}
In \eqref{eq:psi_for}, we can extract the $-\ii$ from the $\log$ to obtain a phase of $\ee^{-\ii\frac\pi4}$. Following the same procedure of the allowed zone, we conclude
\begin{align}
\begin{aligned}
&\exp\left(-\frac12\cosh\left(\frac{\ii\hbar}2\partial_x\right)^{-1}\log\sin D_{\ii\hbar}T\right)\exp\left(\pm\frac{1}{\hbar}T(x,\hbar)\right)=\\
=&\exp\left(-\frac12\cosh\left(\frac{\ii\hbar\partial_x}2\right)^{-1}\log\sinh D_{\ii\hbar}\phi(x,\hbar)\right)\frac{1}{(1-\phi(x,\hbar)^2)^{\frac 14}}\beta_{\pm}(\phi(x,\hbar),-\ii\hbar V'(x)).
\end{aligned}
\end{align}
For the other determinations, as we can recover them from $S\to S+2\pi\ii(x-x_0)$, we have to impose $T\to T-2\pi(x-x_0)$. This we can do by shifting the $\arccos$ in \eqref{eq:for_eq} by $-2\pi n$. Following the same steps, we obtain
\begin{align}
\begin{aligned}
&\exp\left(-\frac12\cosh\left(\frac{\ii\hbar}2\partial_x\right)^{-1}\log\sin D_{\ii\hbar}T\right)\exp\left(\pm\frac{1}{\ii\hbar}T(x,\hbar)\right)q_{x_0}^{\pm n}=\\
=&(q^{(\phi)})^{\mp n}\exp\left(-\frac12\cosh\left(\frac{\ii\hbar\partial_x}2\right)^{-1}\log\sinh D_{\ii\hbar}\phi(x,\hbar)\right)\times\\
&\times\frac{1}{(\phi(x,\hbar)^2-1)^{\frac 14}}\beta_{\pm}(\phi(x,\hbar),-\ii\hbar V'(x)).
\end{aligned}
\label{eq:bridge_2}
\end{align}
There is an important difference between \eqref{eq:bridge_1} and \eqref{eq:bridge_2}: the exponents of $q^{(\phi)}$ have inverted sign. This will be very important in determining the correct transition formulae. Now that we have established the bridges between the two descriptions, we can obtain the formulae.

\subsection{Computation of deformed WKB transition formulae}

\begin{figure}
\centering
\includegraphics[scale=1.2]{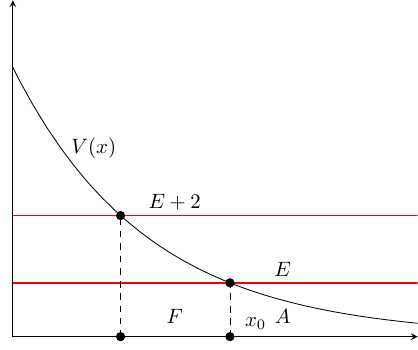}
\caption{Turning point, allowed region and forbidden region in deformed WKB. We also picture (without label) the turning point at $-1$, delimited by the line $V(x)=E+2$.}
\label{fig:connform_en}
\end{figure}

We now come to the main result. We will assume that $x_0$ is a turning point at $1$, with $V'(x_0)<0$ so the classically allowed region is for $x>x_0$ and the classically forbidden region is for $x<x_0$. We picture the arrangement in figure \ref{fig:connform_en}

The wavefunction can be written in the uniform deformed WKB ansatz as
\begin{align}
\begin{aligned}
\psi(x,\hbar)=&\exp\left(-\frac12\cosh\left(\frac{\ii\hbar\partial_x}2\right)^{-1}\log\sinh D_{\ii\hbar}\phi(x,\hbar)\right)\times
\\&\times \left(A(q^{(\phi)})J_{-\frac{\phi(x,\hbar)}{V'(x)\ii\hbar}}\left(-\frac{1}{V'(x)\ii\hbar}\right)+B(q^{(\phi)})Y_{-\frac{\phi(x,\hbar)}{V'(x)\ii\hbar}}\left(-\frac{1}{V'(x)\ii\hbar}\right)\right)
\end{aligned}
\end{align}
We can use the expansions obtained in subsection \ref{sec:phys_reg} to obtain the asymptotic expansions of $J$ and $Y$. For now we will not use explicit matrices, but denote $M_A(q^{(\phi)})$ as the matrix for the expansion in the allowed region, $M_F(q^{(\phi)}$ as the matrix in the forbidden region. The expansion reads
\begin{align}
\begin{aligned}
&\psi(x,\hbar)\simeq\frac{1}{(\phi^2-1)^{\frac14}}\exp\left(-\frac12\cosh\left(\frac{\ii\hbar\partial_x}2\right)^{-1}\log\sinh D_{\ii\hbar}\phi\right)\times\\
&\times \begin{cases}
\left(\begin{array}{cc}
A(q_\phi)&B(q_\phi)
\end{array}\right)
M_A(q_\phi)
\left(\begin{array}{c}
\beta_+(\phi,-V'(x)\ii\hbar)\\\beta_-(\phi,-V'(x)\ii\hbar)
\end{array}\right),\quad &x>x_0,\\
\left(\begin{array}{cc}
A(q_\phi)&B(q_\phi)
\end{array}\right)
M_F(q_\phi)
\left(\begin{array}{c}
\beta_+(\phi,-V'(x)\ii\hbar)\\\beta_-(\phi,-V'(x)\ii\hbar)
\end{array}\right),\quad &x<x_0.
\end{cases}
\end{aligned}
\end{align}
Using formulae \eqref{eq:bridge_1} and \eqref{eq:bridge_2}, we can convert this asymptotic expansion in an expansion in terms of the deformed WKB ansatz as
\begin{align}
\begin{aligned}
&\psi(x,\hbar)\simeq
\exp\left(-\frac12\cosh\left(\frac{\ii\hbar}2\partial_x\right)^{-1}\log\sinh D_{\ii\hbar}S\right)\times\\
\times&\left(\begin{array}{cc}
A\left(q_{x_0}\right)&B\left(q_{x_0}\right)
\end{array}\right)
M_A\left(q_{x_0}\right)
\left(\begin{array}{c}
\exp\left(-\frac1{\ii\hbar}S(x,\hbar)\right)\\\exp\left(\frac1{\ii\hbar}S(x,\hbar)\right)
\end{array}\right)
\end{aligned}\label{eq:allowed_exp}
\end{align}
in the allowed region, and
\begin{align}
\begin{aligned}
&\psi(x,\hbar)\simeq
\exp\left(-\frac12\cosh\left(\frac{\ii\hbar}2\partial_x\right)^{-1}\log\sin D_{\ii\hbar}T\right)\times\\
\times&\left(\begin{array}{cc}
A\left(q_{x_0}^{-1}\right)&B\left(q_{x_0}^{-1}\right)
\end{array}\right)
M_A\left(q_{x_0}^{-1}\right)
\left(\begin{array}{c}
\exp\left(\frac1{\hbar}T(x,\hbar)\right)\\\exp\left(-\frac1{\hbar}T(x,\hbar)\right)
\end{array}\right)\label{eq:forbidden_exp}
\end{aligned}
\end{align}
in the forbidden region. Note the presence of the  $q_{x_0}^{-1}$ in the expansion of the forbidden region, consequence of \eqref{eq:bridge_2}. Obviously, if $V'(x)>0$ (so $x<x_0$ is the allowed region) the two regions have to be inverted.

We now give the connection formulae case by case. We have two parameters to account for: the deformation $\hbar\to\hbar\pm\ii\epsilon$ and the sign of the potential $V'(x_0)$.

\subsubsection{Case 1: $\hbar\to\hbar-\ii\epsilon, V'(x_0)<0$}

In this case the allowed zone is at $x>x_0$, and $\arg (-V'(x_0)\ii\hbar)=\frac{\pi}2^-$, so the expansions to use are \eqref{eq:first_exp} in the allowed region, and \eqref{eq:for_1} for the forbidden region. We write them here\footnote{The additional phase of $\ee^{\ii\frac{\pi}4}$ comes from the $\sqrt{a}$ factor in the expansions, that becomes $\sqrt{\ii\hbar}$ once $-V'(x_0)\hbar$ is plugged in. For $V'(x_0)>0$ we will have an additional phase of $\ee^{-\ii\frac{\pi}4}$.}:
\begin{align}
&M_A\left(q^{(\phi)}\right)=\sqrt{-\frac{V'(x_0)\hbar}{2\pi}}\frac{1}{(\phi^2-1)^\frac14}\left(
\begin{array}{cc}
\ee^{\frac{\ii\pi}4}&0\\
\ee^{-\frac{\ii\pi}4}\frac{q^{(\phi)}+1}{q^{(\phi)}-1}&-2\ee^{\frac{\ii\pi}4}
\end{array}
\right),\\
&M_F\left(q^{(\phi)}\right)=\sqrt{-\frac{V'(x_0)\hbar}{2\pi}}\frac{1}{(1-\phi^2)^\frac14}\left(
\begin{array}{cc}
1&\ii\\
-\ii&-1
\end{array}
\right).
\end{align}
We can now use \eqref{eq:allowed_exp} and \eqref{eq:forbidden_exp} to obtain the expansions for $\psi(x,\hbar)$ and connect the solutions. We can choose $A$ and $B$ in such a way to obtain the connection formulae for the different exponentials. We obtain\footnote{For brevity, we denote
\begin{align}
&\frac1{\ii\hbar}S_{\text o}(x,\hbar)=-\frac12\cosh\left(\frac{\ii\hbar}2\partial_x\right)^{-1}\log\sinh D_{\ii\hbar}S,\\
&\frac1{\ii\hbar}T_o(x,\hbar)=-\frac12\cosh\left(\frac{\ii\hbar}2\partial_x\right)^{-1}\log\sin D_{\ii\hbar}T.
\end{align}
}
\begin{align}
&\begin{aligned}
&\exp\left(\frac{1}{\ii\hbar}S_{\text o}\right)\exp\left(-\frac{1}{\ii\hbar}S\right)\to\\\to&\exp\left(\frac{1}{\ii\hbar}T_o\right)\left(\ee^{-\ii\frac{\pi}4}\exp\left(\frac1\hbar T\right)+\ee^{\ii\frac\pi4}\exp\left(-\frac1\hbar T\right)\right),
\end{aligned}\label{eq:conn_1}\\
&\begin{aligned}
&\exp\left(\frac{1}{\ii\hbar}S_{\text o}\right)\exp\left(\frac{1}{\ii\hbar}S\right)\to\\\to&\exp\left(\frac{1}{\ii\hbar}T_o\right)\left(\frac{q_{x_0}}{q_{x_0}-1}\ee^{\ii\frac{\pi}4}\exp\left(\frac1\hbar T\right)-\frac{1}{q_{x_0}-1}\ee^{-\ii\frac\pi4}\exp\left(-\frac1\hbar T\right)\right).
\end{aligned}\label{eq:conn_2}
\end{align}
Inverse formulae are obtained in the same way:
\begin{align}
&\begin{aligned}
&\exp\left(\frac{1}{\ii\hbar}T_o\right)\exp\left(\frac{1}{\hbar}T\right)\to\\\to&\exp\left(\frac{1}{\ii\hbar}S_{\text o}\right)\left(\frac{q_{x_0}}{q_{x_0}-1}\ee^{\ii\frac{\pi}4}\exp\left(-\frac1{\ii\hbar} S\right)+\ee^{-\ii\frac\pi4}\exp\left(\frac1{\ii\hbar} S\right)\right),
\end{aligned}\label{eq:conn_3}\\
&\begin{aligned}
&\exp\left(\frac{1}{\ii\hbar}T_o\right)\exp\left(-\frac{1}{\hbar}T\right)\to\\\to&\exp\left(\frac{1}{\ii\hbar}S_{\text o}\right)\left(-\frac{1}{q_{x_0}-1}\ee^{-\ii\frac{\pi}4}\exp\left(-\frac1{\ii\hbar} S\right)+\ee^{\ii\frac\pi4}\exp\left(\frac1{\ii\hbar}S\right)\right).
\end{aligned}\label{eq:conn_4}
\end{align}
As the transformations are linear, we can arrange them in matrix form:
\begin{align}
&\begin{aligned}
&\exp\left(\frac{1}{\ii\hbar}S_{\text o}\right)\left(\begin{array}{c}
\exp\left(-\frac{1}{\ii\hbar}S\right)\\
\exp\left(\frac{1}{\ii\hbar}S\right)
\end{array}\right)\to\\
\to&\exp\left(\frac{1}{\ii\hbar}T_o\right)\left(\begin{array}{cc}
\ee^{-\ii\frac\pi4}&\ee^{\ii\frac\pi4}\\
\frac{q_{x_0}}{q_{x_0}-1}\ee^{\ii\frac\pi4}&-\frac{1}{q_{x_0}-1}\ee^{-\ii\frac\pi4}
\end{array}
\right)
\left(\begin{array}{c}
\exp\left(\frac{1}{\hbar}T\right)\\
\exp\left(-\frac{1}{\hbar}T\right)
\end{array}\right),
\end{aligned}\label{eq:mat_conn_1}\\
&\begin{aligned}
&\exp\left(\frac{1}{\ii\hbar}T_o\right)\left(\begin{array}{c}
\exp\left(\frac{1}{\hbar}T\right)\\
\exp\left(-\frac{1}{\hbar}T\right)
\end{array}\right)\to\\
\to&\exp\left(\frac{1}{\ii\hbar}S_{\text o}\right)\left(\begin{array}{cc}
\frac{q_{x_0}}{q_{x_0}-1}\ee^{\ii\frac\pi4}&\ee^{-\ii\frac\pi4}\\
-\frac{1}{q_{x_0}-1}\ee^{-\ii\frac\pi4}&\ee^{\ii\frac\pi4}
\end{array}
\right)
\left(\begin{array}{c}
\exp\left(-\frac{1}{\ii\hbar}S\right)\\
\exp\left(\frac{1}{\ii\hbar}S\right)
\end{array}\right).
\end{aligned}\label{eq:mat_conn_2}
\end{align}
\eqref{eq:mat_conn_1} and \eqref{eq:mat_conn_2} are more compact than \eqref{eq:conn_1} to \eqref{eq:conn_4}, so we will always give results in matrix form. Furthermore, one can verify that the matrix in \eqref{eq:mat_conn_1} is the inverse of the matrix in \eqref{eq:mat_conn_2} evaluated with $q_{x_0}\to q_{x_0}^{-1}$: the necessity of the inversion comes from the $-1$ factors in \eqref{eq:forbidden_exp}, not present in \eqref{eq:allowed_exp}. This will be a general property.

\subsubsection{Case 2: $\hbar\to\hbar-\ii\epsilon, V'(x_0)>0$}

In this case the forbidden region is at $x>x_0$ and the allowed region is at $x<x_0$. As $\arg(-V'(x_0)\ii\hbar)=\left(-\frac{\pi}2\right)^-$, the matrices to use are \eqref{eq:four_exp} in the allowed zone and \eqref{eq:for_4} in the forbidden zone. The matrices are
\begin{align}
&M_A(q^{(\phi)})=\sqrt{\frac{V'(x_0)\hbar}{2\pi}}\frac{1}{(\phi^2-1)^\frac14}\left(\begin{array}{cc}
\ee^{-\ii\frac{\pi}4}(1-(q^{(\phi)})^{-1})&0\\
-\ee^{\ii\frac{\pi}4}(1+(q^{(\phi)})^{-1})&\ee^{-\ii\frac\pi4}\frac{2}{(q^{(\phi)})^{-1}-1)}
\end{array}\right),\\
&M_F(q^{(\phi)})=\sqrt{\frac{V'(x_0)\hbar}{2\pi}}\frac{1}{(1-\phi^2)^\frac14}\left(\begin{array}{cc}
\ii(q^{(\phi)})^{-1}&1\\
-(2+(q^{(\phi)})^{-1})&\ii
\end{array}\right).
\end{align}
The transitions in matrix form are given by
\begin{align}
&\begin{aligned}
&\exp\left(\frac{1}{\ii\hbar}S_{\text o}\right)\left(\begin{array}{c}
\exp\left(-\frac{1}{\ii\hbar}S\right)\\
\exp\left(\frac{1}{\ii\hbar}S\right)
\end{array}\right)\to\\
\to&\exp\left(\frac{1}{\ii\hbar}T_o\right)\left(
\begin{array}{cc}
 -\ee^{-\ii\frac{\pi}{4}}\frac{1}{1-q_{x_0}^{-1}} & \ee^{\ii\frac{\pi}{4}}\frac{q_{x_0}^{-1}}{q_{x_0}^{-1}-1} \\
 \ee^{\ii\frac{\pi}{4}} & \ee^{-\ii\frac{\pi}{4}} \\
\end{array}
\right)
\left(\begin{array}{c}
\exp\left(\frac{1}{\hbar}T\right)\\
\exp\left(-\frac{1}{\hbar}T\right)
\end{array}\right),
\end{aligned}\label{eq:mat_conn_3}\\
&\begin{aligned}
&\exp\left(\frac{1}{\ii\hbar}T_o\right)\left(\begin{array}{c}
\exp\left(\frac{1}{\hbar}T\right)\\
\exp\left(-\frac{1}{\hbar}T\right)
\end{array}\right)\to\\
\to&\exp\left(\frac{1}{\ii\hbar}S_{\text o}\right)\left(
\begin{array}{cc}
 \ee^{\ii\frac{\pi}{4}} & \ee^{-\ii\frac{\pi}{4}}\frac{1}{1-q_{x_0}^{-1}} \\
 \ee^{-\ii\frac{\pi}{4}} & \ee^{\ii\frac{\pi}{4}}\frac{q_{x_0}^{-1}}{q_{x_0}^{-1}-1} \\
\end{array}
\right)
\left(\begin{array}{c}
\exp\left(-\frac{1}{\ii\hbar}S\right)\\
\exp\left(\frac{1}{\ii\hbar}S\right)
\end{array}\right).
\end{aligned}\label{eq:mat_conn_4}
\end{align}

\subsubsection{Case 3: $\hbar\to\hbar+\ii\epsilon,V'(x_0)<0$}

In this case the forbidden region is at $x<x_0$ and the allowed region at $x>x_0$. As $\arg(-V'(x_0)\ii\hbar)=\frac\pi2^+$, we use \eqref{eq:sec_exp} for the allowed region and \eqref{eq:for_2} for the forbidden region. The matrices are
\begin{align}
&M_A\left(q^{(\phi)}\right)=\sqrt{\frac{-V'(x_0)\hbar}{2\pi}}\frac{1}{(\phi^2-1)^\frac14}\left(
\begin{array}{cc}
\ee^{\frac{\ii\pi}4}(1-q^{(\phi)})&0\\
-\ee^{-\frac{\ii\pi}4}(q^{(\phi)}+1)&-\frac{2}{q^{(\phi)}-1}\ee^{\frac{\ii\pi}4}
\end{array}
\right),\\
&M_F\left(q^{(\phi)}\right)=\sqrt{\frac{-V'(x_0)\hbar}{2\pi}}\frac{1}{(1-\phi^2)^\frac14}\left(
\begin{array}{cc}
1&-\ii q^{(\phi)}\\
-\ii&-(2+q^{(\phi)})
\end{array}
\right).
\end{align}
The transitions are
\begin{align}
&\begin{aligned}
&\exp\left(\frac{1}{\ii\hbar}S_{\text o}\right)\left(\begin{array}{c}
\exp\left(-\frac{1}{\ii\hbar}S\right)\\
\exp\left(\frac{1}{\ii\hbar}S\right)
\end{array}\right)\to\\
\to&\exp\left(\frac{1}{\ii\hbar}T_o\right)\left(
\begin{array}{cc}
 \ee^{-\ii\frac{\pi}{4}}\frac{q_{x_0}}{q_{x_0}-1} & -\ee^{\ii\frac{\pi}{4}}\frac{1}{q_{x_0}-1} \\
 \ee^{\ii\frac{\pi}{4}} & \ee^{-\ii\frac{\pi}{4}} \\
\end{array}
\right)
\left(\begin{array}{c}
\exp\left(\frac{1}{\hbar}T\right)\\
\exp\left(-\frac{1}{\hbar}T\right)
\end{array}\right),
\end{aligned}\label{eq:mat_conn_5}\\
&\begin{aligned}
&\exp\left(\frac{1}{\ii\hbar}T_o\right)\left(\begin{array}{c}
\exp\left(\frac{1}{\hbar}T\right)\\
\exp\left(-\frac{1}{\hbar}T\right)
\end{array}\right)\to\\
\to&\exp\left(\frac{1}{\ii\hbar}S_{\text o}\right)\left(
\begin{array}{cc}
 \ee^{\ii\frac{\pi }{4}} & \ee^{-\ii\frac{\pi }{4}}\frac{q_{x_0}}{q_{x_0}-1} \\
 \ee^{-\ii\frac{\pi }{4}} & \ee^{\ii\frac{\pi }{4}}\frac{1}{1-q_{x_0}} \\
\end{array}
\right)
\left(\begin{array}{c}
\exp\left(-\frac{1}{\ii\hbar}S\right)\\
\exp\left(\frac{1}{\ii\hbar}S\right)
\end{array}\right).
\end{aligned}\label{eq:mat_conn_6}
\end{align}

\subsubsection{Case 4: $\hbar\to\hbar+\ii\epsilon,V'(x_0)>0$}

In this case the forbidden region is at $x>x_0$ and the allowed region at $x<x_0$. As $\arg(-V'(x_0)\ii\hbar)=\left(-\frac\pi2\right)^+$, we use \eqref{eq:thir_exp} for the allowed region and \eqref{eq:for_3} for the forbidden region. The matrices are
\begin{align}
&M_A\left(q^{(\phi)}\right)=\sqrt{\frac{V'(x_0)\hbar}{2\pi}}\frac{1}{(\phi^2-1)^\frac14}\left(
\begin{array}{cc}
\ee^{-\frac{\ii\pi}4}&0\\
\ee^{\frac{\ii\pi}4}\frac{(q^{(\phi)})^{-1}+1}{q^{(\phi)})^{-1}-1}&-2\ee^{-\frac{\ii\pi}4}
\end{array}
\right),\\
&M_F\left(q^{(\phi)}\right)=\sqrt{\frac{V'(x_0)\hbar}{2\pi}}\frac{1}{(1-\phi^2)^\frac14}\left(
\begin{array}{cc}
-\ii&1\\
-1&\ii
\end{array}
\right).
\end{align}
The transitions are
\begin{align}
&\begin{aligned}
&\exp\left(\frac{1}{\ii\hbar}S_{\text o}\right)\left(\begin{array}{c}
\exp\left(-\frac{1}{\ii\hbar}S\right)\\
\exp\left(\frac{1}{\ii\hbar}S\right)
\end{array}\right)\to\\
\to&\exp\left(\frac{1}{\ii\hbar}T_o\right)\left(
\begin{array}{cc}
 \ee^{-\ii\frac{\pi }{4}} & \ee^{\ii\frac{\pi }{4}} \\
 \ee^{\ii\frac{\pi }{4}} \frac{1}{1-q^{-1}_{x_0}} & \ee^{-\ii\frac{ \pi }{4}}\frac{q_{x_0}^{-1}}{q_{x_0}^{-1}-1} \\
\end{array}
\right)
\left(\begin{array}{c}
\exp\left(\frac{1}{\hbar}T\right)\\
\exp\left(-\frac{1}{\hbar}T\right)
\end{array}\right),
\end{aligned}\label{eq:mat_conn_7}\\
&\begin{aligned}
&\exp\left(\frac{1}{\ii\hbar}T_o\right)\left(\begin{array}{c}
\exp\left(\frac{1}{\hbar}T\right)\\
\exp\left(-\frac{1}{\hbar}T\right)
\end{array}\right)\to\\
\to&\exp\left(\frac{1}{\ii\hbar}S_{\text o}\right)\left(
\begin{array}{cc}
 \ee^{\ii\frac{ \pi }{4}}\frac{1}{1-q_{x_0}^{-1}} & \ee^{-\ii\frac{\pi }{4}} \\
 -\ee^{\ii\frac{\pi }{4}}\frac{q_{x_0}^{-1}}{q_{x_0}^{-1}-1} & \ee^{\ii\frac{\pi }{4}} \\
\end{array}
\right)
\left(\begin{array}{c}
\exp\left(-\frac{1}{\ii\hbar}S\right)\\
\exp\left(\frac{1}{\ii\hbar}S\right)
\end{array}\right).
\end{aligned}\label{eq:mat_conn_8}
\end{align}

\subsection{Turning points at $-1$}

Up until now, we have studied the turning points at $1$. As anticipated, in order to obtain those transition formulae we can use a trick. If $\psi(x,\hbar)$ solves
\begin{align}
\psi(x+\ii\hbar,\hbar)+\psi(x-\ii\hbar,\hbar)=2Q(x)\psi(x,\hbar),
\end{align}
with a turning point $x_0$ such as $Q(x_0)=-1$, then $\Psi(x,\hbar)=\sqrt{q_{x_0}}\Psi(x,\hbar)$ solves
\begin{align}
\Psi(x+\ii\hbar,\hbar)+\Psi(x-\ii\hbar,\hbar)=-2Q(x)\Psi(x,\hbar),
\end{align}
so $x_0$ is a turning point at $1$ for $\Psi(x,\hbar)$, and we can use the standard connection formulae there.

In order to implement this, we note that we can implement this transformation by changing the $P_0$ term as
\begin{align}
P_0(x)=\arccosh Q(x)\to -\arccosh (-Q(x))+\ii\pi:
\end{align}
the integration from $x_0$ to $x$ will produce the exact $\sqrt{q_{x_0}}$ term that we are looking for\footnote{Our choice is not the only possibility, as there are other determinations of $\arccosh$ that we have ignored. We have made this choice a posteriori, by looking at the example of the Toda lattice that we will study in the next chapter and comparing with the known result. A future line of research is to make this work more rigorous, and justifying this choice among all possibilities could be a future result.}. Obtaining the transition matrices is then very simple: defining
\begin{align}
I^{(pm)}(q_{x_0})=\left(
\begin{array}{cc}0&q^{\mp\frac12}_{x_0}\\
q^{\pm\frac12}_{x_0}&0\end{array}
\right),\label{eq:invmat}
\end{align}
where the upper sign is picked for the deformation $\hbar\to\hbar-\ii\epsilon$ and the lower sign for the deformation $\hbar\to\hbar+\ii\epsilon$ all transition matrices at $-1$ can be obtained by multiplication on the left and the right by this matrix, also considering the fact that as $Q(x)$ is inverted we have to invert the sign of $V'(x_0)$ to pick the right matrix.

\section{Limit to standard WKB}

The connection formulae that we have written have to be intended as deformations of the standard WKB connection formulae. In particular, in the limit to standard quantum mechanics, they reproduce the standard WKB connection formulae. We will show that in this section, studying how the deformed WKB ansatz reduces to the standard one in the appropriate limit.

At first, we have to clarify how to get the limit to standard quantum mechanics. As we have seen in chapter \ref{chap:1}, this can be done by restoring the $\Lambda,m$ dependency through dimensional analysis, and then sending $\Lambda\to\infty$. We start from \eqref{eq:fin_dif_start}, that we rewrite for convenience:
\begin{align}
\begin{aligned}
\exp\left(\frac{1}{\ii\hbar}(S_{\text T}(x+\ii\hbar,\hbar)-S_{\text T}(x,\hbar))\right)+&\exp\left(\frac{1}{\ii\hbar}(S_{\text T}(x-\ii\hbar,\hbar)-S_{\text T}(x,\hbar))\right)=\\
&=2Q(x),
\end{aligned}
\end{align}
with $Q(x)=E-V(x)+1$. Here we have reintroduced the total action $S_{\text T}$, including even and odd parts in $\hbar$. We first have to make $Q(x)$ adimensional: as $\Lambda$ has the dimensions of an energy, this is trivially done by
\begin{align}
Q(x)=\frac{E-V(x)}\Lambda+1.
\end{align}
Let us now turn to the argument of the exponentials. We cannot have linear combinations of $x$ and $\hbar$ as they have different dimensions: exploiting the fact that $\sqrt{m\Lambda}$ has the dimension of momentum, we see that the appropriate rescaling is
\begin{align}
\begin{aligned}
&\Lambda\exp\left(\frac{1}{\ii\hbar}\left(S_{\text T}\left(x+\frac{\ii\hbar}{\sqrt{m \Lambda}},\hbar\right)-S_{\text T}(x,\hbar)\right)\right)+\\
+&\Lambda\exp\left(\frac{1}{\ii\hbar}\left(S_{\text T}\left(x-\frac{\ii\hbar}{\sqrt{m \Lambda}},\hbar\right)-S_{\text T}(x,\hbar)\right)\right)=2(E-V(x)+\Lambda),\label{eq:expansion}
\end{aligned}
\end{align}
We now just have to expand $\Lambda$ around $\infty$ to obtain
\begin{align}
\begin{aligned}
&\exp\left(\frac{1}{\ii\hbar}\left(S_{\text T}\left(x\pm\frac{\ii\hbar}{\sqrt{m \Lambda}},\hbar\right)-S_{\text T}(x,\hbar)\right)\right)=\\
=&1\pm\frac{S_{\text T}'(x,\hbar)}{\sqrt{m\Lambda}}+\frac{S_{\text T}'(x,\hbar)^2+\ii\hbar S_{\text T}''(x,\hbar)}{2m\Lambda}+o\left(\Lambda^{-\frac32}\right).
\end{aligned}
\end{align}
In the sum of \eqref{eq:expansion}, the terms proportional to $\Lambda^{\frac12}$ cancel out, and the term $2\Lambda$ on the LHS is matched by the one on the RHS. We are left with
\begin{align}
S'_T(x,\hbar)+\ii S''_T(x,\hbar)=2m(E-V(x)),
\end{align}
that is precisely the starting point for the WKB analysis of the Schr\"odinger equation. From this we obtain that the $S$ of deformed WKB goes to the $S$ of standard WKB  in the $\Lambda\to\infty$ limit, and also the even-odd terms become
\begin{align}
&\frac1{\ii\hbar}S_{\text o}(x,\hbar)\to\frac{1}{\sqrt{S'(x,\hbar)}}\\
&\frac1{\ii\hbar}T_o(x,\hbar)\to\frac{1}{\sqrt{-T'(x,\hbar)}},
\end{align}
simplifying to the standard WKB expressions.

The exponential term $q_{x_0}$ also has an important limit. Restoring units, we have
\begin{align}
q_{x_0}=\exp\left(-\frac{2\pi\sqrt{m\Lambda}}{\hbar}(x-x_0)\right).
\end{align}
As $\hbar>0$, the limit of $q_{x_0}$ for $\Lambda\to\infty$ entirely depends on the sign of $x-x_0$: $q_{x_0}\to0$ for $x>x_0$, and $q_{x_0}\to\infty$ for $x<x_0$. With this limit, we are ready to take the standard WKB limit of our connection formulae. As always, we work case by case.

We also note that the only relevant connection formulae are the ones at $x_0$ such as $E=V(x_0)$, the turning points at $1$. This is because the other turning points are located at $E+2\Lambda=V(x_0)$: in the $\Lambda\to\infty$ limit, those connection formulae are never encountered.

\subsubsection{Case 1: $\hbar\to\hbar-\ii\epsilon, V'(x_0)<0$}

The transition formulae are \eqref{eq:mat_conn_1} and \eqref{eq:mat_conn_2}. We have $x>x_0$ in the allowed region and $x<x_0$ in the forbidden region. As \eqref{eq:mat_conn_1} mixes terms in the forbidden region, we have to take the limit $q_{x_0}\to\infty$ there and $q_{x_0}\to0$ in the other matrix. The transition formulae become
\begin{align}
&\begin{aligned}
&\frac{1}{\sqrt{S'}}\left(\begin{array}{c}
\exp\left(-\frac{1}{\ii\hbar}S\right)\\
\exp\left(\frac{1}{\ii\hbar}S\right)
\end{array}\right)
\to&\frac{1}{\sqrt{-T'}}\left(\begin{array}{cc}
\ee^{-\ii\frac\pi4}&\ee^{\ii\frac\pi4}\\
\ee^{\ii\frac\pi4}&0
\end{array}
\right)
\left(\begin{array}{c}
\exp\left(\frac{1}{\hbar}T\right)\\
\exp\left(-\frac{1}{\hbar}T\right)
\end{array}\right),
\end{aligned}\\
&\begin{aligned}
&\frac{1}{\sqrt{-T'}}\left(\begin{array}{c}
\exp\left(\frac{1}{\hbar}T\right)\\
\exp\left(-\frac{1}{\hbar}T\right)
\end{array}\right)
\to&\frac{1}{\sqrt{S'}}\left(\begin{array}{cc}
0&\ee^{-\ii\frac\pi4}\\
\ee^{-\ii\frac\pi4}&\ee^{\ii\frac\pi4}
\end{array}
\right)
\left(\begin{array}{c}
\exp\left(-\frac{1}{\ii\hbar}S\right)\\
\exp\left(\frac{1}{\ii\hbar}S\right)
\end{array}\right).
\end{aligned}
\end{align}
The first transition reproduces \eqref{eq:stand_1} and \eqref{eq:stand_2} and the second is the inverse of the first, so we have correctly recovered the standard limit.

\subsubsection{Case 2: $\hbar\to\hbar-\ii\epsilon, V'(x_0)>0$}

The transition formulae are \eqref{eq:mat_conn_3} and \eqref{eq:mat_conn_4}. This time we have to invert the limits, as $x<x_0$ is the allowed region and $x>x_0$ is the forbidden region. We have
\begin{align}
&\begin{aligned}
&\frac{1}{\sqrt{S'}}\left(\begin{array}{c}
\exp\left(-\frac{1}{\ii\hbar}S\right)\\
\exp\left(\frac{1}{\ii\hbar}S\right)
\end{array}\right)
\to&\frac{1}{\sqrt{-T'}}\left(
\begin{array}{cc}
0& \ee^{\ii\frac{\pi}{4}} \\
 \ee^{\ii\frac{\pi}{4}} & \ee^{-\ii\frac{\pi}{4}} \\
\end{array}
\right)
\left(\begin{array}{c}
\exp\left(\frac{1}{\hbar}T\right)\\
\exp\left(-\frac{1}{\hbar}T\right)
\end{array}\right),
\end{aligned}\\
&\begin{aligned}
&\frac{1}{\sqrt{-T'}}\left(\begin{array}{c}
\exp\left(\frac{1}{\hbar}T\right)\\
\exp\left(-\frac{1}{\hbar}T\right)
\end{array}\right)
\to&\frac{1}{\sqrt{S'}}\left(
\begin{array}{cc}
 \ee^{\ii\frac{\pi}{4}} & \ee^{-\ii\frac{\pi}{4}} \\
 \ee^{-\ii\frac{\pi}{4}} & 0\\
\end{array}
\right)
\left(\begin{array}{c}
\exp\left(-\frac{1}{\ii\hbar}S\right)\\
\exp\left(\frac{1}{\ii\hbar}S\right)
\end{array}\right).
\end{aligned}
\end{align}
The first matrix coincides with \eqref{eq:stand_3} and \eqref{eq:stand_4}, and the second matrix is the inverse of the first, so this transition also correctly reduces to the standard WKB transition.

\subsubsection{Case 3: $\hbar\to\hbar+\ii\epsilon, V'(x_0)<0$}

Here we use the transitions \eqref{eq:mat_conn_5} and \eqref{eq:mat_conn_6}, with $x>x_0$ the allowed region and $x<x_0$ the forbidden region. The appropriate limit is
\begin{align}
&\begin{aligned}
&\frac1{\sqrt{S'}}\left(\begin{array}{c}
\exp\left(-\frac{1}{\ii\hbar}S\right)\\
\exp\left(\frac{1}{\ii\hbar}S\right)
\end{array}\right)
\to&\frac1{\sqrt{-T'}}\left(
\begin{array}{cc}
 \ee^{-\ii\frac{\pi}{4}}&0 \\
 \ee^{\ii\frac{\pi}{4}} & \ee^{-\ii\frac{\pi}{4}} \\
\end{array}
\right)
\left(\begin{array}{c}
\exp\left(\frac{1}{\hbar}T\right)\\
\exp\left(-\frac{1}{\hbar}T\right)
\end{array}\right),
\end{aligned}\\
&\begin{aligned}
&\frac1{\sqrt{-T'}}\left(\begin{array}{c}
\exp\left(\frac{1}{\hbar}T\right)\\
\exp\left(-\frac{1}{\hbar}T\right)
\end{array}\right)
\to&\frac1{\sqrt{S'}}\left(
\begin{array}{cc}
 \ee^{\ii\frac{\pi }{4}} & 0 \\
 \ee^{\ii\frac{\pi }{4}} & \ee^{\ii\frac{\pi }{4}} \\
\end{array}
\right)
\left(\begin{array}{c}
\exp\left(-\frac{1}{\ii\hbar}S\right)\\
\exp\left(\frac{1}{\ii\hbar}S\right)
\end{array}\right).
\end{aligned}
\end{align}
The first matrix gives transformations \eqref{eq:stand_5} and \eqref{eq:stand_6}, and the second matrix is the inverse of the first, so the standard WKB transition is correctly reproduced.

\subsubsection{Case 4: $\hbar\to\hbar+\ii\epsilon, V'(x_0)<0$}

The last transitions to use are \eqref{eq:mat_conn_7} and \eqref{eq:mat_conn_8}, with $x<x_0$ the forbidden region and $x>x_0$ the allowed region. We have
\begin{align}
&\begin{aligned}
&\frac1{\sqrt{S'}}\left(\begin{array}{c}
\exp\left(-\frac{1}{\ii\hbar}S\right)\\
\exp\left(\frac{1}{\ii\hbar}S\right)
\end{array}\right)
\to&\frac1{\sqrt{-T'}}\left(
\begin{array}{cc}
 \ee^{-\ii\frac{\pi}{4}}&\ee^{\ii\frac{\pi}{4}} \\
 0 & \ee^{-\ii\frac{\pi}{4}} \\
\end{array}
\right)
\left(\begin{array}{c}
\exp\left(\frac{1}{\hbar}T\right)\\
\exp\left(-\frac{1}{\hbar}T\right)
\end{array}\right),
\end{aligned}\\
&\begin{aligned}
&\frac1{\sqrt{-T'}}\left(\begin{array}{c}
\exp\left(\frac{1}{\hbar}T\right)\\
\exp\left(-\frac{1}{\hbar}T\right)
\end{array}\right)
\to&\frac1{\sqrt{S'}}\left(
\begin{array}{cc}
 \ee^{\ii\frac{\pi }{4}} & \ee^{-\ii\frac{\pi}4 }\\
 0 & \ee^{\ii\frac{\pi }{4}} \\
\end{array}
\right)
\left(\begin{array}{c}
\exp\left(-\frac{1}{\ii\hbar}S\right)\\
\exp\left(\frac{1}{\ii\hbar}S\right)
\end{array}\right).
\end{aligned}
\end{align}
As always, the first matrix coincides with \eqref{eq:stand_7} and \eqref{eq:stand_8} and the second is the inverse of the first. All limits of the connection formulae reproduce the right standard WKB connection formulae. 

%% file: 5applications.tex
\chapter{Applications}
\label{chapter5}

\section{Harmonic potential}

The first example that we examine is the harmonic potential, given by the finite difference equation
\begin{align}
\psi(x+\ii\hbar,\hbar)+\psi(x-\ii\hbar,\hbar)=2(E-x^2+1)\psi(x,\hbar).\label{eq:harm_pot}
\end{align}
An unique feature of this problem is that there is a way to turn this problem in a standard WKB problem via Fourier transform: for the Fourier transform $\tilde\psi(p,\hbar)$ related to $\psi(x,\hbar)$ as
\begin{align}
\psi(x,\hbar)=\int\frac{\dd p}{2\pi\ii}\ee^{\frac{\ii px}{\hbar}}\tilde\psi(p,\hbar),
\end{align}
the term $x^2\psi(x,\hbar)$ gets converted in a double derivative term, so if $\psi(x,\hbar)$ is a normalizable solution to \eqref{eq:harm_pot}, then $\tilde\psi(x,\hbar)$ is a normalizable solution to
\begin{align}
-\hbar^2\tilde\psi''(p,\hbar)=(E-\cosh p+1)\tilde\psi(p,\hbar),\label{eq:mathieu}
\end{align}
that is the modified Mathieu equation, studied as example in \cite{grassi2020non}. The two problems will have the same spectrum. We will examine the problem from the deformed WKB side.

\begin{figure}
\centering
\includegraphics[scale=1.3]{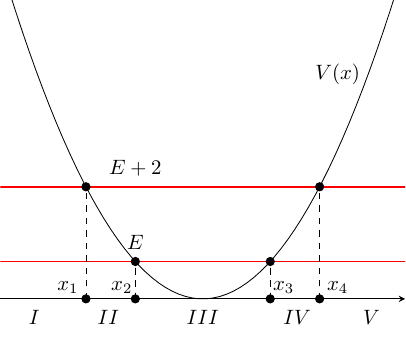}
\caption{Turning points for the harmonic potential, together with the regions. With reference to the terminology of subsection \ref{subsec:clas_an}, $III$ is the classically allowed region, $II$ and $IV$ are classically forbidden regions and $I$ and $V$ are imaginary allowed regions.}
\label{fig:harm_pot}
\end{figure}

\begin{figure}
\centering
\includegraphics[scale=1.3]{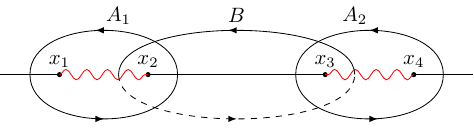}
\caption{Cycles for the harmonic potential. The red wiggly lines indicate square root branch cuts, and the $B$ cycle has a dashed part that lies on the second sheet of the potential.}
\label{fig:harm_cyc}
\end{figure}

In this problem, we will be able to set a method for getting the spectrum in a straightforward way. The first step is to identify the turning points of the problem. By setting $E-x^2+1=\pm1$, we see that the turning points at $1$ are
\begin{align}
x_2=-\sqrt E,\quad x_3=\sqrt E,
\end{align}
and the turning points at $-1$ are
\begin{align}
x_1=-\sqrt{E+2},\quad x_4=\sqrt{E+2}.
\end{align}
We are ordering the turning points from left to right. The denominators of the terms $p_n(x)$ with $n>0$ even will contain integer powers of $(E-x^2)^{\frac12}$ and $(E-x^2+2)^{\frac52}$, so we will set our branch cuts for the square roots to go from $x_1$ to $x_2$ and from $x_3$ to $x_4$. We have three important cycles in the problem: the cycle between $x_1$ and $x_2$ (denoted as $A_1$), the cycle between $x_2$ and $x_3$ (denoted as $B$) and the cycle between $x_3$ and $x_4$ (denoted as $A_2$). The turning points also divide the $x$ line in five different regions, delimited by the turning points themselves. We picture the arrangement of turning points in figure \ref{fig:harm_pot} and the cycles and branch cuts in figure \ref{fig:harm_cyc}.

In order to get the quantization conditions, we have to write a wavefunction ansatz and then impose normalizability in the regions $I$ and $V$, that extend up to $\pm\infty$. In each region $R$, the wavefunction ansatz can be specified by giving two periodic functions $A_R(x,\hbar)$ and $B_R(x,\hbar)$ as
\begin{align}
\psi(x,\hbar)=\exp\left(\frac{1}{\ii\hbar}S_{\text o}\right)\left(A_R\exp\left(-\frac{1}{\ii\hbar}S\right)+B_R\exp\left(\frac{1}{\ii\hbar}S\right)\right).\label{eq:wavefun}
\end{align}
With this convention, we can identify a wavefunction with a vector of periodic functions $(A_R,B_R)$ in each region.

The first step is to identify the asymptotic behaviour of the wavefunctions when in regions $I$ and $V$ (that we will denote as the \textit{asymptotic regions}). In order to have an $\mathbb L^2(\mathbb R)$ function, it is sufficient to look at the imaginary part of $S$, that will dictate the exponential growth or decay of the wavefunction. We have
\begin{align}
\Im S=\int_{x_R}^x\Im(\arccosh (E-V(t)+1))\dd t,
\end{align}
where $x_R$ stands for either $x_1$ in region $I$ and $x_5$ in region $V$.
As $I$ and $V$ are imaginary forbidden regions, $E-V(x)<-2$ there, so according to our conventions we must have $\Im(\arccosh (E-V(t)+1))=\pm\ii\pi$, where also in this case the sign depends on the deformation of $\hbar$: we choose the upper sign for $\hbar\to\hbar-\ii\epsilon$ and the lower sign for $\hbar\to\hbar+\ii\epsilon$. We conclude
\begin{align}
\Im S=\pm\ii\pi(x-x_R),
\end{align}
so the solution $\exp\left(\mp\frac{1}{\ii\hbar}\right)$ is exponentially decaying in region $V$ (as $x>x_5$ there) and conversely $\exp\left(\pm\frac1{\ii\hbar}S\right)$ is exponentially decaying in region $I$. We note already a first contrast with standard WKB, where the asymptotic behaviour is richer, given by
\begin{align}
\int_{x_0}^x\Im\sqrt{2(E-V(t))}\dd t,
\end{align}
so the behaviour at infinity can be arbitrary: in deformed WKB we only have exponential decays and growths.

The coefficients $A_R$ and $B_R$ also influence the asymptotic behaviour. As they are periodic functions, their $x$ dependency can be expanded in a Fourier series, and it will be convenient to normalize the coefficients $a_n(\hbar)$ and $b_n(\hbar)$ of the series to have the expansions
\begin{align}
A_R(x,\hbar)=\sum_{n=-\infty^\infty}a_n(\hbar)q_{x_R}^n,\quad B_R(x,\hbar)=\sum_{n=-\infty^\infty}b_n(\hbar)q_{x_R}^n.
\end{align}
For $\mathbb L^2(\mathbb R)$ wavefunctions, there are constraints to impose, as the $q_R$ factors are exponentially decaying.

We now have to deal with how to connect the coefficients $A_R$ and $B_R$ from different regions. This we can do using the connection formulae of chapter \ref{chapter4}, especially in matrix form: as an example, we take the connection from region $I$ to region $II$ with $\hbar$ modified to $\hbar-\ii\epsilon$. The two regions are separated by the turning point $x_1$, that is a turning point at $-1$ with $V'(x_1)<1$. The appropriate matrix for the transition is $\eqref{eq:mat_conn_7}$ with $q_{x_1}$ instead of $q_{x_0}$, multiplied to the left and right by matrices \eqref{eq:invmat}. We need to normalize the actions at the turning point $x_1$, and we will denote this as $S^{(x_1)}$ ($S_{\text o}$ does not need normalization as it depends on derivatives of $S$). We can write \eqref{eq:wavefun} as a row-column product as
\begin{align}
\psi(x,\hbar)=\exp\left(\frac{1}{\ii\hbar}S_{\text o}\right)\left(\begin{array}{cc}
A_R&B_R
\end{array}\right)
\left(\begin{array}{c}
\exp\left(-\frac{1}{\ii\hbar}S^{(x_1)}\right)\\
\exp\left(\frac{1}{\ii\hbar}S^{(x_1)}\right)
\end{array}\right).
\end{align}
Using the transition matrices, we have
\begin{align}
\left(\begin{array}{c}A_{II}\\B_{II}\end{array}\right)=
I^{(+)}(q_{x_1})^{-1}\left(
\begin{array}{cc}
 \ee^{\ii\frac{ \pi }{4}}\frac{1}{1-q_{x_1}^{-1}} & \ee^{-\ii\frac{\pi }{4}} \\
 -\ee^{\ii\frac{\pi }{4}}\frac{q_{x_1}^{-1}}{q_{x_1}^{-1}-1} & \ee^{\ii\frac{\pi }{4}} \\
\end{array}\right)I^{(+)}(q_{x_1})\left(\begin{array}{c}A_{I}\\B_{I}\end{array}\right).\label{eq:trans_ex}
\end{align}
Lastly, when crossing a turning point, we obtain wavefunctions normalized at the turning point itself, in terms of $S^{(x_1)}$\footnote{We will not use the $T$ functions of chapter \ref{chapter4} anymore, substituting them with $\ii S$.}. In order to cross another turning point, we first have to normalize the action at that turning point. This can be done using the quantum periods, as
\begin{align}
\left(\begin{array}{c}
\exp\left(-\frac{1}{\ii\hbar}S^{(x_2)}\right)\\
\exp\left(\frac{1}{\ii\hbar}S^{(x_2)}\right)
\end{array}\right)=
\left(
\begin{array}{cc}
\exp\left(-\frac{1}{\ii\hbar}\Pi^{(1,2)}\right)&0\\
0&\exp\left(\frac{1}{\ii\hbar}\Pi^{(1,2)}\right)
\end{array}
\right)
\left(\begin{array}{c}
\exp\left(-\frac{1}{\ii\hbar}S^{(x_1)}\right)\\
\exp\left(\frac{1}{\ii\hbar}S^{(x_1)}\right)
\end{array}\right),
\end{align}
where we recall the definition of quantum periods as
\begin{align}
\Pi^{(i,j)}(\hbar)=\int_{x_i}^{x_j}P_0(t)\dd t+\frac12\sum_{n=1}^\infty\oint_{\gamma_{i,j}}\frac{P_{2n}(t)}{(2n)!}(\ii\hbar)^{2n}\dd t,
\end{align}
and $\gamma_{i,j}$ can be identified as one of the cycles of figure \ref{fig:harm_cyc}. The quantization conditions will impose conditions on the coefficients $A_V$ and $B_V$ computed in terms of $A_I$ and $B_I$, and as the transition matrix obtained by following this procedure only contains the cycles as energy dependent factors, quantization conditions will be expressed as equations to be satisfied by the cycles.

\subsection{Quantization conditions}

We now come to the quantization conditions for the harmonic potential. We have to take into account the two different deformations of $\hbar$.

For this section, we will change our notation. It will be convenient to split $q_{x_R}$ in two terms, one dependent only on $x$ and the other depending only on the turning point. We will define
\begin{align}
q=\exp\left(-\frac{2\pi}{\hbar}x\right),\quad q_{R}=\exp\left(-\frac{2\pi}{\hbar}x_R\right).
\end{align}
With this definition, the old $q_{x_R}$ gets rewritten as $q/q_{R}$.

\subsubsection{Case 1: $\hbar\to\hbar-\ii\epsilon$}

We start from region $I$, where $\exp\left(\frac1{\ii\hbar}S\right)$ is exponentially decaying. This does not mean that $\exp\left(\frac1{\ii\hbar}S\right)$ is the only wavefunction in region $I$, as we can use factors of $q$ to obtain other exponentially decaying wavefunctions involving the other exponential too. In each region, we can use a set of two coordinates to write the wavefunction. We will start in region $I$ by indicating the coordinates as $C_I^{(x_1)}=(A_I(q),B_I(q))$, where the upper index indicates that those are coordinates for expanding according to the basis $\left(\exp\left(-\frac{1}{\ii\hbar}S^{(x_1)}\right),\exp\left(\frac{1}{\ii\hbar}S^{(x_1)}\right)\right)$. $A_I(q)$ must be a series in $q$ with $a_0=0$, but $B_I(q)$ can also have a $b_0$ term (both series cannot have terms $a_n$ or $b_n$ with $n<0$, as they would not be normalizable).

We start with the following coordinate vector:
\begin{align}
C_I^{(x_1)}=\left(\begin{array}{c}
\frac{a_0}q\\
b_0+\frac{b_1}q
\end{array}\right).\label{eq:harm_start}
\end{align}
This particular vector indicates a normalizable wavefunction in region $I$. In region $II$, we apply \eqref{eq:trans_ex} to obtain
\begin{align}
C_{II}^{(x_1)}=\left(\begin{array}{c}
\ee^{\ii\frac\pi4}\frac{-\ii a_0q+(b_1+b_0q)q_{1}}{q^2}\\
\ee^{-\ii\frac\pi4}\frac{b_1+b_0q-\ii a_0}{q-q_{1}}
\end{array}\right).
\end{align}
Before passing through the point at $x_2$, we must write coordinates for expanding according to the basis $\left(\exp\left(-\frac{1}{\ii\hbar}S^{(x_2)}\right),\exp\left(\frac{1}{\ii\hbar}S^{(x_2)}\right)\right)$. This is done with the quantum period $\Pi^{(1,2)}$:
\begin{align}
C_{II}^{(x_2)}=\left(\begin{array}{c}
\exp\left(-\frac{1}{\ii\hbar}\Pi^{(1,2)}\right)\ee^{\ii\frac\pi4}\frac{-\ii a_0q+(b_1+b_0q)q_{1}}{q^2}\\
\exp\left(\frac{1}{\ii\hbar}\Pi^{(1,2)}\right)\ee^{-\ii\frac\pi4}\frac{b_1+b_0q-\ii a_0}{q-q_{1}}
\end{array}\right).
\end{align}
We can now cross $x_2$. Proceeding in the same way, using quantum periods to change basis and transition formulas to cross the turning points, we arrive in region $V$ with a complicated expression.

This expression simplifies when we impose normalizability. The coefficients in region $V$ must be $(A_V(q),B_V(q))$ with $A_V$ being a series in $q$ (only positive powers, due to the fact that they decay exponentially in region $V$) with possibly a constant term, while $B_V$ must be a series in $q$ (again only positive powers) with no constant term (as $\exp\left(\frac{1}{\ii\hbar}S^{(x_5)}\right)$ is not normalizable). The transitions generate coefficients of the form
\begin{align}
C_V^{(x_4)}=\left(\begin{array}{c}
\frac1{q^3}\left(-\frac{\ii b_1q_{1}q_{4}\exp\left(-\frac{1}{\ii\hbar}\Pi^{(1,2)}\right)}{\exp\left(-\frac{1}{\ii\hbar}\Pi^{(2,4)}\right)}+o(q)\right),\\
\frac{1}{q^3}(o(q))
\end{array}\right).
\end{align}
The coefficient of $q^3$ must be zero: the only possibility is $b_1=0$. We then set $b_1=0$ and proceed to the next term. The coefficient now reads
\begin{align}
C_V^{(x_4)}=\left(\begin{array}{c}
\frac1{q^2}\left(-\frac{(a_0+\ii b_0q_{1})q_{4}\exp\left(-\frac{1}{\ii\hbar}\Pi^{(1,2)}\right)}{\exp\left(-\frac{1}{\ii\hbar}\Pi^{(2,4)}\right)}+o(q)\right),\\
\frac{1}{q^2}(o(q))
\end{array}\right).
\end{align}
The coefficient of $q^{-2}$ must also vanish, so we must impose $a_0=-\ii b_0 q_{x_1}$. As we are left only with one coefficient, we can set $b_0=1$, so the starting coefficient is
\begin{align}
C_I^{(x_1)}=\left(\begin{array}{c}
-\frac{\ii q_{1}}q\\
1
\end{array}\right),
\end{align}
and after the transitions we are left with
\begin{align}
C_V^{(x_4)}=\left(\begin{array}{c}
\frac1{q}\left(\frac{-\ii q_{4}\left(1+\exp\left(-\frac{2}{\ii\hbar}\Pi^{(2,3)}\right)\right)}{\exp\left(-\frac{1}{\ii\hbar}\Pi^{(1,4)}\right)}+o(q)\right),\\
o(q)
\end{array}\right).
\end{align}
There is only one coefficient left that we have to put to zero, as all other possible coefficients are allowed. We have to impose a condition on the cycle $\Pi^{(2,3)}$, the $B-$cycle: we must have
\begin{align}
\left(\exp\left(\frac{1}{\ii\hbar}\Pi^{(2,3)}\right)+\exp\left(-\frac{1}{\ii\hbar}\Pi^{(2,3)}\right)\right)=0\implies\cos\left(\frac{1}{\hbar}\Pi^{(2,3)}\right)=0.
\end{align}
The quantization condition only involves the $B-$cycle $\Pi^{(2,3)}$, even if we have other $A-$cycles in the problem. This result is the same as the one for the harmonic oscillator, and coincides with \cite{pasquier1992periodic} (where the result was obtained without considering perturbative corrections to the cycle, approximating the quantum momentum with the classical momentum) and with \cite{grassi2019solvable}, where the result was derived to all orders with a different method. The quantization condition can be expressed as
\begin{align}
\exp\left(-\frac{1}{\ii\hbar}\Pi^{(2,3)}\right)=\pm\ii,
\label{eq:parity}
\end{align}
and the $\pm$ sign tracks the parity of states. To see that, we look at the coefficients in the various regions. In the asymptotic regions, we have (again, the $\pm$ sign depends on the $\pm$ choice in \eqref{eq:parity})
\begin{align}
C_I^{(x_1)}=\left(\begin{array}{c}
-\frac{\ii q_1}q\\
1
\end{array}\right),\quad
C_V^{(x_4)}=\left(\begin{array}{c}
\pm\frac{\exp\left(-\frac{1}{\ii\hbar}\Pi^{(3,4)}\right)}{\exp\left(-\frac{1}{\ii\hbar}\Pi^{(1,2)}\right)}\\
\mp\frac{\ii q}{q_4}\frac{\exp\left(-\frac{1}{\ii\hbar}\Pi^{(3,4)}\right)}{\exp\left(-\frac{1}{\ii\hbar}\Pi^{(1,2)}\right)}
\end{array}\right).
\end{align}
For the even potential, $\pm\frac{\exp\left(-\frac{1}{\ii\hbar}\Pi^{(3,4)}\right)}{\exp\left(-\frac{1}{\ii\hbar}\Pi^{(1,2)}\right)}=1$ and $q_4=q_1^{-1}$. Furthermore, $S^{(x_1)}(-x,\hbar)=-S^{(x_5)}(x,\hbar)$, meaning that the parity operation inverts the coefficients of the coordinates, and $q(-x,\hbar)=q(x,\hbar)^{-1}$. Applying parity to the coefficient $C_V^{(x_4)}$ then gives
\begin{align}
C_V^{(x_4)}\to\pm\left(\begin{array}{c}
-\frac{\ii q_1}q\\
1
\end{array}\right),
\end{align}
as would be expected from an even or odd wavefunction. It is then trivial to generalize this line of reasoning to the other regions.

In \eqref{eq:harm_start}, we have made a precise choice of not allowing coefficients of order $q^{-2}$ or more. The reason behind this is simple, as allowing them would generate coefficients of order $q^{-4}$ or more in $C_V^{(x_4)}$, and the only way to remove those exponential growths in the asymptotic region is to remove coefficients of order $q^{-2}$ or more in $C_I^{(x_1)}$. Thus \eqref{eq:parity} is the complete quantization condition for the harmonic oscillator.

\subsubsection{Case 2: $\hbar\to\hbar+\ii\epsilon$}

In this case, the role of $\exp\left(-\frac{1}{\ii\hbar}S\right)$ and $\exp\left(\frac1{\ii\hbar}S\right)$ are reversed: in region $I$, the first one asymptotically decays and the second one asymptotically grows, while in region $V$ the opposite is true. As before, we start with the coordinate vector
\begin{align}
C_I^{(x_1)}=\left(\begin{array}{c}
b_0+\frac{b_1}q\\
\frac{a_0}q
\end{array}\right).
\end{align}
Performing the exact same analysis as with the other deformation, we see that $b_1=0$ and $a_0=-\ii b_0 q_{1}$. The starting vector is then (with the normalization $b_1=1$)
\begin{align}
C_I^{(x_1)}=\left(\begin{array}{c}
-\ii\frac{q_{1}}q\\
1
\end{array}\right).
\end{align}
After performing all the transitions, in region $V$ we are left with
\begin{align}
C_V^{(x_4)}=\left(\begin{array}{c}
\mp\frac{\ii q}{q_4}\frac{\exp\left(-\frac{1}{\ii\hbar}\Pi^{(3,4)}\right)}{\exp\left(-\frac{1}{\ii\hbar}\Pi^{(1,2)}\right)}\\
\pm\frac{\exp\left(-\frac{1}{\ii\hbar}\Pi^{(3,4)}\right)}{\exp\left(-\frac{1}{\ii\hbar}\Pi^{(1,2)}\right)}
\end{array}\right).
\end{align}
Again, normalizability of the overall wavefunction requires
\begin{align}
\exp\left(-\frac{1}{\ii\hbar}\Pi^{(2,3)}\right)=\pm\ii,
\end{align}
that is the same condition as the other deformation, \eqref{eq:parity}.

\subsection{Computation of the quantum $B$ period}

In the case of the harmonic oscillator, we can give a procedure for computing arbitrary coefficients in the quantum period. We remind that the quantum periods can be expanded as
\begin{align}
\Pi^{(i,j)}\simeq\sum_{n=0}^\infty\frac{\Pi_{2n}^{(i,j)}}{(2n)!}(\ii\hbar)^{2n},\label{eq:cyc_exp}
\end{align}
where the coefficients of the series expansion are given by
\begin{align}
\Pi_{0}^{(i,j)}=\int_{x_i}^{x_j}\arccosh(E-t^2+1)\dd t,\quad \Pi_{n}^{(i,j)}=\frac12\oint_{\gamma_{i,j}}P_{2n}(t)\dd t,
\end{align}
as expressed in \eqref{eq:quant_per_gen}. For now, we will focus on the quantum $B-$period $\Pi^{(2,3)}$, that is the important period in the quantization condition. In this case, $x_i=-\sqrt{E}$ and $x_j=\sqrt{E}$.

The. $\hbar^0$ contribution is obtained by evaluating the integral
\begin{align}
\Pi_0^{(2,3)}\int_{-\sqrt{E}}^{\sqrt{E}}\arccosh(E-t^2+1)\dd t=4\sqrt{E+2}\left(\mathbf K\left(\frac{E}{E+2}\right)-\mathbf E\left(\frac{E}{E+2}\right)\right).
\end{align}
Here $\mathbf E$ and $\mathbf K$ are the elliptic functions defined as
\begin{align}
\mathbf K(x)=\int_0^{\frac\pi2}\frac{1}{\sqrt{1-x(\sin\theta)^2}}\dd\theta,\quad \mathbf E(x)=\int_0^{\frac\pi2}\sqrt{1-x(\sin\theta)^2}\dd\theta.
\end{align}
As in \cite{grassi2020non}, we can compute higher-order corrections via the Picard-Fuchs approach (see \cite{kreshchuk2019picard} for a review of its application in ordinary WKB). Up to order $\hbar^{22}$, we have found linear operators $\mathcal O_n(E)$ such as
\begin{align}
\mathcal O_n(E)p_{2n}(x)-p_0(x)=\left(a_n(E)\frac{\partial}{\partial E}+b_n(E)\frac{\partial^2}{\partial^2E}\right)p_{2n}(x)-p_0(x)
\end{align}
is a total derivative. Examples of such operators are
\begin{align}
&\mathcal O_1(E)=\frac1{12}\frac{\partial}{\partial E}+\frac{E+1}{6}\frac{\partial^2}{\partial^2E},\\
&\mathcal O_2(E)=\frac{(1+E)(14-(E+2)E)}{120E^2(E+2)^2}\frac{\partial}{\partial E}+\frac{224-E(E+2)(4(E+2)E-145}{240E^2(E+2)^2}\frac{\partial^2}{\partial^2E}.
\end{align}
The higher order corrections are then related to the $\hbar^0$ corrections as
\begin{align}
\Pi^{(i,j)}_{2n}=\mathcal O_n(E)\Pi^{(i,j)}_0.
\end{align}
We have computed quantum corrections up to $\Pi^{(2,3)}_{18}$, and then compared the results with the ones obtained in \cite{grassi2020non}. It turns out that our periods are equal to the periods obtained for the $B-$cycle in the modified Mathieu potential, as to be expected due to the fact that the two problems are related via Fourier transform. We have used the data from \cite{grassi2020non}\footnote{The author wishes to thank Professor Jie Gu for providing the code for computing the coefficients.} in order to compute quantum corrections up to $\Pi^{(2,3)}_{190}$.

\begin{figure}
\centering
\includegraphics[width=0.75\textwidth]{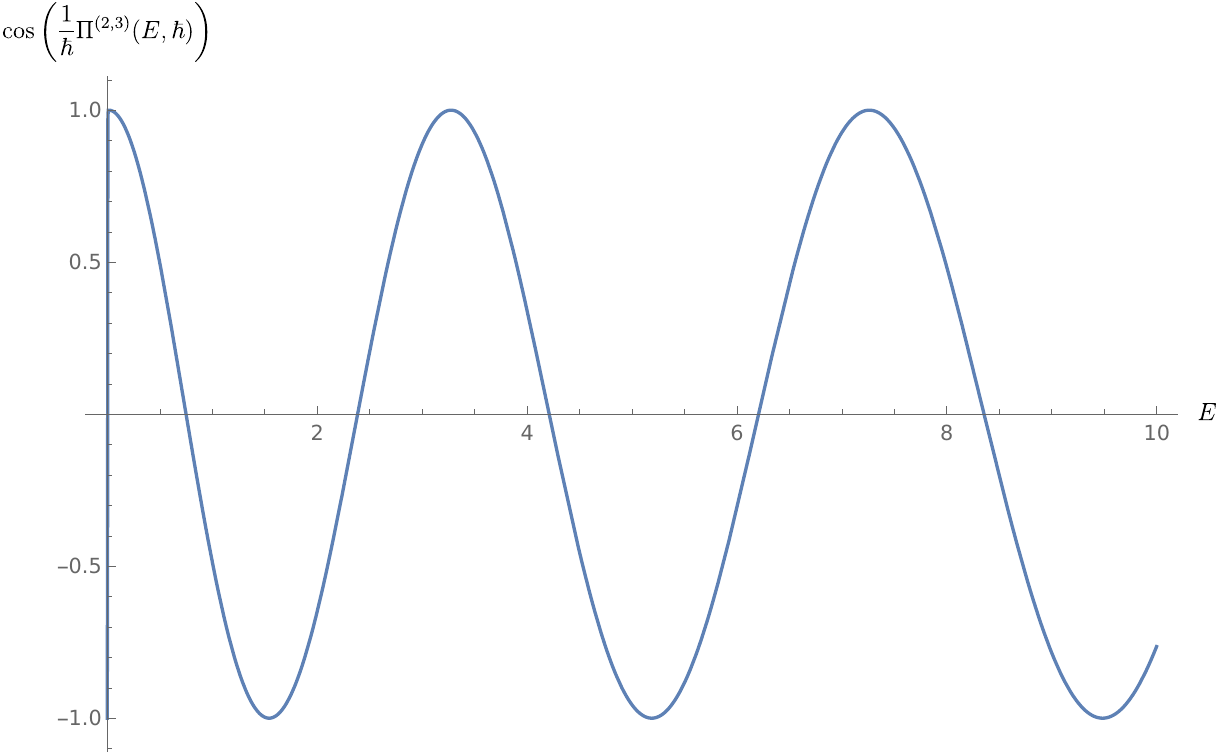}
\caption{Quantization condition for the harmonic potential, $\hbar=1$. We have included terms up to $\hbar^6$ in the computation of the loop. Zeroes of the plotted function correspond to energy levels.}
\label{fig:quant_cond_harm}
\end{figure}

\begin{figure}
\begin{center}
\begin{minipage}{0.49\textwidth}
\includegraphics[width=\columnwidth]{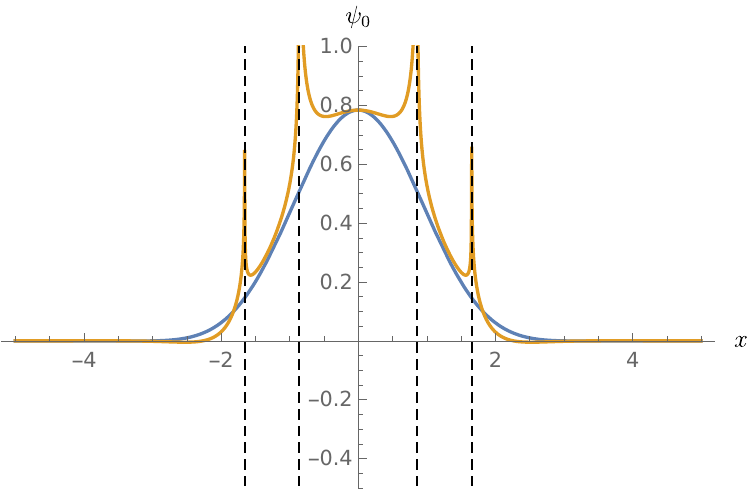}
\includegraphics[width=\columnwidth]{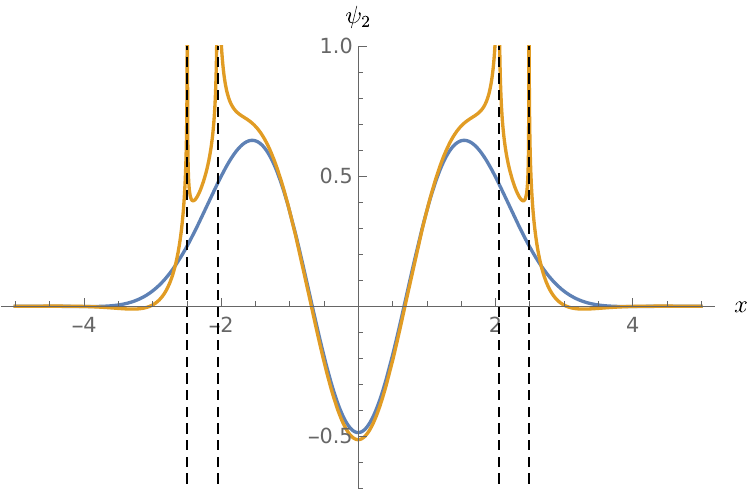}
\end{minipage}
\hfill
\begin{minipage}{0.49\textwidth}
\includegraphics[width=\columnwidth]{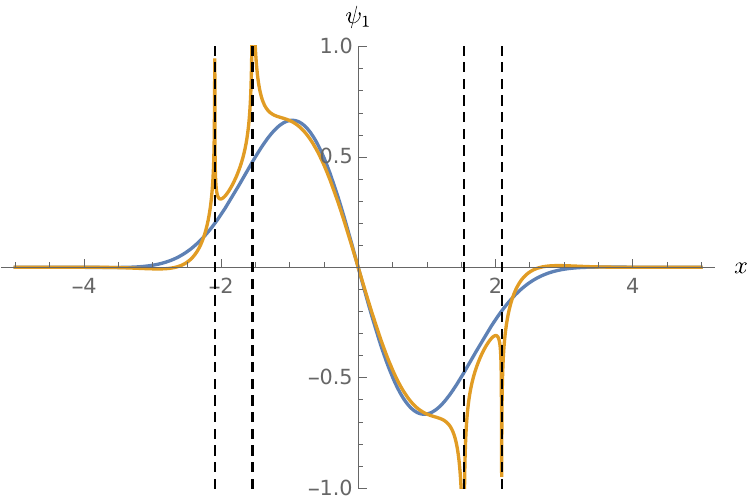}
\includegraphics[width=\columnwidth]{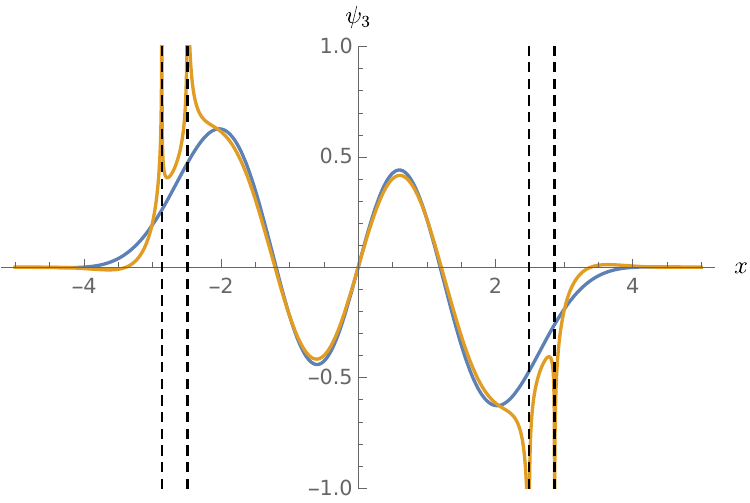}
\end{minipage}
\end{center}
\caption{Wavefunctions $\psi_n$ for the first 4 energy levels, obtained with only the leading correction. In blue we plot the numerical wavefunctions, while in orange we plot the WKB wavefunctions. The vertical dotted lines correspond to $\pm\sqrt{E}$ and $\pm\sqrt{E+2}$, turning points at which the WKB approximation diverges.}
\label{fig:energy_plots}
\end{figure}

\begin{figure}
\begin{minipage}{0.49\textwidth}
\includegraphics[width=\columnwidth]{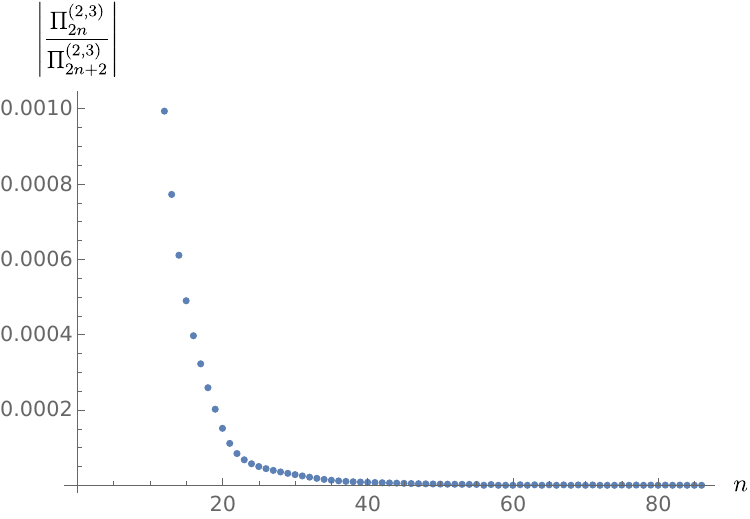}
\end{minipage}
\hfill
\begin{minipage}{0.49\textwidth}
\includegraphics[width=\columnwidth]{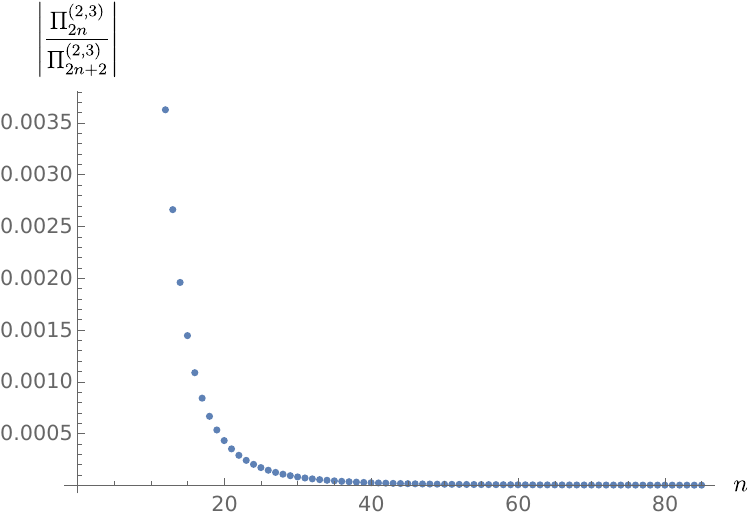}
\end{minipage}
\caption{Estimation of the radius of convergence of the series for $\Pi^{(2,3)}(E,\hbar)$, at $E=1$ (left) and $E=5$ (right). The series has zero radius of convergence.}
\label{fig:as_behaviour}
\end{figure}

The first test is to obtain the energy levels. In figure \ref{fig:quant_cond_harm}, we plot the quantization condition: zeroes of the plotted function indicate energy levels. In table \ref{tab:quant_cond_harm}, we compute zeroes of the quantization condition and compare them with numerical results. In figure \ref{fig:energy_plots}, we plot WKB wavefunctions and compare them with numerical wavefunctions. In addition, we plot the high order behaviour of the coefficients of $\Pi^{(2,3)}_{2n}$ in figure \ref{fig:as_behaviour}, showing that the series has zero radius of convergence.
\begin{table}
\centering
\begin{tabular}{c|c|c|c|c}
&$n=0$&$n=1$&$n=2$&$n=3$\\\hline
$E_n^{(0)}$&$0.7371092220$&$2.374009918$&$4.200969511$&$6.196727659$\\\hline
$E_n^{(2)}$&$0.7651458653$&$2.398079955$&$4.222746426$&$6.216966661$\\\hline
$E_n^{(4)}$&$0.7651570613$&$2.398081389$&$4.222746853$&$6.216966855$\\\hline
$E_n^{(6)}$&$0.7651572398$&$2.398081395$&$4.222746854$&$6.216966856$\\\hline
Numerical&$0.7651572553$&$2.398081395$&$4.222746854$&$6.216966856$
\end{tabular}
\caption{First three energy levels for the harmonic potential with $\hbar=1$. Here $E_n^{(m)}$ is the $n-$th energy level computed with corrections up to $\hbar^m$.}
\label{tab:quant_cond_harm}
\end{table}

\subsection{Computation of the quantum $A_1$ and $A_2$ periods}

We now turn to the computation of the other periods in the harmonic oscillator, the $A_1$ and $A_2$ periods. While the computation of the periods is elementary, comparison with \cite{grassi2020non} will introduce a subtlety.

We start by computing the zero order correction as
\begin{align}
\Pi^{(1,2)}_0=\Pi^{(3,4)}_0=\ii\pi\sqrt{E+2}-2\ii\sqrt{E}\mathbf E\left(-\frac2E\right).
\end{align}
If we take the linear combination of the periods given by
\begin{align}
-\Pi_0^{(1,2)}-\Pi_0^{(3,4)}=4\ii\sqrt{E}\mathbf E\left(-\frac2E\right)-2\ii\pi\sqrt{E+2}
\end{align}
and compare our result with the $A$ period in \cite{grassi2020non}, we notice that there is a difference in our results. By noting the $A$ period of \cite{grassi2020non} as $\tilde\Pi_{\tilde A}^{(0)}$, we have the equality
\begin{align}
-\Pi_0^{(1,2)}-\Pi_0^{(3,4)}=2\ii\pi\sqrt{E+2}+\tilde\Pi_{\tilde A}^{(0)}.
\end{align}
The factor of difference comes from the presence of the turning points at $-1$. In particular, we can observe the following relation between the associated Voros symbols, where the $0$ superscript indicates that the cycles are computed with only the $\Pi_0^{(i,j)}$ contributions to the cycles:
\begin{align}
V^{(0)}_{(1,2)}V^{(0)}_{(3,4)}q_1^{-\frac12}q_4^{\frac12}=\exp\left(\frac{1}{\ii\hbar}\tilde\Pi^{(\tilde A)}_0\right).
\label{eq:voros}
\end{align}
The difference can be explained in the following way. $\tilde\Pi_{\tilde A}^{(0)}$ is a \textit{cycle} integral, while $\Pi_0^{(1,2)}$ and $\Pi_0^{(3,4)}$ are \textit{line} integrals. For a cycle integral around a square root branch cut, if the line integral is convergent then it coincides with half of the cycle integral (as is the case for standard WKB, motivating the choice of regularization for the higher orders of the quantum periods, where only square root branch cuts appear). In our case, we have to consider the additional branch cuts given by the $\arccosh$ function. The additional factors of $q_i$ are what is needed to turn the line integrals into loop integrals, effectively canceling the $\arccosh$ branch cut.

\begin{figure}
\begin{minipage}{0.49\textwidth}
\includegraphics[width=\columnwidth]{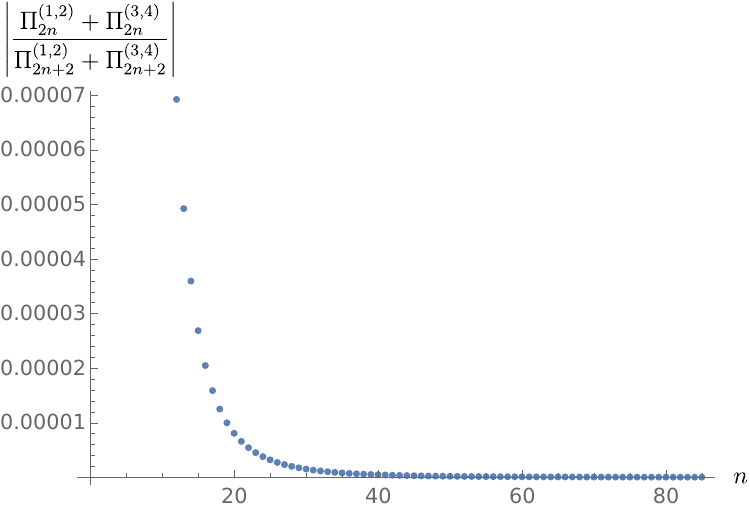}
\end{minipage}
\hfill
\begin{minipage}{0.49\textwidth}
\includegraphics[width=\columnwidth]{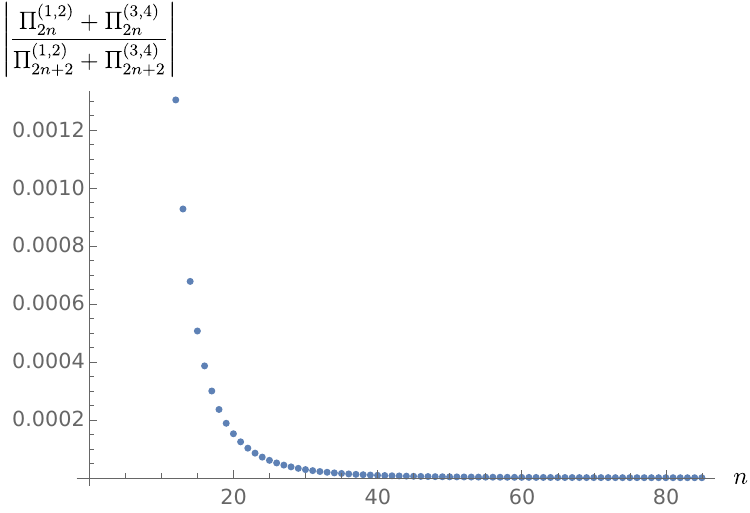}
\end{minipage}
\caption{Estimation of the radius of convergence of the series for $\Pi^{(1,2)}(E,\hbar)+\Pi^{(3,4)}(E,\hbar)$, at $E=1$ (left) and $E=5$ (right). The series has zero radius of convergence.}
\label{fig:aconv}
\end{figure}

There is another reason why the LHS of \eqref{eq:voros} is a better quantity to define the cycle integrals. We can compute the higher order corrections through the quantum operator method, as for the $B-$cycles. We can verify those results via numeric integration, integrating the quantum momenta around the $A_1$ and $A_2$ cycles. We get the correct higher order corrections as
\begin{align}
\Pi_{2n}^{(1,2)}+\Pi_{2n}^{(3,4)}=\mathcal O_n(\Pi_0^{(1,2)}+\Pi_0^{(3,4)}-2\ii\pi\sqrt{E+2}).
\end{align}
This implies that relation \eqref{eq:voros} upgrades to all orders as
\begin{align}
V_{(1,2)}V_{(3,4)}q_1^{-\frac12}q_4^{\frac12}=\exp\left(\frac{1}{\ii\hbar}\tilde\Pi^{(\tilde A)}\right),
\label{eq:vorosfull}
\end{align}
and only the zero order is affected by this difference. As before, the radius of convergence of the series for $\Pi^{(1,2)}(E,\hbar)+\Pi^{(2,3)}(E,\hbar)$ is zero. We estimate the radius of convergence in figure \ref{fig:aconv}.

For the analysis of the resurgent properties of next section, it will be convenient to make some definitions. We define
\begin{align}
&\Pi_B(E,\hbar)=-2\Pi^{(2,3)}(E,\hbar), \\
&\Pi_A(E,\hbar)=-2(\Pi^{(1,2)}(E,\hbar)+\Pi^{(3,4)}(E,\hbar)+2\ii\pi\sqrt{E+2}).
\end{align}
The extra $-2$ factors are used to transform line integrals into cycle integrals. As usual, we will denote as $\Pi_{A,B}^{(0)}$ the terms $\Pi_{A,B}(E,0)$, the order $\hbar^0$ terms of the series.

\subsection{Resurgent properties}
\label{subsec:res}
Due to the fact that the periods coincide with the periods studied in \cite{grassi2020non}, the resurgent properties can be read from the source\footnote{The case that we study here is $E>0$, that is denoted as ``Weak coupling region" in the cited source. A study for the case $-2<E<0$ is also present there (denoted as ``Strongly coupled region"), but we did not analyse it here due to the fact that it is outside of the scope of this thesis.}. Here we list the resurgent properties of the cycles of the harmonic oscillator in deformed quantum mechanics.

\begin{figure}
\centering
\includegraphics[scale=0.7]{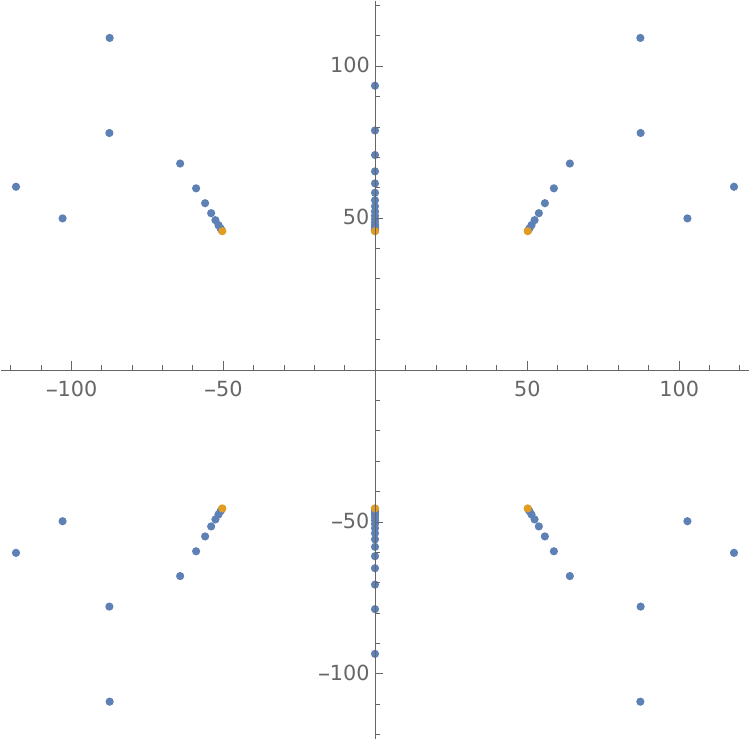}
\caption{Poles for the Padé approximant for the period $\Pi_A$ at $E=15$, with corrections up to $\hbar^{190}$. The poles accumulate on branch points starting at $\ii(\pm\Pi_B^{(0)}+n\Pi_A^{(0)})$ (with $n$ arbitrary integer), as expected from the general theory.}
\label{fig:apoles}
\end{figure}

\begin{figure}
\centering
\includegraphics[scale=0.5]{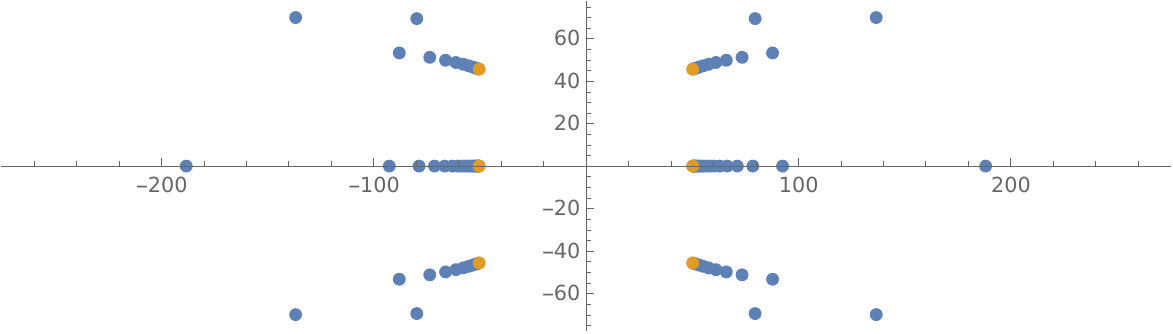}
\caption{Poles for the Padé approximant for the period $\Pi_B$ at $E=15$, with corrections up to $\hbar^{190}$. The poles accumulate on branch points starting at $\ii(\pm\Pi_A^{(0)}+\Pi_B^{(0)})$ (with $n$ arbitrary integer), as expected from the general theory.}
\label{fig:bpoles}
\end{figure}

We will employ the familiar techniques for numerical analysis of resurgent properties, explained in detail in \cite{aniceto2019primer} and in Appendix \ref{app:num} of this thesis. The first step for a numerical analysis is to compute the poles of the Borel-Padé approximation for the two cycles. We plot those in figures \ref{fig:apoles} and \ref{fig:bpoles}.

\begin{figure}[h!]
\centering
\includegraphics[width=\textwidth]{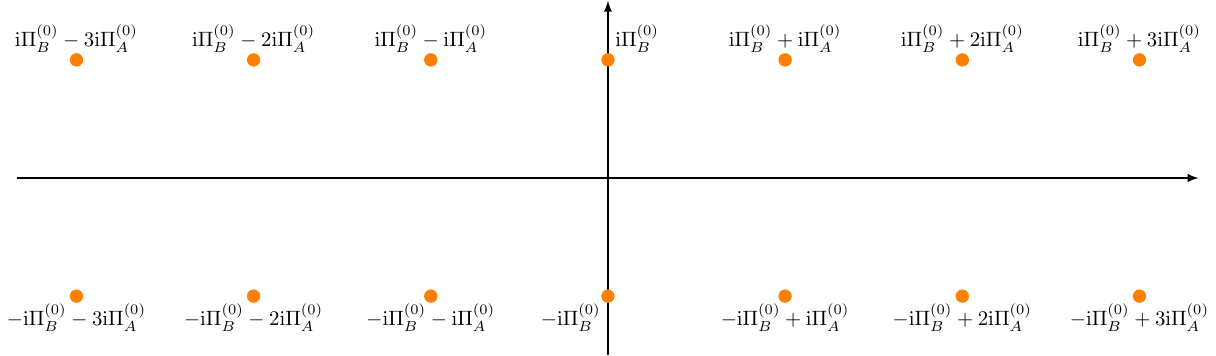}
\vspace{0.5cm}
\\
\includegraphics[width=\textwidth]{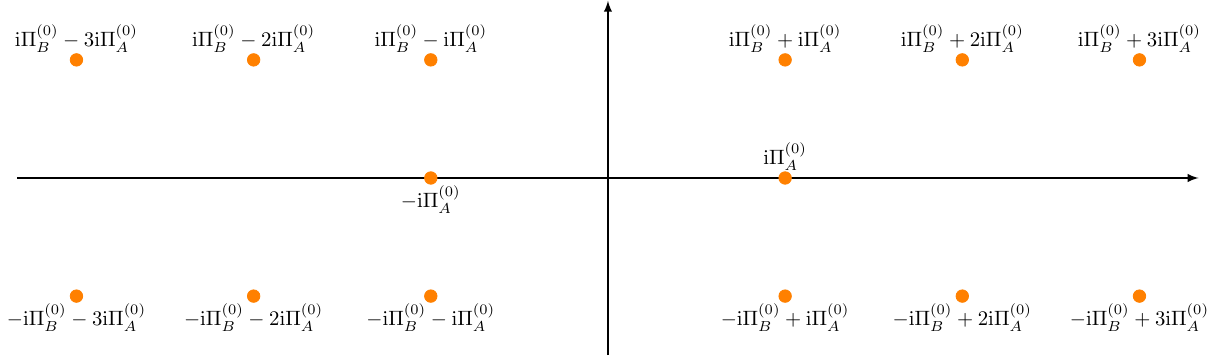}
\caption{Schematic representation of the branch points for the Padé approximations of $\Pi_A$ (above) and $\Pi_B$ (below).}
\label{fig:schemsings}
\end{figure}

The patterns of singularities in \ref{fig:apoles} and \ref{fig:bpoles} can have unexpected generalizations. First, let $\Pi_n=n\Pi_A+\Pi_B$. From \cite{grassi2020non}, we see that we should expect branch points of the Padé approximation to $\Pi_A$ at the points $\ii\Pi_n^{(0)}$. On the other hand, singularities of $\Pi_B$ are expected at points $\pm\ii\Pi_A^{(0)}$ and $\pm\ii\Pi_n^{(0)}$, with $n\neq0$. We plot those singularities schematically in \ref{fig:schemsings}.
From \cite{grassi2020non}, we have the following discontinuities: letting $\alpha_{\pm}=\arg(\ii\Pi^{(0)}_B)$ and $\theta_{n,\pm}=\arg(\ii\Pi_n^{(0)})$, we have for the $A$ cycle
\begin{align}
\left(\mathcal S_{\theta_{0,\pm}^+}-S_{\theta_{0,\pm}^-}\right)\Pi_A=2\hbar\log\left(1+\exp\left(\frac{1}{\ii\hbar}\mathcal S_{\theta_{0,\pm}}(\Pi_B)\right)\right)
\label{eq:test1}
\end{align}
and, for $n\neq0$
\begin{align}
\left(\mathcal S_{\theta_{n,\pm}^+}-S_{\theta_{n,\pm}^-}\right)\Pi_A=-2\hbar\log\left(1+\exp\left(\frac{1}{\ii\hbar}\mathcal S_{\theta_{n,\pm}}(\Pi_n)\right)\right).
\end{align}
Regarding the $B$ cycle, we have
\begin{align}
\left(\mathcal S_{\alpha_{\pm}^+}-S_{\theta_{0,\pm}^-}\right)\Pi_B=4\hbar\log\left(1+\exp\left(\frac{1}{\ii\hbar}\mathcal S_{\alpha_{\pm}}(\Pi_A)\right)\right)
\label{eq:test2}
\end{align}
and, for $n\neq0$
\begin{align}
\left(\mathcal S_{\theta_{n,\pm}^+}-S_{\theta_{n,\pm}^-}\right)\Pi_B=-2n\hbar\log\left(1+\exp\left(\frac{1}{\ii\hbar}\mathcal S_{\theta_{n,\pm}}(\Pi_n)\right)\right).
\end{align}
The resurgent structure is rich, but very hard to probe numerically. With our limited number of coefficients (up to $\hbar^{190}$), we managed to test only formulae $\eqref{eq:test1}$ and $\eqref{eq:test2}$, where we found agreement to at least 9 digits of precision for various values of $E$.

\section{Symmetric double well}

\subsection{Ordinary WKB analysis}
\begin{figure}
\centering
\includegraphics[width=0.6\textwidth]{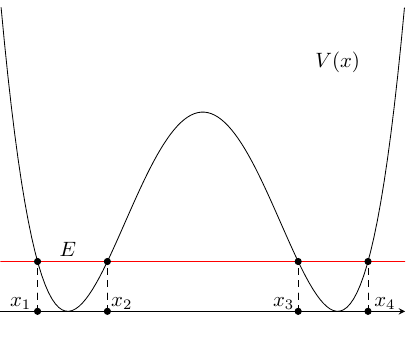}
\caption{Double well analysis for ordinary WKB, with its turning points. $(x_1,x_2)$ and $(x_3,x_4)$ are classically allowed regions, while $(x_2,x_3)$ is classically forbidden.}
\label{fig:doublewellord}
\end{figure}

A classical problem in quantum mechanics is the \textit{symmetric double well} potential, given by
\begin{align}
V(x)=a(x^2-h)^2,
\end{align}
with $a,h>0$. It is worth commenting the ordinary WKB version of this problem. We plot the potential and the turning points for ordinary WKB in figure \ref{fig:doublewellord}.
The analysis of the double well potential is present in many sources: we refer to \cite{marino2021advanced} for the analysis that inspired our method, or to \cite{sueishi2020exact} for a different method involving complexification of the $x$ variable that gives the same result.

Perturbative quantization of the double well potential is straightforward: the quantization condition is given by
\begin{align}
\cos\left(\frac1\hbar\Pi^{(1,2)}\right)=0,
\end{align}
also using the fact that $\Pi^{(3,4)}=\Pi^{(1,2)}$. Wavefunctions can be centered in the $(x_1,x_2)$ interval or the $(x_3,x_4)$ interval, and they would have the same energy even if linearly independent. This poses a problem, as spectra of ordinary QM problems cannot be degenerate. This problem is removed when one considers the WKB quantization condition, given by
\begin{align}
1+\exp\left(\mp\frac{2}{\ii\hbar}\Pi^{(1,2)}\right)=\pm\epsilon\ii\exp\left(-\frac{1}{\ii\hbar}\Pi^{(2,3)}\right),
\end{align}
where the $\pm$ sign depends on the deformation of $\hbar$ and $\epsilon$ keeps track of the parity of states, assuming values of $\pm1$. As $\Pi^{(2,3)}$ is purely imaginary and positive, the term on the RHS is of the form $\ee^{-\frac1\hbar F}$, with $F$ positive: this is a non perturbative correction that is invisible to perturbation theory and breaks the degeneracy of states, creating a non degenerate even ground state and an excited odd state. In fact, neglecting the non perturbative correction we would get exactly the perturbative quantization condition. The semiclassical interpretation of this result is that the degeneracy of the ground state is broken by tunneling through the $(x_2,x_3)$ interval. This is a very important result in ordinary WKB, and it will be worth to check if this result also holds in our deformation of WKB.

\subsection{Deformed WKB analysis}

\begin{figure}
\centering
\includegraphics[width=\textwidth]{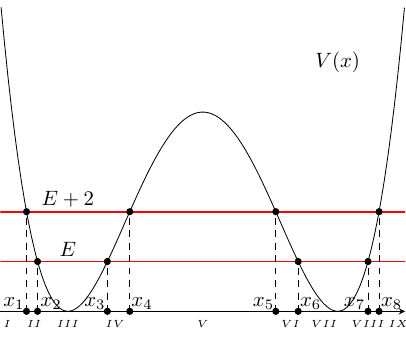}
\caption{Turning points for the symmetric double well potential.}
\label{fig:doublewell}
\end{figure}

\begin{figure}
\centering
\includegraphics[width=\textwidth]{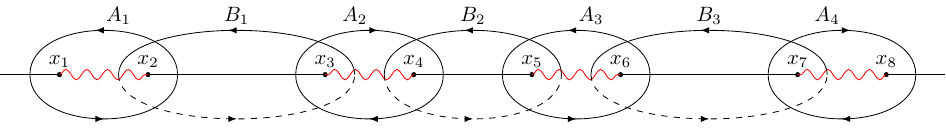}
\caption{Cycles for the double well potential.}
\label{fig:doublewellcycles}
\end{figure}

We now turn to the symmetric double well double well in deformed WKB. In figure \ref{fig:doublewell} we plot the potential together with its turning points, while in figure \ref{fig:doublewellcycles} we report the cycles on the $x$ line. We refer to \ref{fig:doublewell} for the names of the regions and the turning points.

In this case, we will focus on the \textit{minimal chamber}, the part of the spectrum where all turning points are real (so $E+2<V(0)$). We will make use of the symmetry of the potential: for the turning points we have
\begin{align}
x_i=-x_{9-i},\quad i=1,...,4\implies q_{i}=q_{9-i}^{-1},\quad i=1,...,4,
\end{align}
while for the cycles we have
\begin{align}
\Pi^{(x_{i},x_{i+1})}=\Pi^{(x_{9-i},x_{8-i})},\quad i=1,2,3.
\end{align}
Werecall the names to the Voros symbols:
\begin{align}
\begin{aligned}
&{V^{(1,2)}}=\exp\left(-\frac{1}{\ii\hbar}\Pi^{(1,2)}\right),\quad {V^{(2,3)}}=\exp\left(-\frac{1}{\ii\hbar}\Pi^{(2,3)}\right),\\
&{V^{(3,4)}}=\exp\left(-\frac{1}{\ii\hbar}\Pi^{(3,4)}\right),\quad {V^{(4,5)}}=\exp\left(-\frac{1}{\ii\hbar}\Pi^{(4,5)}\right).\label{eq:defs}
\end{aligned}
\end{align}
In our analysis, we will assume that all turning points are real, $E+2<V(0)$. Inspired by subsection \ref{subsec:res}, we interpret periods \eqref{eq:defs} as line integrals. As we have seen, the objects with good resurgent properties are the cycle integrals, in which the branch cut associated to the $\arccosh(x)$ discontinuity at $x=-1$ is removed via multiplication by the appropriate $q$ factor, centered at a turning point at $-1$. The general definition of loop integrals is difficult in this context, as we do not have ways to test if our definition is actually advantageous. Nevertheless, we will give a definition of cycle integrals. As the deformation $\hbar\to\hbar\pm\ii\epsilon$ influences the behaviour of the turning points at $-1$, we have to give different loop definitions. For the $-$ case, we have
\begin{align}
\begin{aligned}
&{V_{A_1}}^{(-)}=\frac1{\sqrt {q_1}}V^{{(1,2)}},\quad {V_{B_2}}^{(-)}=V^{(2,3)},\\
&\quad V_{A_2}^{(-)}=\sqrt{q_4}V^{(3,4)},\quad V_{B_2}^{(-)}=\sqrt\frac{q_4}{q_5}V^{(4,5)}=q_4 V^{(4,5)}.
\end{aligned}
\end{align}
For the $+$ case, we have
\begin{align}
\begin{aligned}
&{V_{A_1}}^{(+)}={\sqrt {q_1}}V^{{(1,2)}},\quad {V_{B_2}}^{(+)}=V^{(2,3)},\\
&\quad V_{A_2}^{(+)}=\frac1{\sqrt{q_4}}V^{(3,4)},\quad V_{B_2}^{(+)}=\sqrt\frac{q_5}{q_4}V^{(4,5)}=\frac{1}{q_4} V^{(4,5)}.
\end{aligned}
\end{align}
Some notes about those definitions:
\begin{itemize}
\item Only $q_1$ and $q_4$ appear in those definitions, as those are the only turning points at $-1$. In particular, we have used $q_4=q_5^{-1}$ for even potentials to get rid of the non independent point.
\item We have redefined via multiplication by $q_i^{\mp\frac12}$ if the turning point at $-1$ is on the left side of the line $(i,j)$, by $q_i^{\pm\frac12}$ if the turning point at $1$ is on the right side of $(i,j)$ (the upper sign refers to the $\hbar\to\hbar-\ii\epsilon$ deformation, the lower sign to the $\hbar\to\hbar+\ii\epsilon$ deformation). This is consistent with subsection \ref{subsec:res}.
\item We have used the nomenclature of figure \ref{fig:doublewellcycles} for the cycles.
\item We will write the quantization conditions both in terms of the line integrals (easier to compute, more relevant to WKB) and the cycle integrals (possibly more relevant for resurgent analysis).
\end{itemize}

Before giving the most general quantization condition, there is a particular quantization condition that we analyse.

\subsubsection{Degenerate spectrum}

We start with the deformation $\hbar\to\hbar-\ii\epsilon$ and with the coordinate vector
\begin{align}
C_I^{(x_1)}=\left(
\begin{array}{c}
-\ii\frac{q_{1}}q\\
1
\end{array}
\right).
\end{align}
Such a state can belong to the discrete spectrum. To see that, we first go through all transitions and examine $C_{IX}^{(x_8)}$. At order $q^{-2}$, this is given by
\begin{align}
C_{IX}^{(x_8)}=\left(
\begin{array}{c}
\frac1{q^2}\left(\frac{\ii q_{4}\left(1+{V^{(2,3)}}^2\right)}{q_1{V^{(1,2)}}^2{V^{(2,3)}}^2}+o(q)\right)\\
-\frac{q_1q_4\left({V^{(1,2)}}^2-1\right)\left({V^{(2,3)}}^2+1\right){V^{(4,5)}}}{{V^{(1,2)}}^2{V^{(2,3)}}^2}+o(q)
\end{array}
\right).
\end{align}
To have a normalizable state, we must have $1+{V^{(2,3)}}^2=0$. By imposing this condition, we can simplify the next term in $C_{IX}^{(x_8)}$ and get
\begin{align}
C_{IX}^{(x_8)}=\left(
\begin{array}{c}
\frac1{q}\left(\frac{\ii V_C\left(1+q_4^2{V^{(4,5)}}^2\right)}{q_1q_4^2{V^{(1,2)}}^2{V^{(4,5)}}}+o(q)\right)\\
o(q)
\end{array}
\right).
\end{align}
Another necessary condition is $1+q_4^2{V^{(4,5)}}^2=0$. The two conditions are solved by
\begin{align}
{V^{(2,3)}}=s_1 \ii,\quad {V^{(4,5)}}=s_2\frac{\ii}{q_4},
\label{eq:quant_toda}
\end{align}
where $s_1$ and $s_2$ are signs, taking values $\pm1$. In terms of the loop integrals, those can be written as
\begin{align}
V_{B_1}^{(-)}=s_1\ii,\quad V_{B_2}^{(-)}=s_2\ii.
\end{align}
At $+\infty$, we have
\begin{align}
C_{IX}^{(x_8)}=s_2\left(
\begin{array}{c}
-q\\
\ii q^2 q_1
\end{array}
\right).
\end{align}
Due to the fact that $q$ and $q^2$ decay exponentially faster than the actions, the wavefunction described by the vectors $C$ is normalizable. But this is not the only wavefunction admissible at this energy: if we start with 
\begin{align}
D_I^{(x_1)}=\left(
\begin{array}{c}
\ii\frac{q_{1}}{q^2}\\
-\frac{1}{q}
\end{array}
\right).
\end{align}
and impose the \textit{same} quantization conditions \eqref{eq:quant_toda}, we get in the $+\infty$ region
\begin{align}
D_{IX}^{(x_8)}=s_2\left(
\begin{array}{c}
1\\
-\ii q_1q^2
\end{array}
\right).
\end{align}
This also indicates a normalizable wavefunction, \textit{at the same energy} of the previous one and \textit{linearly independent} from the previous one. The energy level given by imposing \eqref{eq:quant_toda} is then \textit{degenerate}. This is a stark contrast with respect to the result of standard quantum mechanics where degenerate energy levels are forbidden. Furthermore, the wavefunctions can be organized as even and odd wavefunctions: remembering that the parity operator on vectors in region $IX$ consists in reversing the vector and imposing $q\to q^{-1}$, we can easily verify that for $s_2=1$ the wavefunction obtained from $C_I^{(x_1)}+D_I^{(x_1)}$ is even while the wavefunction obtained from $C_I^{(x_1)}-D_I^{(x_1)}$ is odd, and the opposite is valid for $s_2=-1$. For $\hbar\to\hbar+\ii\epsilon$ the same result holds, with a slightly modified quantization condition:
\begin{align}
{V^{(2,3)}}=s_1 \ii,\quad {V^{(4,5)}}=s_2\ii q_4.
\label{eq:quant_toda2}
\end{align}
In terms of the loop integrals, those quantization conditions become
\begin{align}
V_{B_1}^{(+)}=s_1\ii,\quad V_{B_2}^{(+)}=s_2\ii.
\end{align}
The result that we got coincides with the result in \cite{pasquier1992periodic} at order $\hbar^0$, as at that order the cycles are
\begin{align}
&{V^{(2,3)}}=\exp\left(-\frac{1}{\ii\hbar}\int_{x_2}^{x_3}\arccosh(E-V(t)+1)\dd t\right),\\
&{V^{(4,5)}}=\exp\left(-\frac{1}{\ii\hbar}\int_{x_4}^{x_4}\arccosh(V(t)-E-1)\dd t\right)\exp\left(\pm\frac{2\pi}{\hbar}x_4\right):
\end{align}
in the second line we have used the fact that in the imaginary allowed zone we can use $\arccosh(E-V(t))=\arccosh(V(t)-E)\pm\ii\hbar$, depending on the deformation of $\hbar$. The quantization condition at this level is
\begin{align}
&\cos\left(\frac1\hbar\int_{x_2}^{x_3}\arccosh(E-V(t)+1)\dd t\right)=0,\label{eq:q1}\\
&\cos\left(\frac1\hbar\int_{x_4}^{x_4}\arccosh(V(t)-E-1)\dd t\right)=0.\label{eq:q2}
\end{align}
Our result is an all-order generalization of the result \cite{pasquier1992periodic}, showing that degenerate states exist even if one takes all orders into account, and gives a prescription to compute higher-order corrections of the energy levels.

\begin{figure}
\centering
\includegraphics{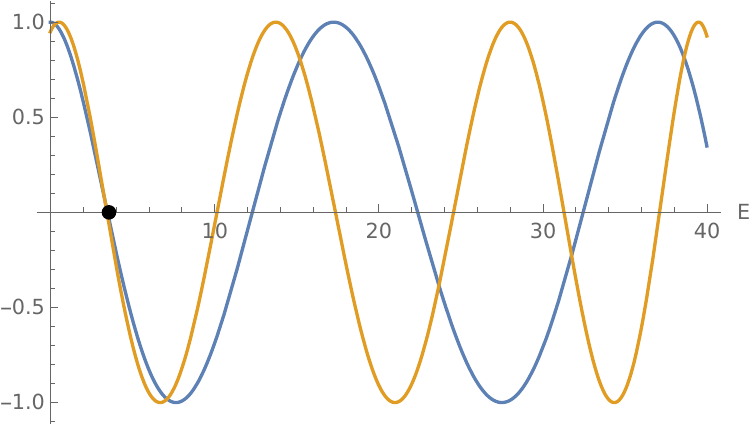}
\caption{Quantization condition \eqref{eq:quant_toda} or \eqref{eq:quant_toda2} at $\hbar=1$, $a=\frac12$, $h=9.11$. The blue line is the LHS of \eqref{eq:q1}, the orange line is the LHS of \eqref{eq:q2}. There is only one energy level from the Toda quantization condition, obtained when both functions are zero at the same energy.}
\label{fig:toda_four}
\end{figure}

\begin{figure}
\centering
\includegraphics{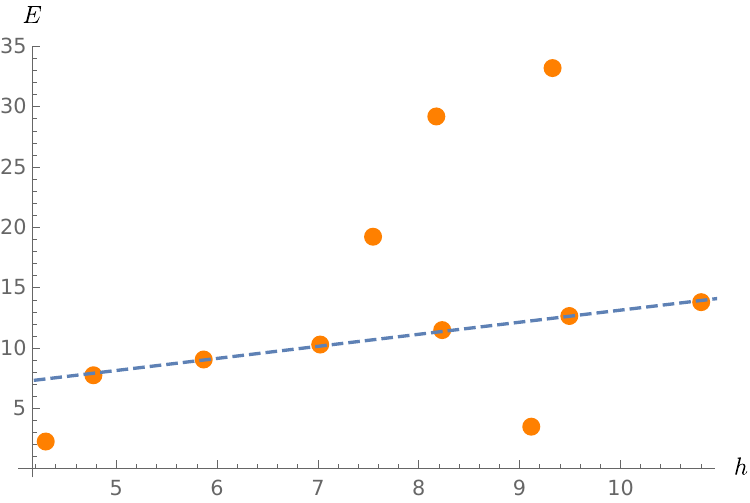}
\caption{A collection of degenerate energy levels arranged in an $(h,E)$ plot, with $\hbar=1$ and $a=\frac12$. Each point corresponds to a degenerate energy level, that exists only for the value of $h$ and takes value $E$. We have also fitted a line to show that some points arrange on a line: this can be interpreted as two distinct energy levels $E_n(h)$ and $E_{n+1}(h)$, joining only at the points $h$ for which we have degeneracy $E_n(h)=E_{n+1}(h)$.}
\label{fig:toda_four2}
\end{figure}

It might be possible that there are no energies such as \eqref{eq:quant_toda} or \eqref{eq:quant_toda2} holds. In fact, this is generally not true for all values of $a$ and $h$, but there are values of the parameters that allow the existence of those energy levels. In \cite{grassi2019solvable} the same was argued, together with the existence of degenerate energy levels. In fact, those energy levels were identified with the energy levels of Toda lattices, a comparison that we will expand in the next section. For now, we give numerical evidence for the existence of such states: in figure \ref{fig:toda_four} we show a combination of $a,h$ and $\hbar$ producing such a state, while in figure \ref{fig:toda_four2} we show a list of possible $h$ at fixed $a$ at which we have the degenerate eigenvalues that we are examining. The quantum mechanics interpretation of those eigenstates is that there are points $(a(\hbar),h(\hbar))$ at which quantum tunnelling is suppressed, so the degeneracy breaking due to tunnelling does not happen.

\subsubsection{General quantization condition}

Degenerate energy levels are not the only admissible energy levels for the double well. It turns out that the spectrum is way richer, and a more general quantization condition can be found with a more general starting wavefunction.

Starting with the deformation $\hbar\to\hbar-\ii\epsilon$, we start in region $I$ with the wavefunction indicated by
\begin{align}
C_I=\left(\begin{array}{c}
\frac{a_1}q+\frac{a_2}{q^2}\\
b_0+\frac{b_1}q+\frac{b_2}{q^2}
\end{array}\right).
\end{align}
Getting the quantization condition is now a very computationally intensive procedure that we performed using a computer algebra system (in our case, Mathematica 12.1.0) and removing all $q^{-n}$ contributions in $C_{IX}^{(x_8)}$ with $n>0$, together with the $q^0$ contribution in the second spot of $C_{IX}^{(x_8)}$. We use four out of the five coefficients to eliminate all terms of order $q^{-n}$ from $C_{IX}^{(x_8)}$, and at the end we are left with a vector whose second component is $o(q)$ (indicating a normalizable contribution), and the first component still has an order $q^{-1}$ component that is proportional to
\begin{align}
\begin{aligned}
&QC_{-}=q_4^6{V^{(4,5)}}^4(1+{V^{(2,3)}}^2)^3\left(1-\frac{q_1^2{V^{(1,2)}}^2}{q_2^2}\right)+\\
+&q_1^{2}{V^{(1,2)}}^{2}({V^{(2,3)}}^2(q_4^3q_3(1+{V^{(2,3)}}^2)^2{V^{(4,5)}}^2+{V^{(3,4)}}^4(1+q_4^2{V^{(4,5)}}^2)^2-\\
-&{V^{(3,4)}}^2(1+{V^{(2,3)}}^2)(1+q_4^2{V^{(4,5)}}^2)^2)).\label{eq:qcm}
\end{aligned}
\end{align}
The quantization condition is $QC_-=0$. In terms of loop integrals, this quantization condition becomes
\begin{align}
\begin{aligned}
&QC_{-}=q_4^2{V_{B_2}^{(-)}}^4(1+{V_{B_1}^{(-)}}^2)^3\left(1-\frac{q_1^4{V_{A_1}^{(-)}}^2}{q_2^2}\right)+\\
+&q_1^{4}{V_{A_1}^{(-)}}^{2}({V_{B_1}^{(-)}}^2(q_4q_3(1+{V_{B_1}^{(-)}}^2)^2{V_{B_2}^{(-)}}^2+\frac{1}{q_4^2}{V_{A_2}^{(-)}}^4(1+{V_{B_2}^{(-)}}^2)^2-\\
-&\frac{1}{{q_4}^2}{V_{A_2}^{(-)}}^2(1+{V_{B_1}^{(-)}}^2)(1+{V_{B_2}^{(-)}}^2)^2)).
\end{aligned}
\end{align}
With this quantization condition the degenerate states are also included, as substituting $\eqref{eq:quant_toda}$ gives $QC_-=0$. In this most general quantization condition, we can see the role of the non perturbative corrections ${V^{(1,2)}}$ and ${V^{(3,4)}}$, that bring to the quantization condition terms of order $\exp\left(-\frac1\hbar P\right)$, with $P$ real positive. The same procedure holds for the positive deformation, where we obtain from the starting vector
\begin{align}
C_I=\left(\begin{array}{c}
a_0+\frac{a_1}q+\frac{a_2}{q^2}\\
\frac{b_1}q+\frac{b_2}{q^2}
\end{array}\right),
\end{align}
the quantization term
\begin{align}
\begin{aligned}
&QC_+=q_2^2q_4^2{V^{(1,2)}}^2(1+{V^{(2,3)}}^2)^3{V^{(3,4)}}^4{V^{(4,5)}}^4-q_1^2q_4^2(1+{V^{(2,3)}}^2)^3{V^{(3,4)}}^4{V^{(4,5)}}^4+\\
&+q_1^2q_2^2{V^{(2,3)}}^2(q_4^4(1-(1+{V^{(2,3)}}^2){V^{(3,4)}}^2)+2q_4^2+\\
&+q_4^2((1+V_B^2){V^{(3,4)}}^2(-2+q_4^2(1+{V^{(2,3)}}^2){V^{(3,4)}}^2)){V^{(4,5)}}^2-\\
&-(-1+(1+{V^{(2,3)}}^2){V^{(3,4)}}^2){V^{(4,5)}}^4),
\end{aligned}
\end{align}
and the quantization condition is $QC_+=0$. In terms of loop integrals, we have
\begin{align}
\begin{aligned}
&QC_+=\frac{q_2^2q_4^2}{q_1}{V_{A_1}^{(+)}}^2(1+q_4^2{V_{B_1}^{(+)}}^2)^3{V_{A_2}^{(+)}}^4{V_{B_2}^{(+)}}^4-\\
&-q_1^2q_4^4(1+{V_{B_1}^{(+)}}^2)^3{V_{A_2}^{(+)}}^4{V_{B_2}^{(+)}}^4+q_1^2q_2^2{V_{B_2}^{(+)}}^2(q_4^4(1-q_4(1+{V_{B_2}^{(+)}}^2){V_{A_{2}}^{(+)}}^2)+2q_4^2+\\
&+q_4^5((1+{V_{B_1}^{(+)}}^2){{V_{A_2}}^{(+)}}^2(-2+q_4^3(1+{V_{B_2}^{(+)}}^2){V_{A_2}^{(+)}}^2)){V_{B_2}^{(+)}}^2-\\
&-(-1+(1+q_4{V_{B_1}^{(+)}}^2){V_{A_2}^{(+)}}^2)q_4^4{V_{B_2}^{(+)}}^4),
\end{aligned}
\end{align}

The two quantization conditions are very complicated, but they do bring to new states. We now compare the spectrum obtained with $\eqref{eq:qcm}$ with numerical results from appendix \ref{app:appbox}. We choose $a=\frac12$ and $h=10$, for which we have $E_0=4.09799$ and $E_1=4.09932$. The results of our quantization condition (with corrections up to $\hbar^4$) are reported in table \ref{tab:quantconddouble}. As we can see, we get a good accuracy, improved by including further corrections. Unfortunately, due to the fact that the energy states are almost degenerate, we are not able to find the two states at the precision that we employed. Improving the precision of this computation is a future line of research.

\begin{table}
\centering
\begin{tabular}{c|c|c}
Order&WKB prediction&\% difference\\\hline
0&3.71163&9.00\\
2&4.11335&0.40\\
4&4.09836&0.01\\
\end{tabular}
\caption{Prediction of the ground state energy level, to compare with the numerical value $E_0=4.09799$. We also list the percentage difference between the predictions and the numerical result.}
\label{tab:quantconddouble}
\end{table}

\section{Successive wells}

\begin{figure}
\centering
\includegraphics[scale=1]{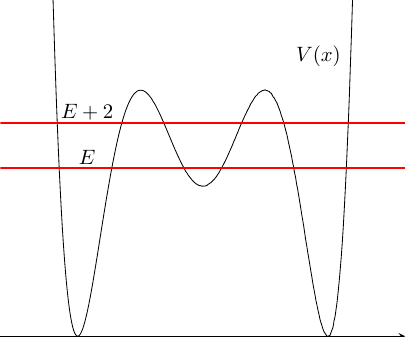}
\caption{An example of the potential in analysis, with $N=6$. We focus on values of $E$ such as the red lines intersect the potential $6$ times each.}
\label{fig:triple_well}
\end{figure}

We conclude this chapter with a simple example, dealing with an arbitrary number of successive potential wells.  The potential for this problem is given by
\begin{align}
V(x)=a\prod_{n=1}^N(x-z_i)^n+V_0,
\end{align}
where $V_0$ is selected in such a way to have the minimum of the potential at $0$, $a$ is real and positive, $N$ is an even integer and $z_i$ are real and positive. The coefficients are such as $E=V(x)$ and $E+2=V(x)$ have exactly $N$ real solutions, all distinct. We plot an example of such a potential in figure \ref{fig:triple_well}. The quantization conditions that we obtain will not be exact, as they neglect contributions from quantum periods in which at least one of the turning point is complex. Nevertheless, those are approximate solutions that have an interesting counterpart in a problem of physical interest.

\begin{figure}
\centering
\includegraphics[scale=1]{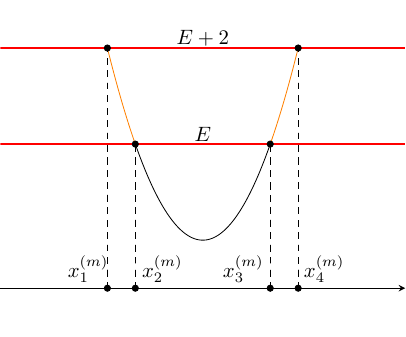}
\includegraphics[scale=1]{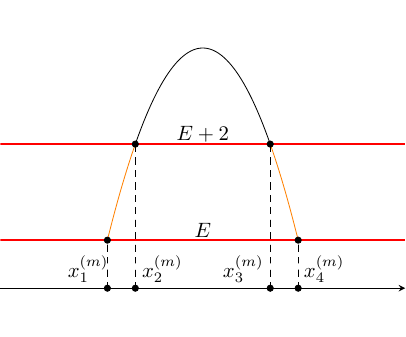}
\caption{Possible wells in a general successive wells problem. The parts colored in orange are common to more than one well.}
\label{fig:wells}
\end{figure}

This problem can be analysed by recognizing that there are two wells to examine separately. Those are pictured in figure \ref{fig:wells}. In a typical problem with potential of order $N$, we encounter $N-1$ wells: of those, $N/2$ will be concave wells (left in figure \ref{fig:wells}), and $N/2-1$ will be convex wells (right in figure \ref{fig:wells}). The concave wells have a classically allowed region in the center, while the convex wells have an imaginary allowed region in the center. From left to right, we always encounter one concave well, then a convex well, then a concave one and so on until we cross all wells.

Referring to figure \ref{fig:wells}, we name the turning points according to the well to which they belong and the order in which they appear, from left to right. There are some turning points that are in common to multiple wells: in particular, $x_3^{(m)}=x_1^{(m+1)}$ and $x_4^{(m)}=x_2^{(m+1)}$ for all wells. We will have associated exponential factors of the form $q_n^{(m)}$, where $n$ ranges from $1$ to $4$ and $m$ from $1$ to $N/2-1$. We will write the exponential factors associated to the turning points as
\begin{align}
q^{(m)}_n=\exp\left(-\frac{2\pi}{\hbar}x_n^{(m)}\right).
\end{align}
We will not get the complete spectrum of the potential, but focus on the degenerate spectrum. For the deformation $\hbar\to\hbar-\ii\epsilon$, we start with a normalizable function at $-\infty$ indicated by the vector (where we use the subscript $L$ or $R$ to indicate whether the coordinate has to be intended for a wavefunction to the left of the turning point or to its right)
\begin{align}
C^{(x_1^{(1)})}_L=\left(\begin{array}{c}
-\ii\frac{q}{q_1^{(1)}}\\
1
\end{array}\right).
\end{align}
We will impose the following quantization conditions in advance, and see that they bring to a normalizable wavefunction. We define
\begin{align}
&V^{(x_2^{(m)},x_3^{(m)})}=\exp\left(-\frac{1}{\ii\hbar}\Pi^{(x_2^{(m)},x_3^{(m)})}\right),\\
&V^{(x_1^{(m)},x_2^{(m)})}=\exp\left(-\frac{1}{\ii\hbar}\Pi^{(x_1^{(m)},x_2^{(m)})}\right),\\
&V^{(x_3^{(m)},x_4^{(m)})}=\exp\left(-\frac{1}{\ii\hbar}\Pi^{(x_3^{(m)},x_4^{(m)})}\right).
\end{align}
the first cycle is the cycle in the allowed region for $m$ even and in the imaginary allowed region for $m$ odd, and the other cycles are the other purely imaginary cycles. The quantization condition that we impose is
\begin{align}
V^{(x_2^{(m)},x_3^{(m)})}=\begin{cases}
s_m\ii,\quad &m\text{ odd},\\
s_m\ii\sqrt{\frac{q_3^{(m)}}{q_2^{(m)}}},\quad &m\text{ even}.
\end{cases}
\end{align}
Here $s_m=\pm1$ and can be chosen arbitrarily. As before, we can define loop integrals $V_{B_m}^{(\mp)}$ as
\begin{align}
V_{B_m}^{(\mp)}=\begin{cases}
V^{(x_2^{(m)},x_3^{(m)})},\quad &m\text{ odd},\\
\left(\frac{{q_2}^{(m)}}{{q_3}^{(m)}}\right)^{\pm\frac12}V^{(x_2^{(m)},x_3^{(m)})},\quad &m\text{ even},
\end{cases}
\end{align}
choosing signs according to the deformation $\hbar\to\hbar\mp\ii\epsilon$. In terms of the loop integral, the quantization condition is simply $V_{B_m}^{(-)}=s_m\ii$ for every $m$, with a definite assignment of signs $s_m$. This quantization condition is a generalization of \eqref{eq:quant_toda}. With this quantization condition, we start crossing the various wells. In the orange region between $x_3^{(1)}$ and $x_4^{(1)}$, the wavefunction will be indicated by the vector
\begin{align}
C^{(x_4^{(1)})}_L=s_1\left(\begin{array}{c}
\frac{\ee^{-\ii\frac\pi4}{V^{(x_3^{(1)},x_4^{(1)})}}}{V^{(x_1^{(1)},x_2^{(1)})}}\\
0
\end{array}\right).
\end{align}
We can now take this wavefunction through the second well, and obtain in the region between $x_2^{(2)}$ and $x_3^{(2)}$
\begin{align}
C^{(x_3^{(2)})}_L=-qs_1s_2\frac{V^{(x_3^{(1)},x_4^{(1)})}}{V^{(x_1^{(1)},x_2^{(1)})}}\frac{1}{\sqrt{q_2^{(2)}q_3^{(2)}}}\left(\begin{array}{c}
-\ii\frac{q}{q_1^{(3)}}\\
1
\end{array}\right),
\end{align}
where we have used $q_3^{(2)}=q_1^{(3)}$. We now are at a concave well, where we can repeat the procedure, as the vector part of $C^{(x_3^{(2)}}$ is equal to the vector part of $C^{(x_1)^{(1)}}$. We can cross an arbitrary number of wells in this way via simple induction: after crossing the last concave well, we arrive in the region containing $+\infty$ with the vector
\begin{align}
\begin{aligned}
&C^{(x_4^{(N-1)})}_R=(-1)^{\frac N2-1}\left(\prod_{m=1}^Ns_m\right)q^{\frac N2-1}\left(\begin{array}{c}
1\\
-\ii\frac{q}{q_4^{(N-1)}}
\end{array}\right)\times\\
&\times\left(\prod_{\stackrel{j=1}{j\text{ odd}}}^{N-1}\frac{{V^{(x_3,x_4)}}^{(j)}}{{V^{(x_1,x_2)}}^{(j)}}\frac1{\sqrt {q_2^{(j+1)}q_3^{(j+1)}}}\right)
\end{aligned}
\end{align}
This wavefunction is normalizable. Not only that, but due to the factor $q^{\frac N2-1}$ that improves normalizability, we can conclude that all wavefunctions indicated by $q^{-j}C_L^{(x_1)^{(1)}}$ are normalizable solutions as long as $j$ is an integer between $0$ and $\frac N2-1$. The degeneracy of each energy level is then $\frac N2$.

For $\hbar\to\hbar+\ii\epsilon$, the situation is the same. We only have to reverse the starting wavefunction as
\begin{align}
C^{(x_1^{(1)})}_L=\left(\begin{array}{c}
1\\
-\ii\frac{q}{q_1^{(1)}}
\end{array}\right),
\end{align}
and impose the quantization condition
\begin{align}
V_{B_m}=\begin{cases}
s_m\ii,\quad &m\text{ odd},\\
s_m\ii\sqrt{\frac{q_2^{(m)}}{q_3^{(m)}}},\quad &m\text{even}:
\end{cases}
\end{align}
note that the coefficients of the square root are inverted here to take the other determination of $\arccosh$ into account. Again, the quantization condition for loop integrals is simply $V_{B_m}^{(+)})=s_m\ii$ for every $m$. The transitions of all wells give rise to
\begin{align}
\begin{aligned}
&C^{(x_4^{(N-1)})}_R=(-1)^{\frac N2-1}\left(\prod_{m=1}^Ns_m\right)q^{\frac N2-1}\left(\begin{array}{c}
-\ii\frac{q}{q_4^{(N-1)}}\\
1
\end{array}\right)\times\\
&\times\left(\prod_{\stackrel{j=1}{j\text{ odd}}}^{N-1}\frac{{V^{(x_3,x_4)}}^{(j)}}{{V^{(x_1,x_2)}}^{(j)}}\frac1{\sqrt {q_2^{(j+1)}q_3^{(j+1)}}}\right)
\end{aligned}
\end{align}
We note that our result is an all-order generalization of \cite{pasquier1992periodic}. In the reference, the intervals $q_2^{(m)}$ and $q_3^{(m)}$ are denoted as intervals of instability, and the quantization condition is
\begin{align}
\cos\frac1\hbar\int_{x_2^{(m)}}^{x_3^{(m)}}\arccosh\abs{E-V(t)+1}\dd t=0.
\end{align}
Our quantization condition coincides with this one at lowest $\hbar$ order, but also includes $\hbar$ corrections. We still underline that for $N\geq4$ this quantization condition is not exact, as we are not in the minimal chamber and there are complex turning points that can bring non perturbative corrections to the quantization condition. Considering those corrections would require a complex WKB approach, that goes outside of the scope of this work. The quantization conditions that we have found coincide with the quantization conditions of the Toda lattice for $N$ particles.

%% file: conclusion.tex
\chapter*{Conclusion}

\addcontentsline{toc}{chapter}{Conclusion}

\markboth{CONCLUSION}{CONCLUSION}

In this thesis, we have presented a method to compute an all-order WKB solution for finite difference equations. In particular, we have studied the resurgent behaviour of the Bessel functions, that coincides with our expectations from general resurgence theory. Using this analysis, we have managed to compute correction formulae analogous to the ones computed in standard WKB, at the cost of having to make some conjectures. We have tested this result against problems that have already been studied, finding exact eigenstates for Toda lattices, studying the effects of quantum tunneling (and its suppression for particular potentials) and also finding extra states in the case of the double well.

This thesis is just a start for the WKB analysis of finite difference equations. From a mathematical point of view, a more rigorous derivation of our connection formulae is a future goal, where the goal will be to prove the conjectures that we have proposed. In particular, the existence of a suitable deformation $\phi$ from section \ref{sec:conn_form_def} that maps the finite difference problem with general potential in the finite difference problem with linear potential is the main conjecture, and it would also be very important to find an equation determining this deformation. In addition, the the factorial growth of the coefficients $S_n(x)$ conjectured in \eqref{eq:fac_growth} is also an important point, as the resurgence techniques used in this thesis depend on such a growth. While we have numerical evidence for the growth of the $S_n$ coefficients in the linear potential and the growth of the coefficients of the cycles $\Pi^{(i,j)}_n$ being factorial, a next step would be to prove that such a growth applies to all potentials, as in the case of standard WKB.

From a more practical point of view, a generalization of this method to complex turning points is also a future goal, that would allow the study of complex instantons and their influence in the quantization formulae that we have found. Computationally, we have been limited by the complexity of our algorithms, and did not manage to get a sufficient number of $\hbar$-corrections to get a full picture of the resurgence of the quantum periods in the simplest case. We are currently studying different implementations of our algorithm to obtain a larger pool of data. Lastly, but maybe most importantly, a more systematic study with more data could lead to the development of resurgence formulae analogous to the Dillinger-Delabaere-Pham formula \cite{delabaere1997exact}, prescribing the full resurgent behaviour of the quantum periods.

What we have here is a solid start, and we are sure that further developments of this topic will be of great interest in the upcoming years.

%% file: appnum.tex
\chapter[Bessel expansions]{Numerical computation of the Bessel expansions}
\label{app:num}

In this appendix, we give a numerical computation of the expansion of the Bessel functions, as computed analytically in chapter \ref{chapter3}. This method is based on the method developed in \cite{garoufalidis2021resurgent} and \cite{garoufalidis2020peacock}, and is a numerical confirmation on the results that we obtained. In order to do so, we will reobtain the form of the Stokes automorphisms for the integrals $I^{(\pm,n)}$ of chapter \ref{chapter3}.

\section{Setting up the problem}

Our goal is to compute the asymptotic expansions of the Bessel functions
\begin{align}
J_{\frac ya}\left(\frac1a\right),\quad Y_{\frac ya}\left(\frac1a\right)
\end{align}
in terms of the asymptotic expansions $\beta_{\pm}(y,a)$ defined in chapter \ref{chapter3}, that we report here for convenience. The asymptotic expansions are based on the polynomials $U_k(p)$, defined by
\begin{align}
\begin{cases}
&U_0(p)=1,\\
&U_{k+1}(p)=\frac12p^2(1-p^2)U_k'(p)+\frac18\int_0^p(1-5t^2)U_k(t)\dd t.
\end{cases}
\end{align}
In terms of those coefficients, the asymptotic expansions are defined in the region $y>1$ as
\begin{align}
\beta_{\pm}(y,a)\simeq\sqrt{\frac a{2\pi}}\frac{1}{(y^2-1)^{\frac14}}\exp\left(\mp\frac1 a\Pi(y)\right)\sum_{k=0}^\infty (\pm y)^{-k}U_k\left(\frac{y}{\sqrt{y^2-1}}\right)a^k
\end{align}
and in the region $0<y<1$ as
\begin{align}
\beta_{\pm}(y,a)\simeq\sqrt{\frac a{2\pi}}\frac{1}{(1-y^2)^{\frac14}}\exp\left(\mp\frac1a\Pi(y)\right)\sum_{k=0}^\infty (\pm y)^{-k}U_k\left(\frac{-\ii y}{\sqrt{y^2-1}}\right)a^k,
\end{align}
where the function $\Pi(y)$ is given by
\begin{align}
\Pi(y)=\begin{cases}
y\arccosh y-\sqrt{y^2-1},\quad &y>1,\\
\ii(y\arccos y-\sqrt{1-y^2}),\quad &0<y<1.
\end{cases}
\end{align}
We will also consider the series $q^n\beta_{\pm}(y,a)$, with
\begin{align}
q=\exp\left(-\frac{2\pi\ii}a y\right).
\end{align}

Consider the series\footnote{The polynomial $U_k$ is evaluated at $\frac{y}{\sqrt{y^2-1}}$ or $\frac{-\ii y}{\sqrt{1-y^2}}$ depending on whether $y>1$ or $0<y<1$.}
\begin{align}
S_{\pm}(y,a)= \sum_{k=0}^\infty(\pm y)^{-k}U_ka^k
\end{align}
appearing in $\beta_{\pm}(y,a)$. Those series are asymptotic, so we will use Borel-Padé resummation to evaluate them. The standard procedure is to define the Borel transform
\begin{align}
\hat S_{\pm}(y,s)=\sum_{k=0}^\infty(\pm y)^{-k-1}U_{k+1}\left(\frac{y}{\sqrt{y^2-1}}\right)\frac{s^k}{\Gamma(k+1)},\label{eq:trans}
\end{align}
and the Borel sum at $\theta=\arg a$
\begin{align}
\mathcal S_\theta S_{\pm}(y,a)=1+\int_0^{\ee^{\ii\theta}\infty}\ee^{-\frac sa}\hat S_{\pm}(y,s)\dd s.\label{eq:borel}
\end{align}
This defines a convergent function up to the Stokes phenomenon. The functions $\beta_{\pm}(y,a)$ are then
\begin{align}
\beta_{\pm}(y,a)=\sqrt{\frac{a}{2\pi}}\frac{1}{(y^2-1)^{\frac14}}\exp\left(\mp\frac{1}{a}\Pi(y)\right)\mathcal S_\theta S_{\pm}(y,a)
\end{align}
for $y>1$ and
\begin{align}
\beta_{\pm}(y,a)=\sqrt{\frac{a}{2\pi}}\frac{1}{(1-y^2)^{\frac14}}\exp\left(\mp\frac{1}{a}\Pi(y)\right)\mathcal S_\theta S_{\pm}(y,a)
\end{align}
for $0<y<1$. This is the standard technique for Borel-Padé resummation, but we will employ a more powerful technique.

\section[Borel transform]{The Borel transform and conformal transformations}

\begin{figure}
\centering
\includegraphics[width=0.6\textwidth]{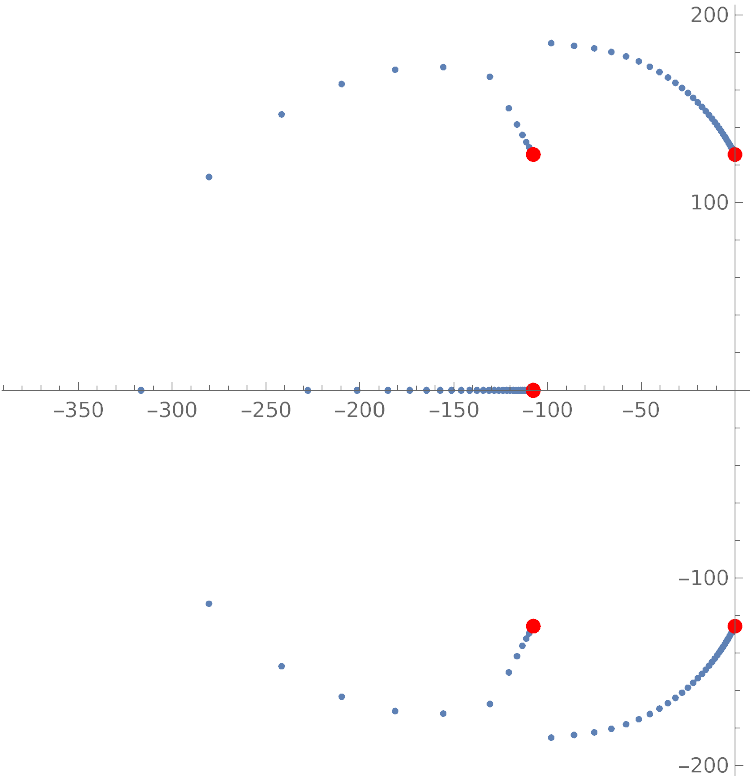}
\includegraphics[width=0.6\textwidth]{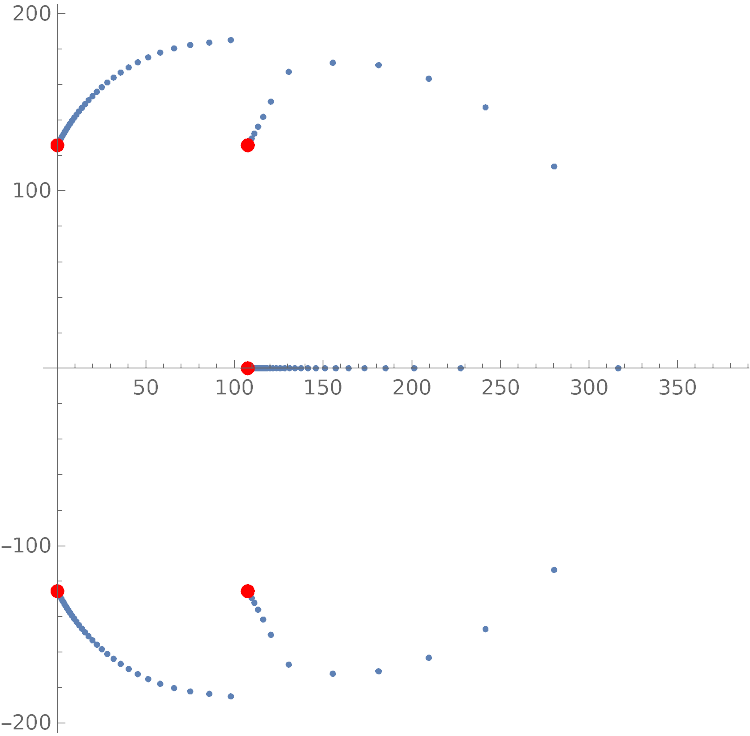}
\caption{Poles for the Padé approximants in the $s-$plane to $\hat S_+$ and $\hat S_-$ (top and bottom respectively), for $y=20$ using $300$ terms in the two series. We have marked in red the points at which the branch cuts are expected to form. We see that the start points of the branch cuts approximated by the poles organize in vertical towers, one on the imaginary axis and the other translated.}
\label{fig:poles_b1}
\end{figure}

\begin{figure}[h!]
\centering
\includegraphics[width=0.6\textwidth]{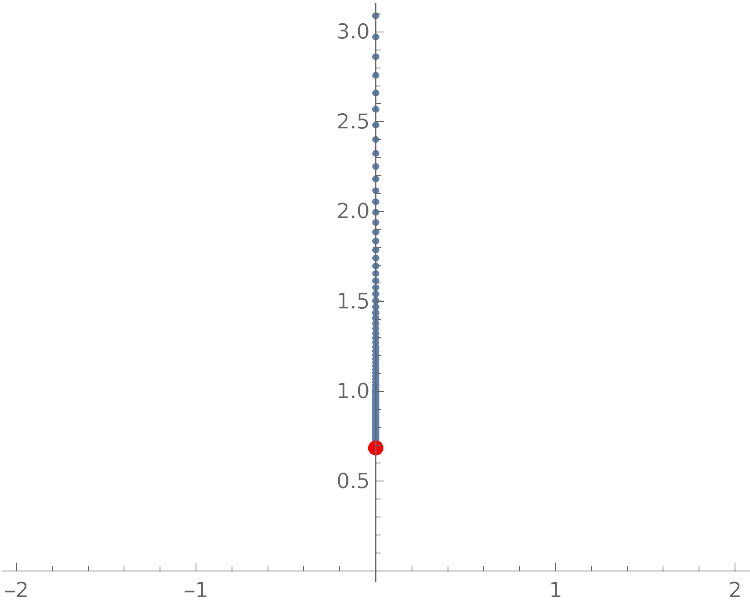}
\includegraphics[width=0.6\textwidth]{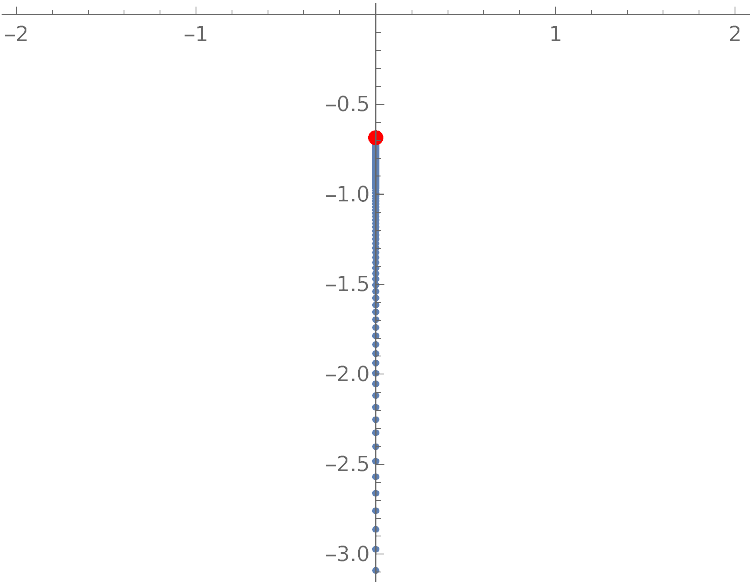}
\caption{Poles for the Padé approximants in the $s-$plane to $\hat S_+$ and $\hat S_-$ (top and bottom respectively), for $y=\frac12$ using $300$ terms in the two series. We have marked in red the points at which the branch cuts are expected to form. In this case we only have poles on the imaginary axis, as expected from the fact that $\Pi\left(\frac12\right)$ is purely imaginary.}
\label{fig:poles_b2}
\end{figure}

Equation \eqref{eq:borel} requires an analytical continuation of the series \eqref{eq:trans} beyond its radius of convergence. The numerical procedure to approximate this the Padé approximant,: the Padé approximant of order $n$ is defined to be the rational function $P(s)/Q(s)$ where $P$ and $Q$ are irreducible polynomials of order $n$ that has a Taylor expansion at $s=0$ coinciding with \eqref{eq:trans} up to order $s^{2n}$. The poles of the Padé approximant can be used to study the singularities of the full analytic expansion, as explained in \cite{aniceto2019primer}. In particular, we expect the singularities to be branch cuts starting at
\begin{align}
-2\Pi(y)+2\pi\ii n y,\quad 2\pi\ii n y
\end{align}
for $\hat S_+$, and
\begin{align}
2\Pi(y)+2\pi\ii n y,\quad 2\pi\ii n y
\end{align}
for $\hat S_-$. We show the array of poles in figure \ref{fig:poles_b1} for $y>1$ and in figure $\ref{fig:poles_b2}$ for $0<y<1$.

In order to improve our precision on the Borel-Padé resummation, we will employ a technique based on \cite{caliceti2007useful,caprini1999accelerated,caprini2000convergence}, consisting in employing the technique of conformal transformations before and after performing the resummation\footnote{We thank Prof. Marcos Mari\~no for suggesting how to implement this technique in our setting.}. At first, we define the transformation
\begin{align}
z^{-1}(y,x)=\frac{\sqrt{1-\frac{x}{2\pi\ii y}}-1}{\sqrt{1-\frac{x}{2\pi\ii y}}+1},
\end{align}
where the factor $2\pi\ii y$ is chosen due to the branches starting at $2\pi\ii y n$. This transformation is inverted by
\begin{align}
z(y,x)=-8\pi\ii y\frac{x}{(1-x)^2}.
\end{align}
We define the conformal-Borel transform of the series $S_{\pm}$ as
\begin{align}
C\hat S_{\pm}(y,s)=\sum_{k=0}^\infty(\pm y)^{-k-1}U_{k+1}\left(\frac{y}{\sqrt{y^2-1}}\right)\frac{z(y,s)^k}{\Gamma(k+1)}:
\end{align}
we can recover the original resummation as
\begin{align}
\mathcal S_\theta S_{\pm}(y,a)=1+\int_0^{\ee^{\ii\theta}\infty}\ee^{-\frac sa}C\hat S_{\pm}(y,z^{-1}(y,s))\dd s.
\end{align}
Even if the final result is the same, the conformal-Borel transform reaches an higher precision with a smaller number of terms.

\section{Numerical expansion of the Bessel functions}

We now turn to the problem of computing the asymptotic expansion of Bessel functions for chosen values of $a$. We will compute those expansions at four values of the argument of $a$ (called rays):
\begin{itemize}
\item ray I, argument of $a$ is small and positive,
\item ray II, argument of $a$ is near $\pi$, but slightly smaller than it,
\item ray III, argument of $a$ is near $-\pi$, but slightly bigger than it,
\item ray IV, argument of $a$ is small and negative.
\end{itemize}
For ray 1, we will choose an argument smaller than $\arg(2\Pi(y)+2\pi\ii y)$, so we won't cross the first non-real singularity in the tower. We will make analogous choices in for all rays.

In each region $R$, we look for matrices $A_R$ to express $J$ and $Y$ as
\begin{align}
\left(
\begin{array}{c}
J_{\frac ya}\left(\frac1a\right)\\
Y_{\frac ya}\left(\frac1a\right)
\end{array}
\right)=A_R(q)\left(
\begin{array}{c}
\beta_+(y,a)\\
\beta_-(y,a)
\end{array}
\right).\label{eq:system}
\end{align}
We take the matrices $A$ to be periodic matrices in $x$ of period $a$, so we assume a dependency on the various $q$ factors. We parametrize the matrices as
\begin{align}
A_R(q)=\left(
\begin{array}{cc}
\sum_{n=-\infty}^\infty a_nq^n&\sum_{n=-\infty}^\infty b_nq^n\\
\sum_{n=-\infty}^\infty c_nq^n&\sum_{n=-\infty}^\infty d_nq^n
\end{array}\label{eq:mat}
\right).
\end{align}
In order to perform the computation, we fix a point $y$ and a phase of $a$: we then choose an array of modules for $a$ to compute the vector parts of \eqref{eq:system} at various values of $a$, sharing the same phase. We impose a cutoff on the number of coefficients in the matrices $A_R(q)$, obtaining a system from which we can extract the values of the various coefficients. In our computations, at each point $y$ and phase of $a$ we have chosen $1/n$ as absolute value of $a$, with $n$ ranging from 10 to 19\footnote{As the expansion is in $a$, it is convenient to choose a small value for its absolute value.}. This allows us to compute $5$ coefficients from each series of \eqref{eq:mat}, and we will always compute the coefficients of $q^{-2},q^{-1},q^0,q^1,q^2$. We always include $40$ coefficients in the resummation.

We now give the results of this analysis. For $y>1$, we have picked $y=\frac32$. For all coefficients, the agreement is always up to at least $11$ digits.
\begin{itemize}
\item For ray $I$, we choose $\arg a=\frac{\pi}{30}$. The resulting matrix is
\begin{align}
A_I=
\left(
\begin{array}{cc}
1&0\\
\ii&-2
\end{array}\right).
\end{align}
\item For ray $II$, we choose $\arg a=\pi-\frac{\pi}{30}$. The resulting matrix is
\begin{align}
A_{II}=
\left(
\begin{array}{cc}
1-q&-\ii q\\
\ii+\ii q&-2-q
\end{array}\right).
\end{align}
\item For ray $III$, we choose $\arg a=\frac{\pi}{30}-\pi$. The resulting matrix is
\begin{align}
A_{III}=
\left(
\begin{array}{cc}
1-\frac1q&-\frac{\ii}q\\
-\ii-\frac{\ii}q&-2-\frac1q
\end{array}\right).
\end{align}
\item For ray $IV$, we choose $\arg a=-\frac{\pi}{30}$. The resulting matrix is
\begin{align}
A_{IV}=
\left(
\begin{array}{cc}
1&0\\
-\ii&-2
\end{array}\right).
\end{align}
\end{itemize}

For $0<y<1$ we have picked $y=\frac12$. For all coefficients, the agreement is always up to at least 31 digits. This higher precision is expected because in this situation we are farther from Stokes lines, that coincide with the imaginary axis.
\begin{itemize}
\item For ray $I$, we choose $\arg a=\frac{\pi}{30}$. The resulting matrix is
\begin{align}
A_I=
\left(
\begin{array}{cc}
\ee^{-\ii\frac\pi4}&\ee^{\ii\frac\pi4}\\
-\ee^{\ii\frac\pi4}&-\ee^{-\ii\frac\pi4}
\end{array}\right).
\end{align}
\item For ray $II$, we choose $\arg a=\pi-\frac{\pi}{30}$. The resulting matrix is
\begin{align}
A_{II}=
\left(
\begin{array}{cc}
\ee^{-\ii\frac\pi4}&-\ee^{\ii\frac\pi4} q\\
-\ee^{\ii\frac\pi4}&-2\ee^{-\ii\frac\pi4}-\ee^{-\ii\frac\pi4}q
\end{array}\right).
\end{align}
\item For ray $III$, we choose $\arg a=\frac{\pi}{30}-\pi$. The resulting matrix is
\begin{align}
A_{III}=
\left(
\begin{array}{cc}
-\ee^{-\ii\frac\pi4}\frac1q&\ee^{\ii\frac\pi4}\\
-2\ee^{\ii\frac\pi4}-\ee^{\ii\frac\pi4}\frac1q&-\ee^{-\ii\frac\pi4}
\end{array}\right).
\end{align}
\item For ray $IV$, we choose $\arg a=-\frac{\pi}{30}$. The resulting matrix is
\begin{align}
A_{IV}=
\left(
\begin{array}{cc}
\ee^{-\ii\frac\pi4}&\ee^{\ii\frac\pi4}\\
-\ee^{\ii\frac\pi4}&-\ee^{-\ii\frac\pi4}
\end{array}\right).
\end{align}
\end{itemize}

\section{Stokes automorphism}

\begin{figure}
\centering
\includegraphics[width=0.7\textwidth]{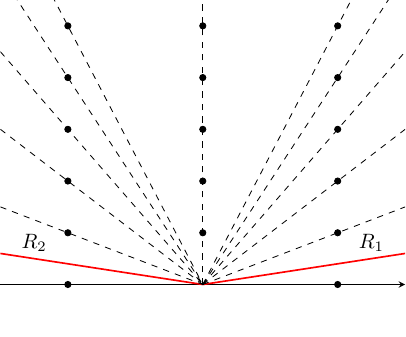}
\caption{Sweeping automorphism $\underline{\mathfrak S}_{R_1\to R_2}$. This automorphism is a product of automorphisms $\underline{\mathfrak S}_{\theta}$ that are represented by the dashed lines.}
\label{fig:rays}
\end{figure}

We can now use the matrices that we just computed to give the Stokes automorphisms for the asymptotic series $\beta_\pm$. In general, we can go from a ray $R_1$ to a ray $R_2$ using a Stokes automorphism $\underline{\mathfrak S}_{R_1\to R_2}$ defined as
\begin{align}
\underline{\mathfrak S}_{R_1\to R_2}=A_{R_2}^{-1}A_{R_1}.
\end{align}
This is a ``sweeping" Stokes automorphism, that encounters all singularities that have their arguments between rays $R_1$ and $R_2$. This sweeping automorphism is made of automorphisms $\underline{\mathfrak S}_\theta$, with $\theta$ between $R_1$ and $R_2$. We give an illustration of this automorphism in figure \ref{fig:rays}.
We start from the region $y>1$. Using the terminology of subsection \ref{sec:phys_reg}, with $\theta_{(\pm,n)}=\arg(2\Pi(y)-2\pi\ii ny)$ we can write the automorphisms as
\begin{align}
\underline{\mathfrak S}_{I\to II}=\left(\prod_{n=1}^\infty\underline{\mathfrak S}_{\theta_{(+,n)}}\right)\underline{\mathfrak S}_{\frac\pi2}\left(\prod_{n=1}^\infty\underline{\mathfrak S}_{\theta_{(-,n)}}\right).\label{eq:system_2}
\end{align}
From the general theory of chapter \ref{chapter3}, we actually know the form of those matrices: those are given by
\begin{align}
&\underline{\mathfrak S}_{\theta_{(+,n)}}=\left(\begin{array}{cc}
1&d_nq^n\\
0&1
\end{array}\right),\\ 
&\underline{\mathfrak S}_{\theta_{\frac\pi2}}=\left(\begin{array}{cc}
\sum_{n=1}^\infty e_nq^n&0\\
0&\sum_{n=1}^\infty f_nq^n
\end{array}\right),\\
&\underline{\mathfrak S}_{\theta_{(+,n)}}=\left(\begin{array}{cc}
1&0\\
g_nq^n&1
\end{array}\right),
\end{align}
Using our computed matrices, we have
\begin{align}
\underline{\mathfrak S}_{I\to II}=\left(\begin{array}{cc}
1+q&\ii q\\
\ii q&1-q
\end{array}\right).
\end{align}
Imposing \ref{eq:system_2}, we obtain the only solution as
\begin{align}
&\underline{\mathfrak S}_{\theta_{(+,n)}}=\left(\begin{array}{cc}
1&\ii q^n\\
0&1
\end{array}\right),\\ 
&\underline{\mathfrak S}_{\frac\pi2}=\left(\begin{array}{cc}
\sum_{n=1}^\infty q^n&0\\
0&1-q
\end{array}\right),\\
&\underline{\mathfrak S}_{\theta_{(+,n)}}=\left(\begin{array}{cc}
1&0\\
 \ii q^n&1
\end{array}\right).
\end{align}
Those matrices reproduce exactly the matrices of subsection \ref{sec:phys_reg}. From the matrix $\underline{\mathfrak S}_{IV\to III}$ (always sweeping counterclockwise) we have that the matrices $\underline{\mathfrak S}_{(+,n)}$ and $\underline{\mathfrak S}_{(-,n)}$ have the same form also for $n<0$, and additionally
\begin{align}
&\underline{\mathfrak S}_{-\frac\pi2}=\left(\begin{array}{cc}
\sum_{n=1}^\infty q^n&0\\
0&1-q
\end{array}\right)
\end{align}
In the region $0<y<1$, the problem is way simpler, as all singularities are concentrated on the imaginary axis. We have
\begin{align}
&\underline{\mathfrak S}_{I\to II}=\left(\begin{array}{cc}
1&\ii(1+q)\\
0&1
\end{array}\right)=\underline{\mathfrak S}_{\frac\pi2},\\
&\underline{\mathfrak S}_{IV\to III}=\left(\begin{array}{cc}
1&0\\
\ii(1+q^{-1})&1
\end{array}\right)=\underline{\mathfrak S}_{-\frac\pi2}.
\end{align}
Again, those matrices coincide with the ones computed in subsection \ref{sec:phys_reg}, confirming our result. The presence of the $q$ terms in those matrices gives us information on figure \ref{fig:poles_b2}: even if in that figure we have marked only one start point of a branch cut, there is actually another such point, covered by the branch cut starting at the marked point.

We have confirmed numerically the correctness of our Stokes automorphisms, and using the expansions at rays $I$ and $II$ we have also confirmed the Debye expansions, also solving the ambiguity in the expansion of $Y$. Using those tools, we can obtain the asymptotic expansions of $J$ and $Y$ for $\arg a$ near $\frac\pi2^{\pm}$, as we have done in subsection \ref{sec:phys_reg}.

%% file: appbox.tex
\chapter[Numerical methods]{Numerical computation of eigenvalues and eigenfunctions}

\label{app:appbox}

In chapter \ref{chapter5}, we have tested our quantization conditions against energy levels obtained through numerics. In this short appendix, we will detail the numerical method that we have used. This is an elementary method based on discretizing the possible momenta by bounding the $x$ coordinate, and then cutting the allowed momenta to a finite set. We will apply this method to the problems examined in chapter \ref{chapter5}.

\section{The method}

The eigenvalue problem that we have to solve is
\begin{align}
\frac12\left(\psi(x+\ii\hbar,\hbar)+\psi(x-\ii\hbar,\hbar)\right)-\psi(x,\hbar)+V(x)\psi(x,\hbar)=E\psi(x,\hbar).\label{eq:problem}
\end{align}
The first step that we take in the numerical method is to bound the value of the $x$ coordinate on the real line: we will assume that $x\in(-L,L)$, for $L$ a large positive number. Furthermore, we will require our wavefunctions to vanish at $L$: we will look for solutions $\psi(x,\hbar)$ of $\eqref{eq:problem}$ such as $\psi(\pm L,\hbar)=0$, mimicking the $\mathbb L^2(\mathbb R)$ requirement in the original problem.

The boundary condition allows us to expand the wavefunctions in terms of a well-known basis of real functions: for $n$ strictly positive integer, we define
\begin{align}
\phi_n(x)=\begin{cases}
\frac{1}{\sqrt L}\cos\left(\frac{\pi}{2L}nx\right),&n\text{ odd},\\
\frac{1}{\sqrt L}\sin\left(\frac{\pi}{2L}nx\right),&n\text{ even}.
\end{cases}
\label{eq:basis_num}
\end{align}
This is an orthonormal basis for the standard product that we use in the thesis:
\begin{align}
\bra{\phi_m}\ket{\phi_n}=\int_{-L}^L(\phi_m(x))^*\phi_n(x)\dd x=\delta_{n,m}.
\end{align}
We now have to compute the action of the Hamiltonian
\begin{align}
H=\cosh p-1+V(x)
\end{align}
on the wavefunction basis $\psi_n(x)$. This is given by
\begin{align}
H\phi_n(x)=\begin{cases}
\frac{1}{\sqrt L}\cos\left(\frac{\pi}{2L}n x\right)\left(2\sinh\left(n\frac{\pi\hbar}{4L}\right)^2+V(x)\right),&n\text{ odd},\\
\frac{1}{\sqrt L}\sin\left(\frac{\pi}{2L}n x\right)\left(2\sinh\left(n\frac{\pi\hbar}{4L}\right)^2+V(x)\right),&n\text{ even}.
\end{cases}
\end{align}
We now define the Hamiltonian matrix $H_{nm}$ as
\begin{align}
H_{n,m}=\bra{\phi_n}H\ket{\phi_m}=\int_{-L}^L(\phi_n(x))^*H\phi_m(x)\dd x,
\end{align}
where we recall that the indices $n$ and $m$ go from $1$ to $\infty$. The last approximation that we make is to cut the matrix $H_{n,m}$ to finite size: we will restrict its indices to go from $1$ to a large positive integer $N_{max}$. With this truncations, we can solve \eqref{eq:problem} by looking for eigenvalues and eigenvectors of $H_{n,m}$ using standard numerical methods that apply to finite size matrices. Our approximation will depend on two parameters: $L$ and $N_{max}$, so we will have to check if the approximation is stable with respect to those parameters. In order to do that, we will obtain different eigenvalues for different values of $L$ and $N_{max}$ and fit a function of those two parameters to those eigenvalues.

We now proceed to apply the method to the two main problems for which we produced numerical results: the harmonic oscillator and the double well potential.

\section{Harmonic oscillator}

\begin{figure}
\centering
\includegraphics[width=0.49\textwidth]{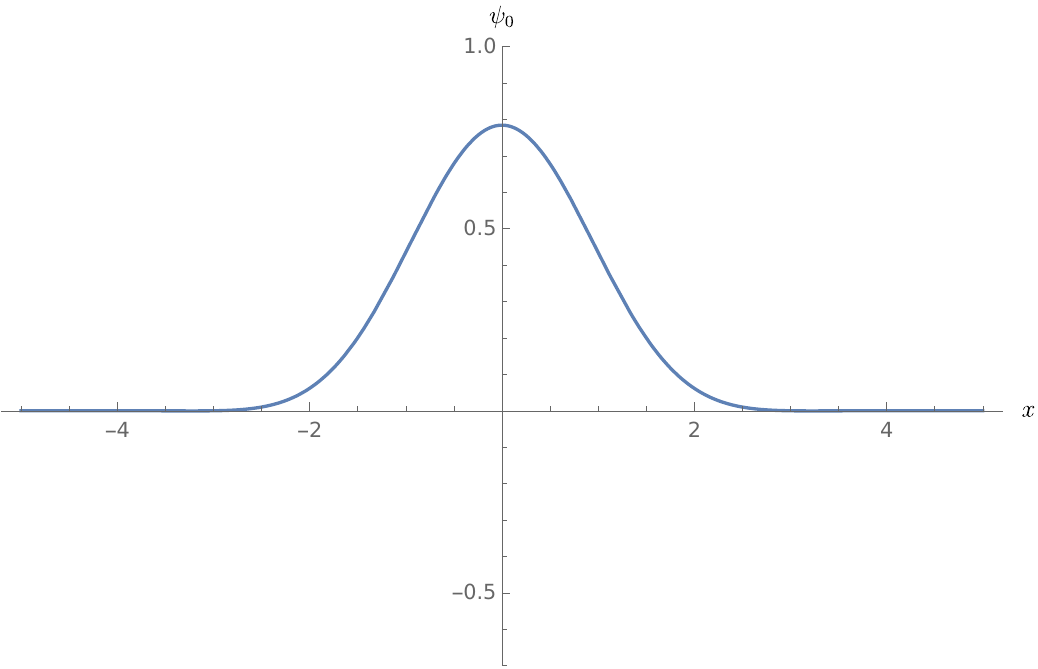}
\includegraphics[width=0.49\textwidth]{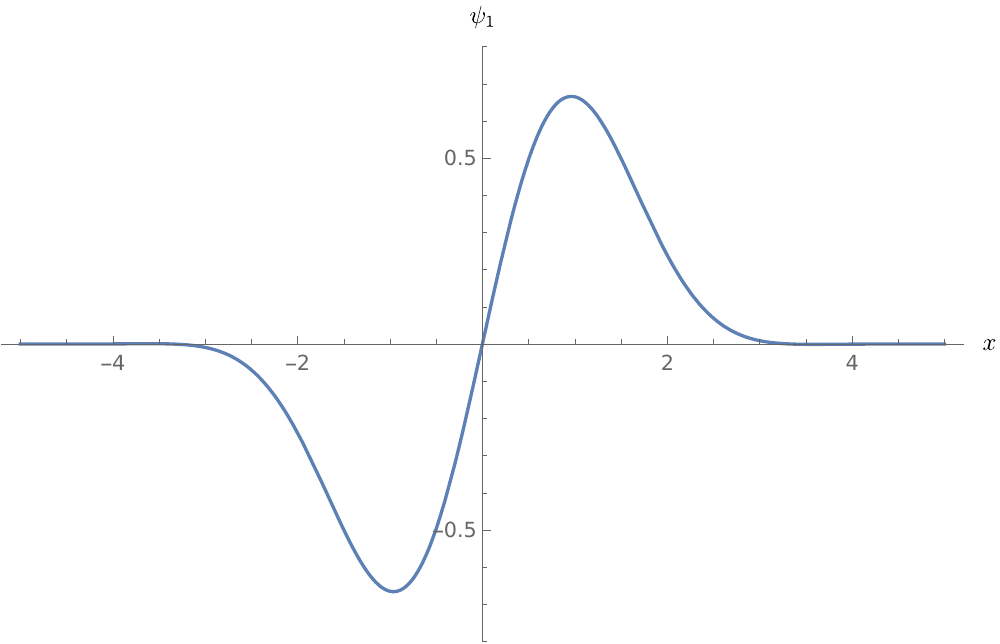}\\
\includegraphics[width=0.49\textwidth]{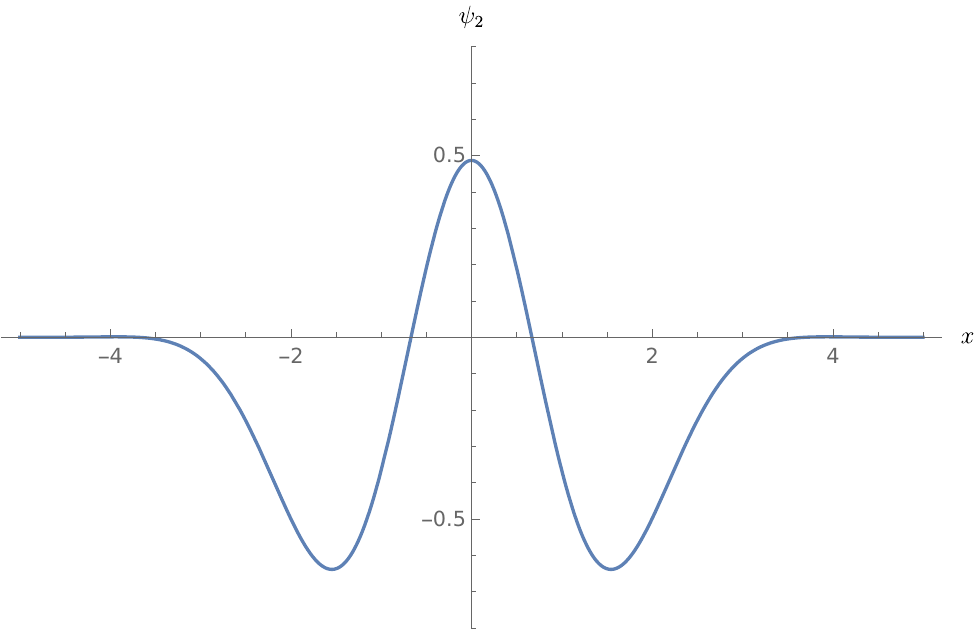}
\includegraphics[width=0.49\textwidth]{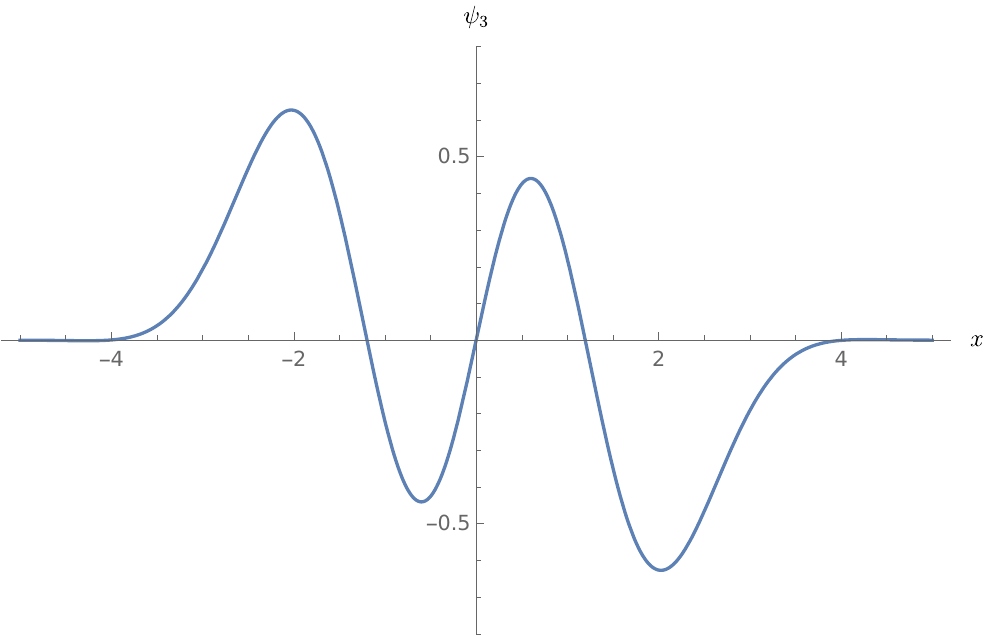}
\caption{First four eigenfunctions of the harmonic oscillator, with $\hbar=1$, $N_{max}=100$ and $L=30$.}
\label{fig:eigenfunctions}
\end{figure}

In this case, the potential is $V(x)=x^2$. For the computations in chapter \ref{chapter5}, we have chosen $N_{max}=100,$ $L=30$ and set $\hbar=1$. By diagonalizing the matrix, we obtain the eigenvalues listed in table \ref{tab:eigen_harm}. As we can see, in this case the energy eigenvalues that we have are non degenerate.
\begin{table}
\centering
\begin{tabular}{c|c}
$n$&$E_n$\\\hline
0&0.765157255\\
1&2.398081395\\
2&4.222746854\\
3&6.216966856\\
4&8.365788262\\
5&10.65819925\\
6&13.08563397\\
7&15.64117299\\
8&18.31906963\\
9&21.11444856
\end{tabular}
\caption{First energy levels $E_n$ as functions of $n$ for the harmonic oscillator, with $\hbar=1$, $N_{max}=100$ and $L=30$.}
\label{tab:eigen_harm}
\end{table}
Due to the fact that the potential is even, the eigenfunctions related to the $n-$th energy eigenvalue are even if $n$ is even and odd if $n$ is odd. By computing the eigenvectors of the Hamiltonian matrix, we obtain the coefficients that we can use to combine the basis functions into the eigenfunctions for the $n-$th energy level. We denote the normalized eigenfunction for the $n-$th energy level as $\psi_n(x)$. To normalize the phase of the eigenfunctions, we will impose $\psi_n(0)>0$ for $n$ even and $\psi_n'(0)>0$ for $n$ odd. We plot the first 4 eigenfunctions in figure \ref{fig:eigenfunctions}.

\section{Double well}

\begin{figure}
\centering
\includegraphics[width=0.49\textwidth]{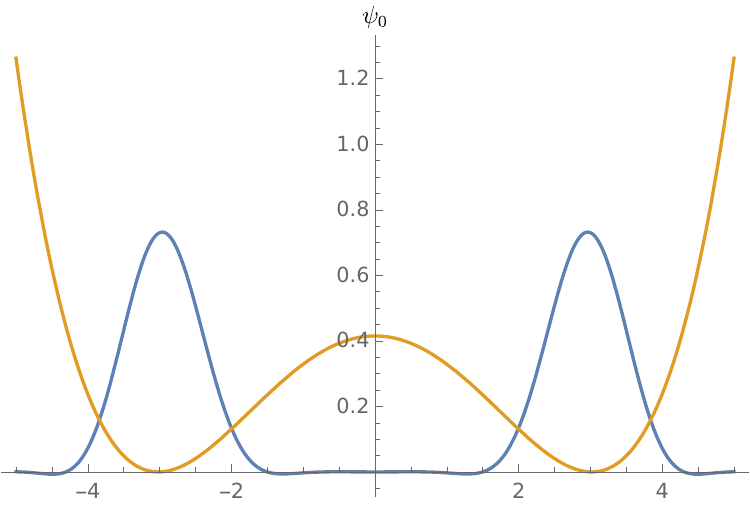}
\includegraphics[width=0.49\textwidth]{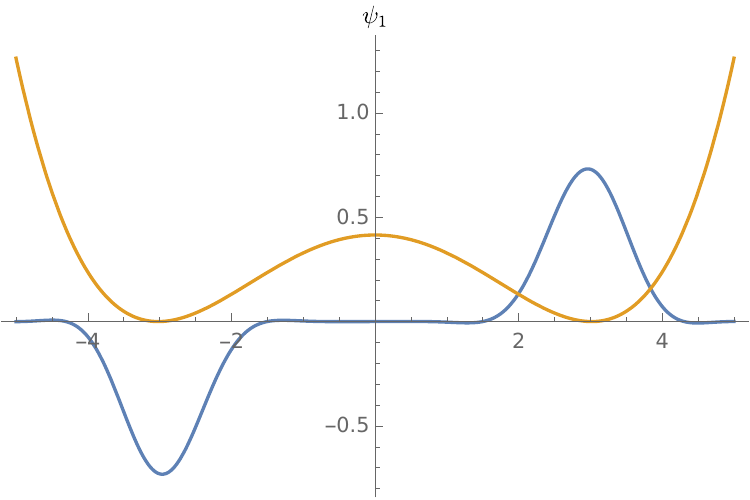}\\
\includegraphics[width=0.49\textwidth]{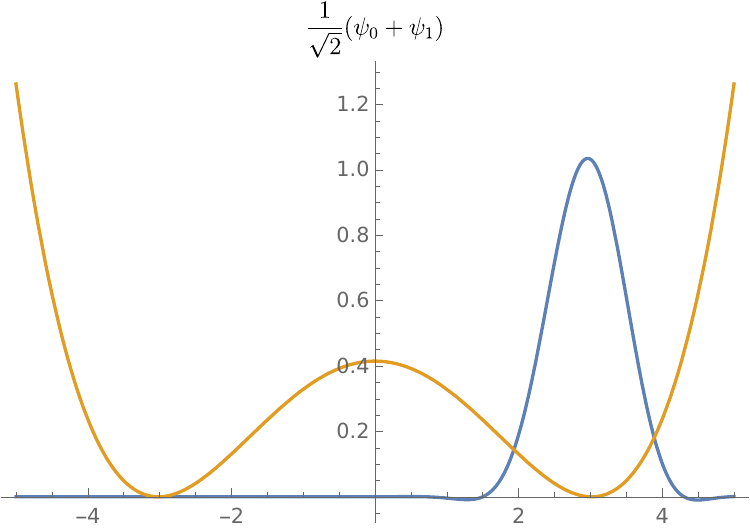}
\includegraphics[width=0.49\textwidth]{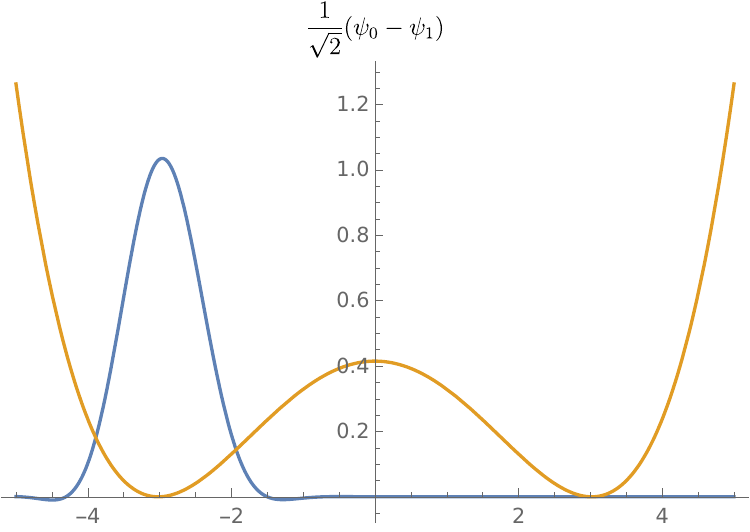}
\caption{First two eigenfunctions of the harmonic oscillator (on top), with $\hbar=1$, $N_{max}=100$, $L=30$, $a=\frac12$ and $h=9.11$. We plot in orange the (rescaled by a factor of $1/100$) potential in use. As the two energy levels share the same energy, we also plot the linear combinations $\psi_0+\psi_1$ and $\psi_0-\psi_1$, showing that in this case it is possible to have eigenfunctions centered around a single minimum of the potential.}
\label{fig:doublewelleigen}
\end{figure}

For the potential $V(x)=a(x^2-h)^2$, we choose two tests. We first test a point in moduli space where we expect a degeneracy, from figure \ref{fig:toda_four}. With $a=\frac12$ and $h=9.11$, we compute the first two energy levels as $E_0=3.8702$ and $E_1=3.8702$: as we can see, those levels are degenerate up to the numerical precision that we are using. In figure \ref{fig:doublewelleigen} we plot the first two energy eigenstates $\psi_0$ and $\psi_1$. As they are degenerate, any linear combination will be an eigenstate, so we also plot $\frac{1}{\sqrt2}(\psi_0+\psi_1)$ and $\frac1{\sqrt2}(\psi_0-\psi_1)$. Those two eigenfunctions are very similar to the eigenfunctions that can be obtained using standard perturbation theory around the energy minima, but they are exact eigenstates. We thus see that at some points of the moduli space quantum tunneling is suppressed, so we can have degenerate eigenstates centered around the various minima of the potential.

Next, we report that for the parameters $a=\frac12$ and $h=10$ we obtain the ground state energy level as $E_0=4.09799$ and the first excited state as $E_1=4.09932$. We will use those parameters in chapter \ref{chapter5} in order to assess the accuracy of our quantization conditions.

%% file: main.bbl
\begin{thebibliography}{10}

\bibitem{aniceto2019primer}
In{\^e}s Aniceto, G{\"o}k{\c{c}}e Ba{\c{s}}ar, and Ricardo Schiappa.
\newblock A primer on resurgent transseries and their asymptotics.
\newblock {\em Physics Reports}, 809:1--135, 2019.

\bibitem{balian2005quartic}
R~Balian, G~Parisi, and A~Voros.
\newblock Quartic oscillator.
\newblock In {\em Feynman Path Integrals: Proceedings of the International
  Colloquium Held in Marseille, May 1978}, pages 337--360. Springer, 2005.

\bibitem{berry1991hyperasymptotics}
Michael~Victor Berry and Christopher~J Howls.
\newblock Hyperasymptotics for integrals with saddles.
\newblock {\em Proceedings of the Royal Society of London. Series A:
  Mathematical and Physical Sciences}, 434(1892):657--675, 1991.

\bibitem{brillouin1926mecanique}
L{\'e}on Brillouin.
\newblock La m{\'e}canique ondulatoire de schr{\"o}dinger; une m{\'e}thode
  g{\'e}n{\'e}rale de r{\'e}solution par approximations successives.
\newblock {\em CR Acad. Sci}, 183(11):24--26, 1926.

\bibitem{caliceti2007useful}
Emanuela Caliceti, M~x Meyer-Hermann, Paolo Ribeca, A~x Surzhykov, and Ulrich~D
  Jentschura.
\newblock From useful algorithms for slowly convergent series to physical
  predictions based on divergent perturbative expansions.
\newblock {\em Physics reports}, 446(1-3):1--96, 2007.

\bibitem{caprini1999accelerated}
Irinel Caprini and Jan Fischer.
\newblock Accelerated convergence of perturbative qcd by optimal conformal
  mapping of the borel plane.
\newblock {\em Physical Review D}, 60(5):054014, 1999.

\bibitem{caprini2000convergence}
Irinel Caprini and Jan Fischer.
\newblock Convergence of the expansion of the laplace-borel integral in
  perturbative qcd improved by conformal mapping.
\newblock {\em Physical Review D}, 62(5):054007, 2000.

\bibitem{cedzich2021almost}
Christopher Cedzich, Jake Fillman, and Darren~C Ong.
\newblock Almost everything about the unitary almost mathieu operator.
\newblock {\em arXiv preprint arXiv:2112.03216}, 2021.

\bibitem{childs2009universal}
Andrew~M Childs.
\newblock Universal computation by quantum walk.
\newblock {\em Physical review letters}, 102(18):180501, 2009.

\bibitem{comtet2012advanced}
Louis Comtet.
\newblock {\em Advanced Combinatorics: The art of finite and infinite
  expansions}.
\newblock Springer Science \& Business Media, 2012.

\bibitem{costin2008asymptotics}
Ovidiu Costin.
\newblock {\em Asymptotics and Borel summability}.
\newblock Chapman and Hall/CRC, 2008.

\bibitem{delabaere1997exact}
Eric Delabaere, Herv{\'e} Dillinger, and Fr{\'e}d{\'e}ric Pham.
\newblock Exact semiclassical expansions for one-dimensional quantum
  oscillators.
\newblock {\em Journal of Mathematical Physics}, 38(12):6126--6184, 1997.

\bibitem{delabaere1999resurgent}
Eric Delabaere and Fr{\'e}d{\'e}ric Pham.
\newblock Resurgent methods in semi-classical asymptotics.
\newblock In {\em Annales de l'IHP Physique th{\'e}orique}, volume~71, pages
  1--94, 1999.

\bibitem{dingle1968wkb}
RB~Dingle and GJ~Morgan.
\newblock Wkb methods for difference equations i.
\newblock {\em Applied Scientific Research}, 18(1):221--237, 1968.

\bibitem{dingle1968wkb2}
RB~Dingle and GJ~Morgan.
\newblock Wkb methods for difference equations ii.
\newblock {\em Applied Scientific Research}, 18:238--245, 1968.

\bibitem{NIST:DLMF}
{\it NIST Digital Library of Mathematical Functions}.
\newblock http://dlmf.nist.gov/, Release 1.1.5 of 2022-03-15.
\newblock F.~W.~J. Olver, A.~B. {Olde Daalhuis}, D.~W. Lozier, B.~I. Schneider,
  R.~F. Boisvert, C.~W. Clark, B.~R. Miller, B.~V. Saunders, H.~S. Cohl, and
  M.~A. McClain, eds.

\bibitem{dunham1932wentzel}
JL~Dunham.
\newblock The wentzel-brillouin-kramers method of solving the wave equation.
\newblock {\em Physical Review}, 41(6):713, 1932.

\bibitem{ecalle1981fonctions}
J~Ecalle.
\newblock Les fonctions resurgentes (in french), vol. i--iii.
\newblock {\em Publ. Math. Orsay, France}, 1981.

\bibitem{farhi1998quantum}
Edward Farhi and Sam Gutmann.
\newblock Quantum computation and decision trees.
\newblock {\em Physical Review A}, 58(2):915, 1998.

\bibitem{fedotov2021wkb}
Alexander Fedotov and Fr{\'e}d{\'e}ric Klopp.
\newblock Wkb asymptotics of meromorphic solutions to difference equations.
\newblock {\em Applicable Analysis}, 100(7):1557--1573, 2021.

\bibitem{flaschka1976canonically}
H~Flaschka and DW~McLaughlin.
\newblock Canonically conjugate variables for the korteweg-de vries equation
  and the toda lattice with periodic boundary conditions.
\newblock {\em Progress of Theoretical Physics}, 55(2):438--456, 1976.

\bibitem{garoufalidis2020peacock}
Stavros Garoufalidis, Jie Gu, and Marcos Marino.
\newblock Peacock patterns and resurgence in complex chern-simons theory.
\newblock {\em arXiv preprint arXiv:2012.00062}, 2020.

\bibitem{garoufalidis2021resurgent}
Stavros Garoufalidis, Jie Gu, and Marcos Mari{\~n}o.
\newblock The resurgent structure of quantum knot invariants.
\newblock {\em Communications in Mathematical Physics}, 386(1):469--493, 2021.

\bibitem{gorsky1995integrability}
A~Gorsky, I~Krichever, A~Marshakov, A~Mironov, and A~Morozov.
\newblock Integrability and seiberg-witten exact solution.
\newblock {\em Physics Letters B}, 355(3-4):466--474, 1995.

\bibitem{grassi2020non}
Alba Grassi, Jie Gu, and Marcos Marino.
\newblock Non-perturbative approaches to the quantum seiberg-witten curve.
\newblock {\em Journal of High Energy Physics}, 2020(7):1--51, 2020.

\bibitem{grassi2019solvable}
Alba Grassi, Mari{\~n}o Marcos, et~al.
\newblock A solvable deformation of quantum mechanics.
\newblock {\em SIGMA. Symmetry, Integrability and Geometry: Methods and
  Applications}, 15:025, 2019.

\bibitem{howls1999resurgence}
CJ~Howls and AB~Olde Daalhuis.
\newblock On the resurgence properties of the uniform asymptotic expansion of
  bessel functions of large order.
\newblock {\em Proceedings of the Royal Society of London. Series A:
  Mathematical, Physical and Engineering Sciences}, 455(1991):3917--3930, 1999.

\bibitem{howls2004higher}
CJ~Howls, PJ~Langman, and AB~Olde~Daalhuis.
\newblock On the higher--order stokes phenomenon.
\newblock {\em Proceedings of the Royal Society of London. Series A:
  Mathematical, Physical and Engineering Sciences}, 460(2048):2285--2303, 2004.

\bibitem{kac1975explicitly}
M~Kac and Pierre van Moerbeke.
\newblock On an explicitly soluble system of nonlinear differential equations
  related to certain toda lattices.
\newblock {\em Advances in Mathematics}, 16(2):160--169, 1975.

\bibitem{kac2002quantum}
Victor~G Kac and Pokman Cheung.
\newblock {\em Quantum calculus}, volume 113.
\newblock Springer, 2002.

\bibitem{kashani2016quantization}
Amir-Kian Kashani-Poor.
\newblock Quantization condition from exact wkb for difference equations.
\newblock {\em Journal of High Energy Physics}, 2016(6):1--34, 2016.

\bibitem{kawai2005algebraic}
Takahiro Kawai and Yoshitsugu Takei.
\newblock {\em Algebraic analysis of singular perturbation theory}, volume 227.
\newblock American Mathematical Soc., 2005.

\bibitem{klemm1996nonperturbative}
Albrecht Klemm, Wolfgang Lerche, and Stefan Theisen.
\newblock Nonperturbative effective actions of (n= 2)-supersymmetric gauge
  theories.
\newblock {\em International Journal of Modern Physics A}, 11(11):1929--1973,
  1996.

\bibitem{klemm1995simple}
Albrecht Klemm, Wolfgang Lerche, Shimon Yankielowicz, and Stefan Theisen.
\newblock Simple singularities and n= 2 supersymmetric yang-mills theory.
\newblock {\em Physics Letters B}, 344(1-4):169--175, 1995.

\bibitem{kramers1926wellenmechanik}
Hendrik~Anthony Kramers.
\newblock Wellenmechanik und halbzahlige quantisierung.
\newblock {\em Zeitschrift f{\"u}r Physik}, 39(10-11):828--840, 1926.

\bibitem{kreshchuk2019picard}
Michael Kreshchuk and Tobias Gulden.
\newblock The picard--fuchs equation in classical and quantum physics:
  application to higher-order wkb method.
\newblock {\em Journal of Physics A: Mathematical and Theoretical},
  52(15):155301, 2019.

\bibitem{langer1937connection}
Rudolph~E Langer.
\newblock On the connection formulas and the solutions of the wave equation.
\newblock {\em Physical Review}, 51(8):669, 1937.

\bibitem{levy1992finite}
Hyman Levy and Freda Lessman.
\newblock {\em Finite difference equations}.
\newblock Courier Corporation, 1992.

\bibitem{marino2004houches}
Marcos Marino.
\newblock Les houches lectures on matrix models and topological strings.
\newblock {\em arXiv preprint hep-th/0410165}, 2004.

\bibitem{marino2021advanced}
Marcos Mari{\~n}o.
\newblock {\em Advanced Topics in Quantum Mechanics}.
\newblock Cambridge University Press, 2021.

\bibitem{martinec1996integrable}
Emil~J Martinec and Nicholas~P Warner.
\newblock Integrable systems and supersymmetric gauge theory.
\newblock {\em Nuclear Physics B}, 459(1-2):97--112, 1996.

\bibitem{olver1997asymptotics}
Frank Olver.
\newblock {\em Asymptotics and special functions}.
\newblock AK Peters/CRC Press, 1997.

\bibitem{pasquier1992periodic}
V~Pasquier and M~Gaudin.
\newblock The periodic toda chain and a matrix generalization of the bessel
  function recursion relations.
\newblock {\em Journal of Physics A: Mathematical and General}, 25(20):5243,
  oct 1992.

\bibitem{sauzin2007resurgent}
David Sauzin.
\newblock Resurgent functions and splitting problems.
\newblock {\em arXiv preprint arXiv:0706.0137}, 2007.

\bibitem{sauzin2014introduction}
David Sauzin.
\newblock Introduction to 1-summability and resurgence.
\newblock {\em arXiv preprint arXiv:1405.0356}, 2014.

\bibitem{seiberg1994electric}
Nathan Seiberg and Edward Witten.
\newblock Electric-magnetic duality, monopole condensation, and confinement in
  n= 2 supersymmetric yang-mills theory.
\newblock {\em Nuclear Physics B}, 426(1):19--52, 1994.

\bibitem{silverstone1985jwkb}
Harris~J Silverstone.
\newblock Jwkb connection-formula problem revisited via borel summation.
\newblock {\em Physical review letters}, 55(23):2523, 1985.

\bibitem{simon2000schrodinger}
Barry Simon.
\newblock Schr{\"o}dinger operators in the twenty-first century.
\newblock {\em Mathematical physics}, 2000:283--288, 2000.

\bibitem{sueishi2020exact}
Naohisa Sueishi, Syo Kamata, Tatsuhiro Misumi, and Mithat {\"U}nsal.
\newblock On exact-wkb analysis, resurgent structure, and quantization
  conditions.
\newblock {\em Journal of High Energy Physics}, 2020(12):1--51, 2020.

\bibitem{toda2012theory}
Morikazu Toda.
\newblock {\em Theory of nonlinear lattices}, volume~20.
\newblock Springer Science \& Business Media, 2012.

\bibitem{voros1983return}
Andr{\'e} Voros.
\newblock The return of the quartic oscillator. the complex wkb method.
\newblock In {\em Annales de l'IHP Physique th{\'e}orique}, volume~39, pages
  211--338, 1983.

\bibitem{watrous2001quantum}
John Watrous.
\newblock Quantum simulations of classical random walks and undirected graph
  connectivity.
\newblock {\em Journal of computer and system sciences}, 62(2):376--391, 2001.

\bibitem{wentzel1926verallgemeinerung}
Gregor Wentzel.
\newblock Eine verallgemeinerung der quantenbedingungen f{\"u}r die zwecke der
  wellenmechanik.
\newblock {\em Zeitschrift f{\"u}r Physik}, 38(6-7):518--529, 1926.

\bibitem{zwillinger2018crc}
Dan Zwillinger.
\newblock {\em CRC standard mathematical tables and formulas}.
\newblock chapman and hall/CRC, 2018.

\end{thebibliography}
